
%
%
%
\input harvmac.tex
\input pictex.tex
\let\includefigures=\iftrue
\includefigures
\def\fig#1#2{\topinsert\epsffile{#1}\noindent{#2}\endinsert}
\else
\def\fig#1#2{}
\fi
\def\Title#1#2{\rightline{#1}
\ifx\answ\bigans\nopagenumbers\pageno0\vskip1in%
\baselineskip 15pt plus 1pt minus 1pt
\else
\def\listrefs{\footatend\vskip 1in\immediate\closeout\rfile\writestoppt
\baselineskip=14pt\centerline{{\bf References}}\bigskip{\frenchspacing%
\parindent=20pt\escapechar=` \input
refs.tmp\vfill\eject}\nonfrenchspacing}
\pageno1\vskip.8in\fi \centerline{\titlefont #2}\vskip .5in}

\ifx\answ\bigans\def\tcbreak#1{}\else\def\tcbreak#1{\cr&{#1}}\fi
%

\def\inbar{\,\vrule height1.5ex width.4pt depth0pt}
\def\IB{\relax{\rm I\kern-.18em B}}
\def\IC{\relax\hbox{$\inbar\kern-.3em{\rm C}$}}
\def\IP{\relax{\rm I\kern-.18em P}}
\def\IR{\relax{\rm I\kern-.18em R}}

\def\sgn{{\rm sgn~}}

\def\D{\Delta}

\def\barint{-\hskip -11pt\int}
\def\D{\Delta}
\def\sgn{{\rm sgn}}
\def\cut{\hskip 2pt{\cal /}\hskip -6.8pt}
\def\sn{{\rm sn}}
\Title{\vbox{\baselineskip12pt
\vbox{\hbox{
\hfill\hbox{LPTENS-95/24}}}}}{Almost Flat Planar
Diagrams}
\vskip -120pt {\it in memoriam Claude Itzykson} \vskip 80pt
\bigskip
\bigskip
\centerline{Vladimir A. Kazakov}
\smallskip
\centerline{Matthias Staudacher $^\dagger$}
\smallskip
\centerline{{\it and}}
\smallskip
\centerline{Thomas Wynter $^\dagger$}\footnote~{
\hskip -11.5pt $^\dagger$ \hskip 4pt
This work is supported by funds provided by the European Community,
Human Capital and Mobility  Programme.}
\bigskip
\centerline{Laboratoire de Physique Th\'eorique de}
\centerline{l'\'Ecole Normale Sup\'erieure \footnote*{
Unit\'e Propre du
Centre National de la Recherche Scientifique,
associ\'ee \`a l'\'Ecole Normale Sup\'erieure et \`a
l'Universit\'e de Paris-Sud.}}
\bigskip
\noindent
We continue our study of matrix models of dually weighted graphs.
Among the attractive features of these models is the possibility to
interpolate between ensembles of regular and random two-dimensional
lattices, relevant for the study of the crossover from two-dimensional
flat space to two-dimensional quantum gravity.  We further develop the
formalism of large $N$ character expansions. In particular, a general
method for determining the large $N$ limit of a character is
derived. This method, aside from being potentially useful for a far
greater class of problems, allows us to exactly solve the matrix
models of dually weighted graphs, reducing them to a well-posed
Cauchy-Riemann problem.  The power of the method is illustrated by
explicitly solving a new model in which only positive curvature
defects are permitted on the surface, an arbitrary amount of negative
curvature being introduced at a single insertion.
\Date{June 1995}
\nref\BIPZ{E.~Br\'ezin, C.~Itzykson, G.~Parisi \& J.-B.~Zuber, Commun.
Math. Phys. 59 (1978), 35.}
\nref\IDiF{P.~Di~Francesco \& C. Itzykson, Ann. Inst. Henri.
Poincar\'e Vol. 59, no. 2 (1993) 117.}
\nref\ITZUB{C.~Itzykson \& J.-B.~Zuber, J.~Math.~Phys.~21(3) (1980) 411.}
\nref\KSW{V.A.~Kazakov, M.~Staudacher \& T.~Wynter, \'Ecole Normale
preprint LPTENS-95/9, accepted for publication in Commun.~Math.~Phys.}
\nref\DAVID{F.~David, Nucl. Phys. B257 (1985) 45.}
\nref\VOL{V.A.~Kazakov, Phys. Lett. B150 (1985) 282.}
\nref\FRO{J.~Fr\" ohlich, in: Lecture Notes in Physics, Vol. 216,
Springer, Berlin, 1985; \hfill\break
J.~Ambj{\o }rn, B.~Durhuus and J.~Fr\" ohlich,
Nucl. Phys. B257[FS14](1985) 433.}
\nref\BYRD{P.F.~Byrd \& M.D.~Friedman, ``Handbook of Elliptic
Integrals for Engineers and Physicists'', Springer, Berlin, 1954.}
\nref\LAWD{D.F.~Lawden, ``Elliptic Functions and Applications'',
Springer, New York, 1989.}

\newsec{Introduction}

Hermitian one matrix models were introduced and for the first time
solved in the large $N$ limit in the seminal paper by
Br\'ezin, Itzykson, Parisi and Zuber \BIPZ. These models generate
ensembles of planar, random graphs whose vertex coordination numbers
are controlled by the matrix potential. By varying the potential,
different classes of diagrams may be obtained, e.g.~random
square or random triangular lattices. However, despite this freedom,
there is a class of physically important lattices that can not
be generated by simply tuning the potential: {\it regular}, flat lattices
with fixed coordination numbers of both vertices and faces.
To attain them it is necessary to study planar
graphs having coordination number dependent weights for both the
vertices and faces. It is straightforward to define modified
hermitian matrix models producing such graphs, but they can no longer
be treated with the methods of \BIPZ. In fact, until very recently
this class of models of dually weighted graphs seemed
intractable. However, an important but little
noticed observation due to Itzykson and Di~Francesco \IDiF\ has made
possible the explicit treatment of dually weighted graphs.
The number of degrees of freedom of these models is crucially
reduced by rewriting the model in the language of group theory.
It should also be noted that this method, based on expanding the
matrix model potential in Weyl characters, was already used
presciently in a special case in another early paper by
Itzykson and Zuber \ITZUB.
In a recent work \KSW\ we demonstrated that this new approach leads
indeed to a problem amenable to mathematical analysis once the
large $N$ limit is taken.

The physical importance of matrix models has been elucidated through
a large body of work over the last ten years.
In \DAVID\ \VOL\ matrix models were first introduced to furnish a description
of two-dimensional quantum gravity and non-critical bosonic
strings and successfully used to calculate the critical properties
of these theories.
This approach is based on the representation of the sum over
world-sheet metrics as a sum over dynamical triangulations as originally
proposed in \DAVID\ \VOL\ \FRO. Studying the crossover from random, dynamical
graphs to regular, static graphs, then, will correspond to suppressing
the curvature fluctuations of the world-sheet metric and result in
a flat two-dimensional metric. Our work, in conjunction with \KSW,
should thus be seen as representing a first attempt towards establishing
a connection between integrable two-dimensional models both coupled to
and decoupled from quantum gravity.

To be precise, let us consider general planar graphs
and introduce a set of couplings
$t^*_1,t^*_2,...t^*_q,...\;$, namely the weights of vertices with
$1,2,...,q,...$
neighbours, and a dual set $t_1,t_2,...t_q,...\,$,
the weights of the dual
vertices (or faces) with appropriate coordination numbers.
The partition function of closed planar graphs $G$ is defined to be
\eqn\DWG{
Z(t^*,t)=
\sum_{G} \prod_{v^*_q,v_q \in G} {t_q^*}^{\# v^*_q}\ {t_q}^{ \# v_q}}
where $v_q^*,v_q$ are the vertices with $q$ neighbours on the original
and dual graph, respectively, and $\# v_q^*,\# v_q$ are the numbers of
such vertices in the given graph $G$.
Choosing $t_q^*=t_q=\delta_{q,4}$ the only
allowed graphs are regular square lattices (see Fig.~1.a).
\vskip 20pt
\hskip 15pt
\beginpicture
\setcoordinatesystem units <1.00000cm,1.00000cm>
\linethickness=1pt
\ellipticalarc axes ratio  0.025:0.025  360 degrees
	from  4.354 25.785 center at  4.329 25.785
\plot  0.737 25.737  1.223 25.737 /
\plot  0.978 25.495  0.978 25.980 /
\plot  1.259 24.424  2.409 24.147 /
\plot  1.099 24.306  2.246 24.014 /
\plot  0.919 24.172  2.081 23.874 /
\plot  1.465 24.445  0.707 23.948 /
\plot  0.711 24.031  1.899 23.719 /
\plot  1.759 24.361  1.024 23.859 /
\plot  2.076 24.280  1.346 23.770 /
\plot  2.379 24.196  1.657 23.677 /
\plot  1.672 25.976  2.159 25.976 /
\plot  1.668 25.491  1.668 25.976 /
\plot  2.163 25.487  2.163 25.971 /
\plot  1.672 25.491  2.159 25.491 /
\plot  4.578 25.785  4.094 25.785 /
\plot  4.341 24.467  3.954 24.003 /
\plot  4.348 24.462  4.756 23.990 /
\plot  5.652 24.115  5.961 24.481 /
\plot  6.045 24.477  6.390 24.081 /
\plot  9.059 25.432  9.059 26.094 /
\plot  9.036 24.242  8.835 23.781 /
\plot  9.042 24.238  9.836 24.403 /
\plot  9.042 24.229  9.114 24.299 /
\plot  9.036 24.229  8.410 24.581 /
\plot  8.594 23.603  8.454 23.273 /
\plot  9.356 25.929  8.782 25.599 /
\plot  9.356 25.595  8.782 25.929 /
\plot 11.081 26.158 10.839 25.741 /
\plot 10.842 25.737 11.083 25.315 /
\plot 11.817 25.741 11.576 26.162 /
\plot 11.572 26.160 11.085 26.160 /
\plot 11.091 25.311 11.576 25.311 /
\plot 11.578 25.315 11.819 25.737 /
\plot 10.751 23.243 10.939 23.616 /
\plot  4.208 24.295 	 4.233 24.232
	 4.307 24.136
	 4.407 24.024
	 4.488 23.934
	 4.597 23.812
	/
\plot  4.494 24.295 	 4.483 24.246
	 4.420 24.162
	 4.299 24.037
	 4.226 23.970
	 4.168 23.918
	 4.089 23.848
	/
\plot  4.053 24.122 	 4.060 24.060
	 4.104 23.991
	 4.157 23.929
	 4.265 23.823
	/
\plot  4.646 24.115 	 4.631 24.037
	 4.583 23.969
	 4.528 23.908
	 4.424 23.812
	/
\plot  5.967 26.041 	 5.941 25.940
	 5.931 25.885
	 5.928 25.802
	 5.931 25.718
	 5.941 25.666
	 5.967 25.569
	/
\plot  5.971 26.039 	 5.994 25.937
	 6.003 25.883
	 6.007 25.799
	 6.003 25.715
	 5.994 25.663
	 5.971 25.565
	/
\plot  5.846 24.346 	 5.840 24.333
	 5.844 24.312
	 5.855 24.291
	 5.944 24.194
	 6.013 24.135
	 6.083 24.079
	 6.158 24.020
	 6.216 23.973
	 6.295 23.910
	/
\plot  5.994 24.471 	 6.013 24.481
	 6.028 24.486
	 6.041 24.479
	 6.047 24.464
	 6.037 24.448
	 6.018 24.420
	 5.980 24.382
	 5.906 24.289
	 5.842 24.205
	 5.779 24.119
	 5.690 24.003
	/
\plot  5.709 24.183 	 5.702 24.170
	 5.704 24.149
	 5.723 24.113
	 5.800 24.016
	 5.870 23.949
	 5.944 23.884
	 6.016 23.833
	 6.075 23.794
	 6.155 23.741
	/
\plot  6.058 23.721 	 6.136 23.821
	 6.193 23.899
	 6.248 23.980
	 6.308 24.079
	 6.322 24.111
	 6.329 24.134
	 6.329 24.149
	 6.322 24.158
	/
\plot  5.992 24.471 	 5.977 24.483
	 5.971 24.486
	 5.961 24.479
	 5.958 24.471
	 5.965 24.458
	 5.977 24.441
	 6.079 24.359
	 6.147 24.297
	 6.215 24.238
	 6.310 24.149
	 6.392 24.079
	/
\plot  5.850 23.857 	 5.931 23.960
	 5.973 24.014
	 6.032 24.089
	 6.090 24.164
	 6.153 24.259
	 6.170 24.293
	 6.176 24.308
	 6.176 24.325
	 6.172 24.335
	/
\plot  8.443 23.281 	 8.465 23.362
	 8.485 23.431
	 8.522 23.540
	 8.559 23.617
	 8.598 23.671
	 8.704 23.754
	 8.770 23.784
	 8.835 23.793
	 8.901 23.774
	 8.964 23.732
	 9.059 23.633
	 9.097 23.562
	 9.130 23.465
	 9.145 23.403
	 9.160 23.330
	 9.175 23.245
	 9.191 23.146
	/
\plot  8.513 23.554 	 8.560 23.655
	 8.598 23.728
	 8.661 23.815
	 8.767 23.894
	 8.833 23.925
	 8.898 23.935
	 8.963 23.915
	 9.026 23.873
	 9.121 23.772
	 9.157 23.715
	 9.192 23.634
	 9.231 23.519
	 9.252 23.446
	 9.275 23.362
	/
\plot  8.598 23.743 	 8.642 23.827
	 8.677 23.887
	 8.733 23.959
	 8.840 24.038
	 8.907 24.070
	 8.970 24.081
	 9.059 24.026
	 9.137 23.933
	 9.203 23.832
	 9.251 23.714
	 9.281 23.619
	 9.299 23.559
	 9.320 23.489
	/
\plot  9.042 24.221 	 9.131 24.113
	 9.194 24.031
	 9.265 23.920
	 9.287 23.866
	 9.310 23.791
	 9.337 23.687
	 9.353 23.621
	 9.370 23.544
	/
\plot  9.047 23.639 	 9.083 23.740
	 9.104 23.793
	 9.155 23.904
	 9.191 23.992
	 9.275 24.026
	 9.347 24.044
	 9.419 24.060
	 9.541 24.085
	 9.663 24.111
	 9.746 24.125
	 9.812 24.137
	 9.902 24.153
	/
\plot  9.146 23.427 	 9.182 23.547
	 9.203 23.609
	 9.254 23.732
	 9.286 23.766
	 9.404 23.821
	 9.491 23.844
	 9.561 23.858
	 9.657 23.874
	 9.718 23.885
	 9.789 23.897
	 9.873 23.911
	 9.970 23.927
	/
\plot  9.330 24.354 	 9.320 24.293
	 9.358 24.215
	 9.401 24.118
	 9.440 24.020
	 9.475 23.922
	 9.504 23.821
	 9.512 23.748
	 9.515 23.689
	 9.519 23.609
	/
\plot  9.191 23.146 	 9.219 23.227
	 9.241 23.286
	 9.269 23.362
	 9.307 23.449
	 9.358 23.531
	 9.434 23.577
	 9.512 23.609
	 9.602 23.637
	 9.692 23.658
	 9.793 23.678
	 9.874 23.689
	 9.984 23.705
	/
\plot  9.781 24.439 	 9.840 24.403
	 9.894 24.283
	 9.925 24.160
	 9.946 24.046
	 9.963 23.931
	 9.970 23.855
	 9.975 23.794
	 9.980 23.711
	/
\plot  9.567 24.397 	 9.593 24.348
	 9.631 24.259
	 9.657 24.188
	 9.680 24.117
	 9.702 24.002
	 9.720 23.889
	 9.730 23.811
	 9.737 23.750
	 9.747 23.666
	/
\plot  9.030 24.227 	 8.956 24.174
	 8.902 24.134
	 8.835 24.081
	 8.788 24.033
	 8.704 23.937
	/
\plot  8.289 24.388 	 8.372 24.340
	 8.433 24.304
	 8.511 24.259
	 8.604 24.201
	 8.697 24.145
	 8.782 24.092
	 8.814 24.060
	 8.810 24.020
	 8.769 23.944
	 8.725 23.872
	 8.697 23.817
	 8.642 23.715
	/
\plot  8.211 24.172 	 8.289 24.126
	 8.347 24.091
	 8.420 24.047
	 8.513 23.995
	 8.604 23.937
	 8.647 23.893
	 8.659 23.865
	/
\plot  8.160 23.954 	 8.248 23.909
	 8.312 23.875
	 8.393 23.832
	 8.472 23.794
	 8.547 23.743
	 8.560 23.711
	 8.541 23.639
	 8.510 23.516
	 8.484 23.419
	 8.467 23.358
	 8.448 23.288
	/
\plot  8.892 24.376 	 8.820 24.348
	 8.740 24.265
	 8.674 24.172
	 8.651 24.091
	 8.632 24.009
	 8.607 23.929
	 8.587 23.864
	 8.560 23.777
	/
\plot  8.687 24.488 	 8.615 24.475
	 8.538 24.390
	 8.484 24.293
	 8.444 24.183
	 8.416 24.071
	 8.398 23.994
	 8.387 23.932
	 8.371 23.848
	/
\plot  8.587 24.543 	 8.671 24.501
	 8.732 24.470
	 8.810 24.431
	 8.916 24.374
	 9.025 24.320
	 9.119 24.299
	 9.226 24.316
	 9.330 24.342
	 9.450 24.368
	 9.567 24.392
	 9.644 24.408
	 9.704 24.421
	 9.785 24.439
	/
\plot  8.587 24.549 	 8.476 24.576
	 8.410 24.570
	 8.324 24.496
	 8.266 24.403
	 8.226 24.288
	 8.198 24.172
	 8.185 24.096
	 8.176 24.036
	 8.164 23.954
	/
\plot 11.396 23.108 	11.429 23.202
	11.454 23.271
	11.489 23.357
	11.533 23.440
	11.585 23.518
	11.692 23.594
	11.760 23.623
	11.830 23.645
	11.944 23.673
	12.059 23.696
	12.146 23.715
	12.217 23.729
	12.313 23.747
	/
\plot 11.352 23.385 	11.380 23.475
	11.402 23.540
	11.434 23.620
	11.483 23.702
	11.540 23.777
	11.640 23.846
	11.708 23.878
	11.777 23.906
	11.892 23.932
	12.006 23.954
	12.098 23.971
	12.171 23.985
	12.270 24.003
	/
\plot 10.541 23.963 	10.554 24.047
	10.565 24.109
	10.583 24.185
	10.613 24.303
	10.657 24.416
	10.719 24.471
	10.820 24.515
	10.869 24.511
	10.930 24.492
	/
\plot 10.742 23.836 	10.759 23.926
	10.771 23.993
	10.789 24.075
	10.816 24.177
	10.854 24.278
	10.924 24.350
	11.013 24.390
	11.079 24.392
	11.138 24.373
	/
\plot 10.657 24.412 	10.739 24.367
	10.798 24.335
	10.871 24.295
	10.958 24.253
	11.043 24.206
	11.115 24.126
	11.134 24.073
	11.125 23.999
	11.093 23.933
	11.055 23.872
	11.023 23.805
	10.960 23.681
	/
\plot 11.256 23.650 	11.303 23.765
	11.328 23.825
	11.360 23.890
	11.396 23.952
	11.472 24.052
	11.572 24.107
	11.686 24.138
	11.808 24.163
	11.930 24.185
	12.028 24.207
	12.105 24.224
	12.211 24.246
	/
\plot 11.282 24.295 	11.220 24.285
	11.138 24.246
	11.047 24.153
	11.008 24.062
	10.977 23.967
	10.947 23.878
	10.924 23.806
	10.892 23.709
	/
\plot 10.901 23.633 	10.950 23.722
	10.977 23.768
	11.034 23.848
	11.064 23.876
	11.151 23.893
	11.250 23.880
	11.335 23.840
	11.377 23.777
	11.408 23.697
	11.432 23.616
	11.455 23.529
	11.473 23.460
	11.496 23.366
	/
\plot 10.577 24.196 	10.658 24.151
	10.717 24.118
	10.789 24.075
	10.868 24.026
	10.943 23.971
	10.990 23.912
	11.002 23.853
	10.981 23.781
	10.954 23.725
	10.901 23.620
	/
\plot 10.869 24.511 	10.951 24.468
	11.011 24.436
	11.087 24.397
	11.175 24.351
	11.263 24.308
	11.326 24.280
	11.409 24.271
	11.491 24.270
	11.604 24.280
	11.723 24.305
	11.845 24.333
	11.939 24.349
	12.014 24.363
	12.116 24.382
	/
\plot 11.542 24.280 	11.578 24.238
	11.606 24.196
	11.642 24.113
	11.673 24.037
	11.695 23.964
	11.716 23.891
	11.734 23.797
	11.749 23.722
	11.769 23.620
	/
\plot 11.828 24.337 	11.839 24.333
	11.900 24.270
	11.934 24.191
	11.950 24.125
	11.963 24.060
	11.991 23.948
	12.007 23.859
	12.019 23.787
	12.035 23.688
	/
\plot 10.933 24.492 	11.059 24.423
	11.153 24.372
	11.271 24.310
	11.343 24.278
	/
\plot 12.093 24.388 	11.984 24.367
	11.891 24.348
	11.811 24.332
	11.743 24.319
	11.636 24.298
	11.557 24.282
	11.491 24.272
	/
\plot 10.541 23.961 	10.623 23.913
	10.683 23.877
	10.755 23.827
	10.825 23.779
	10.886 23.715
	10.894 23.641
	10.875 23.575
	10.820 23.461
	/
\plot 10.748 23.243 	10.777 23.334
	10.799 23.401
	10.831 23.484
	10.885 23.587
	10.958 23.681
	11.074 23.709
	11.174 23.690
	11.250 23.645
	11.284 23.589
	11.313 23.514
	11.336 23.438
	11.352 23.376
	11.370 23.281
	11.379 23.204
	11.390 23.099
	/
\plot 10.903 23.664 	10.940 23.744
	10.968 23.803
	11.005 23.876
	11.049 23.943
	11.102 24.009
	11.206 24.050
	11.274 24.050
	11.343 24.037
	11.447 23.956
	11.489 23.865
	11.519 23.772
	11.547 23.685
	11.567 23.615
	11.593 23.520
	/
\plot 12.093 24.390 	12.150 24.365
	12.200 24.289
	12.243 24.168
	12.257 24.088
	12.270 24.009
	12.283 23.917
	12.294 23.845
	12.308 23.747
	/
\put{$t_4^*$} [lB] at  0.845 24.989
\put{$t_4$} [lB] at  1.812 24.989
\put{(a) flat space} [lB] at  0.358 22.864
\put{(b) positive curvature} [lB] at  3.319 22.693
\put{(c) negative curvature } [lB] at  8.162 22.589
\put{$t_2^*$} [lB] at  4.096 25.066
\put{$t_2$} [lB] at  5.836 25.066
\put{$t_6^*$} [lB] at  8.896 25.082
\put{$t_6$} [lB] at 11.246 25.002
\linethickness=0pt
\putrectangle corners at  0.358 26.187 and 12.330 22.513
\endpicture
\vskip 15pt
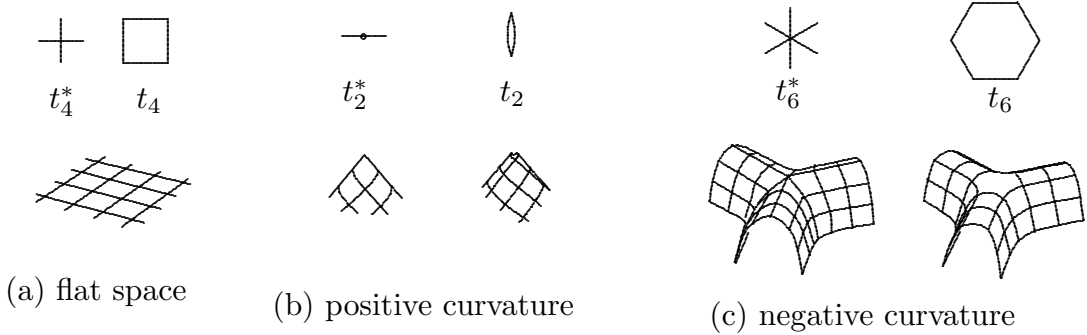
\centerline{{\bf Fig. 1:} Flat space and curvature defects}
\vskip 20pt
However, it is easy to see that a regular square lattice cannot be of
spherical (i.e.~planar) topology.  Positive curvature defects have to
be added in order to be able to close the surface. Considering for the
moment only even couplings, we must therefore ``turn on''
couplings $t_2$ or $t^*_2$, or both (see Fig.~1.b).  Exactly four such
defects are needed to close the square lattice.  Adding more
defects, then requires balancing the total curvature by also adding
negative curvature defects.  The simplest examples for such negative
defects, corresponding to the couplings $t_6$ and $t_6^*$, are shown
in Fig.~1.c.  Allowing for an arbitrary number of positive and
negative curvature defects we expect to generate random graphs which,
at critical values of the couplings, corresponding to very big graphs
dominating the sum in the partition function \DWG , allow us to reach
a continuum limit lying in the universality class of pure
two-dimensional quantum gravity \DAVID\ \VOL. On the other hand, having
``tuned away'' the negative curvature couplings $t_q$, $t_q^*$ with $q >
4$, no such continuum limit is possible. Then, only a small, finite number
of positive curvature defects are allowed; this brings us back to the phase of
essentially flat surfaces. The main physical motivation for studying
the models of dually weighted graphs, then, is to understand the
transition between these two very distinct phases.

In the present paper we continue to develop powerful techniques which
permit us to address this physical problem. Furthermore, we will
present the full and explicit solution of a non-trivial problem: the
case of flat, planar graphs with an arbitrary number of positive
curvature defects and a single negative curvature defect (see
Fig.~2(a)) adjusted to balance the total curvature.  We
call the resulting lattice surfaces ``almost flat planar diagrams''. A
typical surface of this type is shown in Fig. 2(b).

This model illustrates a non-trivial example that can be solved by the
method presented in this paper. This model cannot currently be solved
by standard matrix model techniques
\vskip 20pt
\hskip 35pt
\beginpicture
\setcoordinatesystem units <1.00000cm,1.00000cm>
\linethickness=1pt
\plot  3.114 23.755  2.487 22.655 /
\plot  2.411 23.732  3.224 22.661 /
\plot  1.966 22.816  3.613 23.597 /
\plot  1.899 23.533  3.759 22.885 /
\plot  1.691 23.148  3.909 23.281 /
\plot  2.805 23.209  2.841 22.437 /
\plot  2.845 22.441  3.014 22.210 /
\plot  3.198 22.559  3.406 22.282 /
\plot  2.843 22.441  2.709 22.202 /
\plot  2.498 22.555  2.309 22.250 /
\plot  2.362 22.339  2.400 22.559 /
\plot  2.421 22.437  2.453 22.606 /
\plot  2.955 23.057  3.306 22.670 /
\plot  2.680 23.040  2.373 22.598 /
\plot  3.112 23.125  3.787 22.896 /
\plot  2.587 23.122  1.918 22.807 /
\plot  2.595 23.216  1.683 23.180 /
\plot  1.941 23.535  2.614 23.290 /
\plot  3.890 23.307  3.004 23.237 /
\plot  3.562 23.601  2.978 23.307 /
\plot  2.707 23.385  2.534 23.660 /
\plot  2.870 23.396  2.963 23.611 /
\plot  3.664 22.646  3.857 22.574 /
\plot  3.854 22.746  3.926 22.750 /
\plot  3.410 22.447  3.613 22.377 /
\plot  2.095 22.329  2.305 22.428 /
\plot  1.740 22.623  1.854 22.631 /
\plot  1.865 22.473  2.051 22.568 /
\plot  1.700 22.845  1.721 22.847 /
\plot  3.935 22.915  3.816 22.915 /
\plot  1.740 22.792  1.869 22.803 /
\plot  1.825 23.360  2.176 23.230 /
\plot  2.297 23.592  2.447 23.381 /
\plot  3.234 23.635  3.118 23.410 /
\plot  3.700 23.453  3.346 23.290 /
\plot  3.310 22.418  3.287 22.585 /
\plot  3.243 22.502  3.224 22.585 /
\plot  2.024 22.559  2.030 22.610 /
\plot  3.700 22.638  3.685 22.701 /
\plot  2.328 25.055  3.238 25.055 /
\plot  2.347 24.915  3.217 25.195 /
\plot  2.413 24.788  3.154 25.322 /
\plot  2.515 24.685  3.052 25.425 /
\plot  2.642 24.621  2.925 25.489 /
\plot  2.781 24.600  2.786 25.512 /
\plot  2.923 24.621  2.644 25.491 /
\plot  3.050 24.685  2.515 25.425 /
\plot  3.154 24.786  2.415 25.326 /
\plot  3.217 24.911  2.350 25.197 /
\plot  6.911 22.991  7.806 23.717 /
\plot 10.513 24.077 10.274 22.682 /
\plot  9.569 23.679  9.569 23.605 /
\plot 10.566 23.796 10.645 23.743 /
\plot  6.541 23.743  7.184 24.970 /
\plot  8.213 23.851  8.213 23.554 /
\plot  8.365 23.679  8.365 23.400 /
\plot  8.534 23.476  8.560 23.247 /
\plot  8.702 23.311  8.750 23.133 /
\plot  7.700 25.394  8.778 25.453 /
\plot  8.879 23.334  8.956 23.347 /
\plot  9.292 25.749 10.406 25.620 /
\plot 10.490 25.034 11.119 24.024 /
\plot  2.301 22.261 	 2.313 22.364
	 2.322 22.420
	 2.347 22.525
	 2.398 22.625
	 2.487 22.646
	 2.578 22.602
	 2.621 22.510
	 2.648 22.418
	 2.670 22.340
	 2.685 22.280
	 2.705 22.197
	/
\plot  3.018 22.206 	 3.021 22.284
	 3.025 22.343
	 3.035 22.418
	 3.067 22.523
	 3.120 22.623
	 3.219 22.665
	 3.304 22.665
	 3.341 22.603
	 3.357 22.536
	 3.377 22.449
	 3.388 22.379
	 3.401 22.284
	/
\plot  1.858 22.483 	 1.866 22.593
	 1.873 22.653
	 1.888 22.729
	 1.922 22.803
	 1.969 22.816
	 2.011 22.765
	 2.049 22.664
	 2.074 22.562
	 2.085 22.480
	 2.090 22.416
	 2.095 22.329
	/
\plot  3.613 22.377 	 3.618 22.466
	 3.622 22.531
	 3.632 22.610
	 3.650 22.714
	 3.681 22.816
	 3.747 22.885
	 3.789 22.896
	 3.831 22.820
	 3.838 22.739
	 3.846 22.682
	 3.857 22.574
	/
\plot  1.736 22.623 	 1.736 22.697
	 1.735 22.751
	 1.729 22.818
	 1.723 22.930
	 1.710 23.040
	 1.702 23.122
	 1.691 23.180
	 1.683 23.122
	 1.691 22.998
	 1.694 22.945
	 1.698 22.847
	/
\plot  3.255 22.727 	 3.158 22.807
	 3.101 22.822
	 3.046 22.786
	 3.010 22.710
	 2.989 22.631
	 2.971 22.510
	 2.964 22.415
	 2.961 22.354
	 2.957 22.284
	/
\plot  3.128 22.864 	 3.035 22.957
	 2.982 22.976
	 2.957 22.945
	 2.929 22.822
	 2.922 22.754
	 2.915 22.661
	 2.912 22.602
	 2.909 22.533
	 2.907 22.452
	 2.904 22.358
	/
\plot  3.545 22.983 	 3.478 22.989
	 3.435 22.911
	 3.423 22.832
	 3.421 22.752
	 3.431 22.639
	 3.446 22.549
	 3.467 22.426
	/
\plot  3.304 23.059 	 3.224 23.082
	 3.192 23.082
	 3.175 23.050
	 3.189 22.978
	 3.213 22.911
	 3.229 22.858
	 3.260 22.761
	/
\plot  2.138 22.911 	 2.239 22.926
	 2.265 22.871
	 2.276 22.783
	 2.280 22.697
	 2.270 22.592
	 2.258 22.510
	 2.242 22.399
	/
\plot  2.371 23.017 	 2.460 23.025
	 2.472 22.976
	 2.464 22.871
	 2.455 22.826
	 2.436 22.744
	/
\plot  2.024 23.194 	 2.047 23.171
	 2.047 23.091
	 2.024 22.991
	 1.977 22.871
	/
\plot  2.328 23.205 	 2.358 23.188
	 2.362 23.122
	 2.322 23.025
	/
\plot  2.214 23.436 	 2.184 23.436
	 2.093 23.385
	 2.050 23.331
	 1.988 23.218
	/
\plot  2.474 23.343 	 2.443 23.347
	 2.343 23.294
	 2.275 23.226
	/
\plot  2.635 23.497 	 2.574 23.571
	 2.523 23.578
	 2.464 23.550
	 2.424 23.504
	 2.364 23.410
	/
\plot  2.940 23.559 	 2.978 23.588
	 3.020 23.588
	 3.076 23.571
	 3.152 23.506
	 3.192 23.448
	/
\plot  2.874 23.400 	 2.904 23.431
	 2.938 23.436
	 3.035 23.410
	 3.065 23.391
	/
\plot  3.315 23.472 	 3.363 23.476
	 3.450 23.431
	 3.550 23.311
	/
\plot  3.101 23.364 	 3.137 23.368
	 3.232 23.324
	 3.287 23.283
	/
\plot  3.537 23.277 	 3.558 23.262
	 3.615 23.156
	 3.643 23.092
	 3.687 22.970
	/
\plot  3.217 23.252 	 3.238 23.239
	 3.300 23.163
	 3.368 23.074
	/
\plot  1.952 23.527 	 1.897 23.531
	 1.865 23.480
	 1.830 23.402
	 1.801 23.321
	 1.789 23.197
	/
\plot  2.242 23.444 	 2.265 23.535
	 2.298 23.623
	 2.347 23.707
	 2.409 23.734
	 2.474 23.721
	 2.563 23.611
	/
\plot  3.550 23.592 	 3.613 23.597
	 3.664 23.537
	 3.698 23.459
	 3.720 23.395
	 3.747 23.307
	/
\plot  2.940 23.565 	 2.994 23.677
	 3.037 23.730
	 3.109 23.755
	 3.179 23.717
	 3.229 23.643
	 3.257 23.582
	 3.291 23.497
	/
\plot  3.886 23.311 	 3.909 23.283
	 3.920 23.218
	 3.926 23.121
	 3.926 23.023
	 3.928 22.928
	 3.928 22.853
	 3.926 22.750
	/
\plot  2.625 22.511 	 2.692 22.638
	 2.769 22.731
	 2.826 22.752
	 2.915 22.722
	 2.995 22.659
	 3.076 22.543
	/
\plot  2.671 22.354 	 2.726 22.447
	 2.777 22.532
	 2.834 22.574
	 2.908 22.538
	 2.963 22.464
	 3.027 22.377
	/
\plot  3.649 22.716 	 3.563 22.743
	 3.500 22.763
	 3.421 22.786
	 3.327 22.815
	 3.232 22.839
	 3.203 22.835
	/
\plot  3.624 22.562 	 3.498 22.597
	 3.433 22.614
	 3.372 22.631
	/
\plot  3.935 23.084 	 3.828 23.080
	 3.750 23.076
	 3.653 23.070
	 3.587 23.064
	 3.463 23.050
	/
\plot  2.060 22.642 	 2.142 22.679
	 2.202 22.706
	 2.280 22.739
	 2.365 22.776
	 2.453 22.807
	 2.487 22.811
	/
\plot  2.089 22.485 	 2.208 22.541
	 2.271 22.572
	 2.341 22.608
	/
\plot  1.721 22.976 	 1.836 22.982
	 1.921 22.987
	 2.028 22.993
	 2.118 23.000
	 2.191 23.005
	 2.290 23.012
	/
\plot  2.870 23.391 	 2.830 23.319
	 2.798 23.298
	 2.758 23.319
	 2.711 23.381
	 2.673 23.421
	 2.627 23.431
	 2.559 23.400
	 2.523 23.360
	/
\plot  2.498 22.773 	 2.582 22.811
	 2.654 22.752
	 2.692 22.649
	 2.711 22.543
	 2.726 22.450
	 2.735 22.375
	 2.747 22.274
	/
\plot  2.589 22.907 	 2.663 22.964
	 2.720 22.896
	 2.750 22.785
	 2.771 22.672
	 2.781 22.565
	 2.788 22.479
	 2.796 22.360
	/
\plot  7.857 23.755 	 7.924 23.674
	 7.983 23.604
	 8.077 23.494
	 8.147 23.415
	 8.202 23.360
	 8.285 23.287
	 8.393 23.199
	 8.504 23.114
	 8.596 23.053
	 8.665 23.020
	 8.759 22.982
	 8.820 22.959
	 8.891 22.934
	 8.974 22.905
	 9.070 22.873
	/
\plot  9.097 22.873 	 9.181 22.977
	 9.254 23.066
	 9.318 23.142
	 9.372 23.206
	 9.461 23.304
	 9.531 23.372
	 9.599 23.428
	 9.684 23.492
	 9.782 23.560
	 9.885 23.630
	 9.990 23.699
	10.089 23.765
	10.177 23.823
	10.249 23.872
	10.341 23.938
	10.415 23.992
	10.516 24.064
	/
\plot  9.110 22.862 	 9.141 22.784
	 9.165 22.727
	 9.201 22.655
	 9.263 22.557
	 9.341 22.466
	 9.432 22.418
	 9.531 22.388
	 9.637 22.403
	/
\plot 10.454 23.694 	10.389 23.717
	10.274 23.679
	10.192 23.636
	10.091 23.577
	 9.992 23.515
	 9.914 23.463
	 9.839 23.407
	 9.745 23.333
	 9.653 23.257
	 9.582 23.194
	 9.511 23.118
	 9.426 23.018
	 9.342 22.917
	 9.277 22.837
	 9.172 22.710
	 9.135 22.619
	 9.097 22.530
	 9.037 22.450
	 8.973 22.375
	 8.896 22.321
	 8.816 22.274
	 8.754 22.256
	 8.636 22.233
	/
\plot 10.401 23.410 	10.300 23.400
	10.195 23.358
	10.092 23.311
	10.027 23.274
	 9.944 23.226
	 9.862 23.176
	 9.798 23.133
	 9.695 23.025
	 9.597 22.915
	 9.499 22.812
	 9.404 22.710
	 9.339 22.626
	 9.277 22.543
	 9.210 22.450
	 9.136 22.363
	 9.036 22.305
	 8.928 22.261
	 8.864 22.248
	 8.738 22.233
	/
\plot  8.086 23.463 	 8.174 23.345
	 8.240 23.259
	 8.329 23.156
	 8.396 23.098
	 8.485 23.029
	 8.575 22.964
	 8.647 22.915
	 8.704 22.883
	 8.778 22.847
	 8.851 22.809
	 8.907 22.773
	 8.999 22.660
	 9.083 22.543
	 9.131 22.477
	 9.186 22.415
	 9.277 22.369
	 9.351 22.344
	 9.453 22.314
	/
\plot  7.650 23.643 	 7.662 23.525
	 7.741 23.418
	 7.832 23.321
	 7.917 23.225
	 8.028 23.105
	 8.143 22.987
	 8.240 22.900
	 8.319 22.847
	 8.423 22.786
	 8.528 22.728
	 8.611 22.682
	 8.673 22.646
	 8.752 22.601
	 8.830 22.555
	 8.890 22.515
	 8.985 22.431
	 9.083 22.350
	 9.162 22.321
	 9.226 22.305
	 9.313 22.286
	/
\plot  7.474 23.514 	 7.449 23.423
	 7.501 23.308
	 7.576 23.209
	 7.654 23.117
	 7.758 23.007
	 7.867 22.902
	 7.959 22.824
	 8.084 22.741
	 8.163 22.694
	 8.247 22.647
	 8.332 22.600
	 8.413 22.558
	 8.487 22.521
	 8.547 22.492
	 8.606 22.467
	 8.681 22.439
	 8.757 22.412
	 8.816 22.388
	 8.929 22.329
	 9.047 22.274
	 9.146 22.261
	/
\plot  7.279 23.372 	 7.255 23.285
	 7.258 23.233
	 7.330 23.120
	 7.419 23.027
	 7.498 22.945
	 7.599 22.842
	 7.703 22.742
	 7.794 22.670
	 7.916 22.602
	 7.993 22.566
	 8.075 22.530
	 8.158 22.495
	 8.237 22.464
	 8.307 22.437
	 8.365 22.415
	 8.448 22.390
	 8.553 22.362
	 8.657 22.336
	 8.738 22.314
	 8.828 22.283
	 8.898 22.257
	 8.994 22.223
	/
\plot  7.106 23.194 	 7.091 23.078
	 7.119 23.015
	 7.167 22.947
	 7.269 22.837
	 7.371 22.762
	 7.457 22.710
	 7.576 22.644
	/
\plot  6.911 22.991 	 7.016 22.936
	 7.107 22.889
	 7.185 22.848
	 7.252 22.813
	 7.358 22.759
	 7.436 22.718
	 7.558 22.659
	 7.633 22.622
	 7.713 22.583
	 7.794 22.544
	 7.870 22.508
	 7.995 22.454
	 8.122 22.409
	 8.200 22.384
	 8.283 22.358
	 8.366 22.334
	 8.445 22.310
	 8.515 22.290
	 8.572 22.274
	 8.643 22.255
	 8.700 22.242
	 8.778 22.223
	/
\plot  6.934 23.002 	 7.007 23.082
	 7.070 23.150
	 7.172 23.258
	 7.247 23.333
	 7.305 23.385
	 7.378 23.442
	 7.480 23.513
	 7.546 23.556
	 7.623 23.606
	 7.715 23.664
	 7.821 23.730
	/
\plot  7.180 22.849 	 7.258 22.849
	 7.364 22.914
	 7.461 22.991
	 7.521 23.032
	 7.597 23.078
	 7.681 23.129
	 7.769 23.186
	 7.854 23.248
	 7.931 23.315
	 7.993 23.387
	 8.035 23.463
	 8.020 23.542
	/
\plot  7.461 22.710 	 7.525 22.710
	 7.607 22.742
	 7.704 22.793
	 7.797 22.851
	 7.870 22.900
	 7.940 22.955
	 8.026 23.028
	 8.107 23.108
	 8.164 23.182
	 8.184 23.243
	 8.202 23.360
	/
\plot  7.751 22.566 	 7.862 22.580
	 7.921 22.593
	 7.984 22.621
	 8.060 22.661
	 8.134 22.706
	 8.191 22.748
	 8.247 22.805
	 8.311 22.883
	 8.369 22.965
	 8.405 23.040
	 8.412 23.091
	 8.405 23.182
	/
\plot  7.995 22.439 	 8.114 22.461
	 8.177 22.479
	 8.269 22.531
	 8.354 22.593
	 8.479 22.711
	 8.543 22.783
	 8.585 22.849
	 8.603 22.921
	 8.608 22.977
	 8.611 23.053
	/
\plot  8.255 22.375 	 8.345 22.390
	 8.395 22.403
	 8.460 22.429
	 8.539 22.469
	 8.616 22.513
	 8.674 22.557
	 8.761 22.664
	 8.829 22.786
	 8.842 22.843
	 8.854 22.951
	/
\plot  8.534 22.286 	 8.632 22.331
	 8.703 22.365
	 8.788 22.415
	 8.898 22.503
	 8.994 22.608
	 9.034 22.694
	 9.056 22.765
	 9.083 22.862
	/
\plot 10.361 23.133 	10.272 23.146
	10.219 23.144
	10.134 23.107
	10.039 23.047
	 9.948 22.981
	 9.876 22.928
	 9.797 22.860
	 9.700 22.770
	 9.605 22.680
	 9.531 22.608
	 9.454 22.524
	 9.377 22.439
	 9.316 22.374
	 9.250 22.314
	 9.189 22.277
	 9.070 22.223
	/
\plot 10.312 22.915 	10.211 22.900
	10.115 22.853
	10.004 22.781
	 9.897 22.705
	 9.813 22.644
	 9.745 22.592
	 9.659 22.523
	 9.574 22.455
	 9.504 22.403
	 9.424 22.349
	 9.341 22.299
	 9.222 22.250
	/
\plot 10.287 22.670 	10.200 22.658
	10.136 22.649
	10.058 22.634
	 9.979 22.607
	 9.881 22.570
	 9.785 22.529
	 9.711 22.492
	 9.614 22.415
	 9.519 22.337
	 9.471 22.313
	 9.377 22.274
	/
\plot 10.287 23.884 	10.221 23.805
	10.174 23.746
	10.122 23.666
	10.090 23.604
	10.054 23.524
	10.023 23.441
	10.003 23.372
	 9.994 23.279
	 9.992 23.162
	 9.993 23.044
	 9.995 22.951
	 9.987 22.837
	 9.995 22.718
	10.044 22.634
	/
\plot 10.044 23.730 	 9.939 23.627
	 9.887 23.565
	 9.843 23.479
	 9.797 23.367
	 9.758 23.252
	 9.737 23.156
	 9.732 23.043
	 9.734 22.976
	 9.739 22.905
	 9.748 22.834
	 9.761 22.766
	 9.798 22.655
	 9.906 22.566
	/
\plot  9.608 23.448 	 9.578 23.346
	 9.559 23.269
	 9.546 23.171
	 9.551 23.091
	 9.568 22.994
	 9.589 22.899
	 9.608 22.824
	 9.625 22.763
	 9.648 22.687
	 9.677 22.613
	 9.711 22.557
	 9.775 22.515
	/
\plot  9.404 23.233 	 9.390 23.157
	 9.381 23.101
	 9.377 23.027
	 9.393 22.957
	 9.421 22.871
	 9.453 22.786
	 9.481 22.718
	 9.540 22.595
	 9.622 22.479
	 9.686 22.454
	/
\plot  9.237 23.027 	 9.222 22.915
	 9.263 22.808
	 9.313 22.710
	 9.374 22.617
	 9.442 22.530
	 9.504 22.476
	 9.569 22.426
	 9.622 22.403
	/
\plot 10.325 24.765 	10.312 24.653
	10.337 24.561
	10.380 24.453
	10.430 24.349
	10.478 24.270
	10.536 24.197
	10.616 24.113
	10.700 24.033
	10.772 23.975
	10.873 23.925
	/
\plot  9.929 24.191 	 9.921 24.084
	 9.929 24.024
	 9.955 23.976
	10.018 23.897
	/
\plot  9.737 23.937 	 9.734 23.860
	 9.735 23.804
	 9.749 23.730
	 9.826 23.643
	/
\plot  7.051 24.793 	 7.140 24.782
	 7.217 24.740
	 7.307 24.680
	 7.393 24.616
	 7.461 24.564
	 7.560 24.479
	 7.681 24.368
	 7.801 24.256
	 7.895 24.166
	 7.978 24.084
	 8.072 23.988
	 8.181 23.875
	 8.242 23.811
	 8.309 23.741
	 8.380 23.666
	 8.458 23.584
	 8.542 23.494
	 8.632 23.398
	 8.730 23.293
	 8.835 23.180
	 8.948 23.059
	 9.008 22.994
	 9.070 22.928
	/
\plot  7.180 25.011 	 7.229 24.913
	 7.266 24.841
	 7.319 24.754
	 7.412 24.643
	 7.491 24.559
	 7.599 24.445
	/
\plot  6.945 24.566 	 7.061 24.553
	 7.133 24.519
	 7.215 24.470
	 7.302 24.410
	 7.390 24.344
	 7.477 24.276
	 7.558 24.210
	 7.629 24.150
	 7.688 24.100
	 7.762 24.038
	 7.846 23.965
	 7.942 23.878
	 8.054 23.774
	 8.116 23.715
	 8.183 23.650
	 8.256 23.580
	 8.334 23.504
	 8.419 23.422
	 8.510 23.333
	 8.607 23.236
	 8.712 23.133
	/
\plot  6.797 24.344 	 6.893 24.362
	 6.949 24.359
	 7.014 24.333
	 7.081 24.296
	 7.151 24.248
	 7.227 24.187
	 7.310 24.110
	 7.403 24.015
	 7.508 23.902
	 7.565 23.838
	 7.626 23.768
	/
\plot  6.691 24.113 	 6.799 24.132
	 6.860 24.128
	 6.975 24.079
	 7.038 24.041
	 7.108 23.992
	 7.186 23.929
	 7.273 23.853
	 7.373 23.762
	 7.487 23.654
	/
\plot  6.617 23.961 	 6.718 23.946
	 6.814 23.902
	 6.933 23.828
	 7.005 23.777
	 7.088 23.715
	 7.183 23.641
	 7.292 23.554
	/
\plot  6.551 23.730 	 6.663 23.703
	 6.745 23.680
	 6.845 23.643
	 6.968 23.560
	 7.059 23.487
	 7.180 23.385
	/
\plot  7.307 24.771 	 7.345 24.673
	 7.355 24.617
	 7.353 24.529
	 7.336 24.421
	 7.310 24.315
	 7.281 24.232
	 7.224 24.124
	 7.185 24.061
	 7.142 23.996
	 7.049 23.876
	 6.962 23.787
	 6.893 23.747
	 6.837 23.725
	 6.761 23.698
	/
\plot  7.576 24.460 	 7.578 24.360
	 7.576 24.306
	 7.551 24.198
	 7.534 24.133
	 7.513 24.065
	 7.489 23.998
	 7.464 23.935
	 7.408 23.836
	 7.344 23.767
	 7.254 23.695
	 7.159 23.633
	 7.076 23.590
	 6.972 23.578
	/
\plot  7.781 24.270 	 7.792 24.161
	 7.794 24.100
	 7.772 24.011
	 7.748 23.943
	 7.713 23.851
	/
\plot  8.009 24.050 	 8.020 23.934
	 8.020 23.872
	 8.019 23.827
	 8.009 23.743
	/
\plot  8.572 25.444 	 8.522 25.341
	 8.516 25.240
	 8.527 25.116
	 8.544 24.993
	 8.560 24.896
	 8.571 24.835
	 8.586 24.762
	 8.605 24.680
	 8.626 24.593
	 8.647 24.507
	 8.668 24.425
	 8.686 24.353
	 8.702 24.293
	 8.725 24.198
	 8.753 24.088
	 8.769 24.027
	 8.787 23.960
	 8.806 23.888
	 8.826 23.810
	 8.849 23.725
	 8.873 23.632
	 8.900 23.533
	 8.929 23.425
	 8.960 23.308
	 8.994 23.182
	 9.012 23.115
	 9.031 23.046
	 9.050 22.974
	 9.070 22.900
	/
\plot  8.816 25.444 	 8.788 25.350
	 8.769 25.279
	 8.750 25.186
	 8.747 25.126
	 8.747 25.051
	 8.750 24.968
	 8.754 24.881
	 8.760 24.793
	 8.766 24.711
	 8.773 24.637
	 8.778 24.577
	 8.783 24.514
	 8.791 24.437
	 8.801 24.352
	 8.812 24.261
	 8.823 24.171
	 8.835 24.085
	 8.845 24.009
	 8.854 23.946
	 8.864 23.883
	 8.879 23.797
	 8.900 23.678
	 8.913 23.602
	 8.928 23.514
	/
\plot  8.255 25.432 	 8.233 25.332
	 8.225 25.279
	 8.229 25.168
	 8.235 25.100
	 8.242 25.028
	 8.250 24.957
	 8.259 24.890
	 8.278 24.782
	 8.302 24.687
	 8.337 24.574
	 8.380 24.449
	 8.402 24.383
	 8.426 24.317
	 8.450 24.252
	 8.473 24.187
	 8.517 24.063
	 8.556 23.952
	 8.585 23.861
	 8.602 23.802
	 8.622 23.723
	 8.650 23.613
	 8.667 23.543
	 8.687 23.463
	/
\plot  7.984 25.404 	 7.973 25.279
	 7.976 25.213
	 7.983 25.133
	 7.993 25.044
	 8.005 24.951
	 8.019 24.858
	 8.033 24.770
	 8.047 24.691
	 8.060 24.627
	 8.087 24.522
	 8.106 24.456
	 8.129 24.380
	 8.157 24.291
	 8.191 24.187
	 8.231 24.065
	 8.254 23.997
	 8.278 23.925
	/
\plot  7.700 25.379 	 7.721 25.284
	 7.735 25.213
	 7.751 25.125
	 7.780 25.004
	 7.797 24.929
	 7.816 24.850
	 7.834 24.771
	 7.852 24.697
	 7.882 24.577
	 7.916 24.451
	 7.945 24.353
	 7.963 24.291
	 7.984 24.219
	/
\plot  7.751 25.163 	 7.844 25.089
	 7.918 25.070
	 8.011 25.061
	 8.103 25.060
	 8.177 25.061
	 8.273 25.065
	 8.394 25.074
	 8.515 25.089
	 8.611 25.110
	 8.663 25.139
	 8.750 25.201
	/
\plot  7.806 24.936 	 7.895 24.845
	 7.976 24.816
	 8.079 24.796
	 8.182 24.785
	 8.266 24.782
	 8.340 24.784
	 8.433 24.793
	 8.524 24.809
	 8.596 24.831
	 8.655 24.866
	 8.750 24.945
	/
\plot  7.870 24.640 	 7.991 24.560
	 8.060 24.524
	 8.116 24.509
	 8.188 24.496
	 8.259 24.487
	 8.316 24.483
	 8.392 24.491
	 8.484 24.505
	 8.576 24.526
	 8.647 24.551
	 8.697 24.582
	 8.778 24.653
	/
\plot  7.959 24.369 	 8.054 24.291
	 8.113 24.255
	 8.232 24.227
	 8.354 24.219
	 8.416 24.225
	 8.491 24.240
	 8.566 24.260
	 8.623 24.280
	 8.684 24.317
	 8.788 24.397
	/
\plot  8.177 24.039 	 8.259 23.998
	 8.321 23.971
	 8.405 23.946
	 8.530 23.963
	 8.647 24.001
	 8.711 24.040
	 8.816 24.128
	/
\plot  8.522 23.717 	 8.613 23.731
	 8.681 23.743
	 8.765 23.768
	 8.867 23.861
	/
\plot  8.727 23.527 	 8.818 23.549
	 8.867 23.565
	 8.907 23.590
	/
\plot  9.277 25.753 	 9.285 25.683
	 9.290 25.623
	 9.298 25.527
	 9.301 25.404
	 9.293 25.330
	 9.279 25.240
	 9.260 25.139
	 9.237 25.033
	 9.215 24.927
	 9.193 24.827
	 9.174 24.737
	 9.161 24.663
	 9.149 24.587
	 9.134 24.493
	 9.119 24.387
	 9.103 24.276
	 9.088 24.164
	 9.076 24.058
	 9.066 23.963
	 9.061 23.884
	 9.058 23.817
	 9.056 23.740
	 9.056 23.649
	 9.058 23.543
	 9.061 23.417
	 9.064 23.347
	 9.066 23.270
	 9.070 23.188
	 9.074 23.099
	 9.078 23.003
	 9.083 22.900
	/
\plot 10.401 25.612 	10.341 25.502
	10.298 25.421
	10.249 25.317
	10.212 25.211
	10.192 25.145
	10.172 25.075
	10.153 25.005
	10.135 24.939
	10.107 24.831
	10.089 24.747
	10.075 24.679
	10.058 24.587
	/
\plot  9.646 25.703 	 9.582 25.624
	 9.556 25.525
	 9.540 25.400
	 9.530 25.274
	 9.519 25.176
	 9.494 25.052
	 9.478 24.975
	 9.461 24.895
	 9.444 24.815
	 9.428 24.738
	 9.404 24.615
	 9.392 24.540
	 9.379 24.448
	 9.364 24.345
	 9.350 24.237
	 9.336 24.129
	 9.322 24.026
	 9.310 23.935
	 9.301 23.861
	 9.288 23.749
	 9.277 23.661
	 9.263 23.542
	/
\plot  9.887 25.673 	 9.836 25.637
	 9.808 25.577
	 9.787 25.504
	 9.770 25.422
	 9.758 25.336
	 9.748 25.249
	 9.740 25.167
	 9.732 25.094
	 9.724 25.034
	 9.709 24.950
	 9.691 24.844
	 9.674 24.737
	 9.660 24.653
	 9.643 24.544
	 9.632 24.475
	 9.619 24.395
	 9.605 24.301
	 9.588 24.190
	 9.579 24.128
	 9.569 24.061
	 9.558 23.989
	 9.546 23.912
	/
\plot 10.156 25.650 	10.092 25.586
	10.057 25.500
	10.025 25.389
	 9.999 25.276
	 9.980 25.186
	 9.966 25.126
	 9.952 25.057
	 9.937 24.975
	 9.920 24.879
	 9.900 24.765
	 9.890 24.701
	 9.879 24.631
	 9.867 24.556
	 9.854 24.475
	 9.840 24.387
	 9.826 24.293
	/
\plot  9.290 25.472 	 9.369 25.434
	 9.428 25.408
	 9.504 25.379
	 9.565 25.366
	 9.640 25.354
	 9.723 25.343
	 9.811 25.334
	 9.900 25.326
	 9.984 25.320
	10.059 25.317
	10.122 25.317
	10.174 25.323
	10.274 25.341
	/
\plot  9.277 25.216 	 9.364 25.140
	 9.433 25.106
	 9.522 25.071
	 9.612 25.041
	 9.686 25.021
	 9.776 25.010
	 9.889 25.003
	10.002 25.002
	10.092 25.011
	10.156 25.034
	/
\plot  9.222 24.972 	 9.301 24.922
	 9.366 24.883
	 9.450 24.837
	 9.536 24.793
	 9.608 24.765
	 9.693 24.753
	 9.800 24.748
	 9.908 24.748
	 9.995 24.754
	10.080 24.782
	/
\plot  9.201 24.740 	 9.250 24.640
	 9.328 24.590
	 9.430 24.549
	 9.536 24.518
	 9.622 24.498
	 9.695 24.490
	 9.788 24.488
	 9.880 24.490
	 9.955 24.498
	10.018 24.524
	/
\plot  9.136 24.473 	 9.222 24.359
	 9.294 24.318
	 9.386 24.280
	 9.480 24.250
	 9.557 24.232
	 9.624 24.226
	 9.749 24.232
	/
\plot  9.070 23.912 	 9.150 23.824
	 9.201 23.783
	 9.265 23.754
	 9.392 23.717
	/
\plot  9.061 23.643 	 9.146 23.554
	 9.263 23.501
	/
\plot  9.097 24.219 	 9.150 24.131
	 9.186 24.088
	 9.312 24.029
	 9.384 24.005
	 9.442 23.988
	 9.491 23.976
	 9.582 23.961
	/
\plot 10.107 24.445 	10.103 24.347
	10.107 24.293
	10.144 24.187
	10.182 24.108
	10.236 24.001
	/
\plot 10.490 25.021 	10.421 24.921
	10.361 24.834
	10.309 24.760
	10.265 24.697
	10.196 24.597
	10.145 24.524
	10.097 24.454
	10.030 24.360
	 9.988 24.299
	 9.938 24.228
	 9.880 24.146
	 9.813 24.050
	/
\plot 10.566 24.881 	10.474 24.827
	10.427 24.793
	10.315 24.675
	10.252 24.599
	10.187 24.518
	10.124 24.436
	10.065 24.358
	 9.970 24.232
	 9.918 24.166
	 9.855 24.086
	 9.786 23.996
	 9.713 23.901
	 9.641 23.806
	 9.573 23.715
	 9.513 23.633
	 9.466 23.565
	 9.407 23.473
	 9.330 23.346
	 9.281 23.265
	 9.225 23.169
	 9.159 23.057
	 9.122 22.995
	 9.083 22.928
	/
\plot 10.696 24.714 	10.577 24.626
	10.516 24.577
	10.398 24.462
	10.328 24.390
	10.254 24.313
	10.181 24.235
	10.113 24.161
	10.003 24.039
	 9.961 23.986
	 9.904 23.912
	 9.828 23.808
	 9.779 23.743
	 9.724 23.666
	/
\plot 10.795 24.513 	10.694 24.432
	10.620 24.372
	10.528 24.293
	10.429 24.194
	10.352 24.113
	10.249 24.001
	/
\plot 11.119 24.039 	11.006 23.989
	10.924 23.951
	10.823 23.897
	10.712 23.814
	10.628 23.742
	10.516 23.643
	/
\plot 10.952 24.270 	10.873 24.233
	10.816 24.206
	10.746 24.166
	10.695 24.124
	10.604 24.039
	/
\put{$t_{2q}$} [lB] at  2.656 24.066
\put{(a)} [lB] at  2.557 21.400
\put{(b)} [lB] at  8.867 21.421
\linethickness=0pt
\putrectangle corners at  1.657 25.775 and 11.144 21.323
\endpicture
\vskip 20pt
\centerline{{\bf Fig. 2}
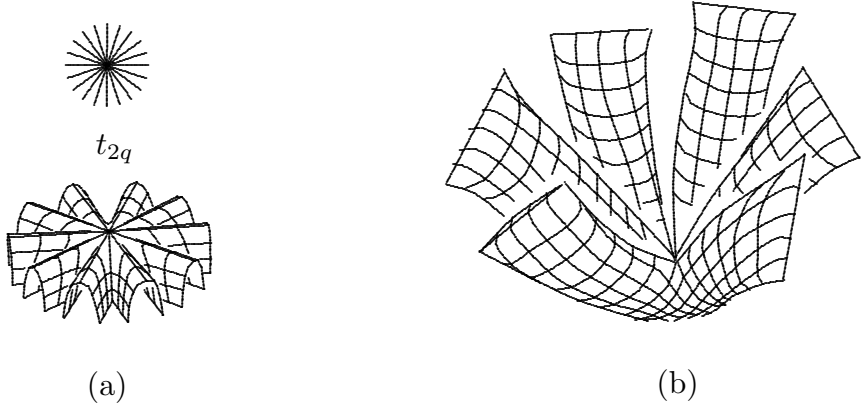
     (a) Negative curvature defect of angle $(2-q)\pi$ and
     (b) a typical surface.}
\vskip 20pt

It should be stressed that the methods we develop here are general and
could have applications going beyond the problem under
investigation. Given that the model of dually weighted graphs
seemed entirely inaccessible even a short while ago, we regard the
present approach to be an important step in extending current large
$N$ techniques.

We will quickly recall in the next section some of the results of our
previous paper \KSW\ and precisely define the class of models we are
studying.  Then, in section 3, we demonstrate how to derive the large
$N$ limit of group theoretical characters. The model of almost flat
planar diagrams will be solved and interpreted in section 4. The full
model capturing the transition from flat to random graphs
will be briefly discussed in section 5. We demonstrate how to
reformulate it as a well-posed Cauchy-Riemann problem.  We conclude in
section 6 and present an outlook on how our
approach might be put to further use in the near future.  Technical
details and additional illustrations are included in two
appendices.

\newsec{Review of the character expansion method for matrix models
of dually weighted graphs}

The partition function \DWG\ for dually weighted graphs can be
formulated as the following matrix model:
\eqn\DWGmatrix{
Z(t^*,t)=\int\,{\cal D}M\ e^{-{N\over 2} \Tr~M^2\ +\ \Tr~V_B(M A)},}
with
\eqn\potential{
V_B(M A)=
\sum_{k=1}^{\infty}{1\over k}~\Tr B^k\ (M A)^k .}
The matrices $A$ and $B$ are fixed, external matrices encoding the
coupling constants through
\eqn\tqAB{
t_q^*= {1\over N}\ \Tr\ B^q
{\rm \hskip 20pt and \hskip 20pt}
t_q=  {1\over N}\ \Tr\ A^q.}
The model generalizes, for $A \neq 1$, the standard one matrix model
first solved by Br\'ezin, Itzykson, Parisi and Zuber \BIPZ.
It can no longer be solved by changing to
eigenvalue variables; a reduction to $N$ variables is nevertheless
possible. An expansion of the potential into a sum over invariant
group characters allows all integrations to be performed and \DWGmatrix\
to be reformulated as a statistical mechanics model in ``Young-tableau
weight space''. This reformulation should be called, after its discoverers,
the ``Itzykson-Di~Francesco formula'' \IDiF\ and reads
\eqn\IzDiFr{
Z(t,t^*)=c\,\sum_{\{h^e,h^o\}}
{\prod_i(h^e_i-1)!!h^o_i!!\over
\prod_{i,j}(h^e_i-h^o_j)}~\chi_{\{h\}}(A)~\chi_{\{h\}}(B)}
Here $c$ is a constant that we can drop, the weights
$\{h^e\}$ are a set of $N/2$ even, increasing, non-negative
integers while the weights $\{h^o\}$ are $N/2$ odd, increasing, positive
integers, and the sum is taken over all such sets. The characters can be
defined through two equivalent formulae. The first is the Weyl formula:
\eqn\weylchar{
\chi_{\{h\}}(A)={det_{_{\hskip -2pt (k,l)}}(a_k^{h_l})\over
\Delta(a)},}
where the $a_k$ are the eigenvalues of the matrix $A$ and
$\D(a)$ is the Vandermonde determinant. The second definition makes
use of Schur polynomials, $P_n(\theta)$, defined by
\eqn\schupol{
e^{\Sigma_{i=1}^{\infty}z^i\theta_i}=\sum_{n=0}^{\infty}z^n
P_n (\theta)\quad {\rm with}\quad \theta_i={1\over i}~\Tr[A^i],}
in terms of which the character is
\eqn\schuchar{
\chi_{\{h\}}(A)=det_{_{\hskip -2pt (k,l)}}
                 \bigl(P_{h_k+1-l}(\theta)\bigr).}

It was demonstrated in \KSW\ how to take the large $N$ limit of this
expansion. In this limit, the weights ${1 \over N} h_i$ condense to
give a smooth, stationary distribution $dh~\rho(h)$, where $\rho(h)$
is a probability density normalized to one.  For technical reasons we
restrict our attention to models in which the matrices $A$ and $B$ are
such that traces of all odd powers of $A$ and $B$ are zero. This means
that the our random surfaces are made from vertices and faces with
even coordination numbers only. As was discussed in \KSW, this ensures
that the support of the density $\rho(h)$ lies entirely on the real
axis, and thus simplifies the solution of the problem\foot{ We do not
want to suggest that models with odd coordination numbers cannot be
treated with our methods.}.  The matrix $A$ will
satisfy this condition if we introduce an ${N\over 2}\times{N\over 2}$
matrix $\sqrt{a}$ in terms of which $A$ and the character
$\chi_{\{h\}}(A)$ are given by
\eqn\Aa{
A=\left[\matrix{\sqrt{a}&0\cr 0&-\sqrt{a}\cr}\right]\quad {\rm and} \quad
\chi_{\{h\}}(A)=
    \chi_{\{{h^e\over 2}\}}(a)\chi_{\{{h^o-1\over 2}\}}(a)
    \,\,\sgn\bigl[\prod_{i,j}(h^e_i-h^o_j)\bigr],}

We now focus our attention on three intimately related models which
capture the transition from flat to random graphs.
\eqn\models{\eqalign{
& {\rm I. }~V_A(MA)
=\sum_{k=1}^\infty {1\over 2k}~\Tr[A^{2k}]~(MA)^{2k}. \cr
& {\rm II. }~V_{A_4}(MA) ={1\over 4}~(MA)^4. \cr
& {\rm III. }~V_A(MA_4)
=\sum_{k=1}^\infty{1\over 2k}~\Tr[A^{2k}]~(MA_4)^{2k} \cr }}
Here $A_4$ is defined to satisfy $\Tr[(A_4)^k]=N \delta_{k,4}$ and $A$
is as defined in \Aa.  The first model is self-dual, i.e.~vertices
and faces having the same coordination number have the same
weights. The second and third models are dual to each other (the
lattice of one corresponds to the dual lattice of the other) and are
in turn related to model I by a simple line map. That is, we place the
diagonal of a square belonging to model III (or alternatively a
four-vertex belonging to II) onto each propagator of model I. Thus the
vertices and face centres of model I become the vertices of model III
(or alternatively the faces of II). We illustrate this in Fig.~3
below\foot{ Note that this line-map is only valid on the sphere. The
${1 \over N}$ corrections of I and III will thus be different.  A
careful analysis shows that the spherical free energy of model I is
precisely twice the free energy of models II and III (since there are
two ways of choosing the diagonal of a square in III, or alternatively
two ways of splitting a four-vertex of model II). Note also that this
non-trivial correspondence is {\it predicted} from our formalism,
since we indeed obtain the same $N=\infty$ equations in all three
cases.}.
\vskip 20pt
\hskip 15pt
\beginpicture
\setcoordinatesystem units <1.00000cm,1.00000cm>
\linethickness=1pt
\plot  8.289 25.049  9.559 25.049 /
\setdashes < 0.0677cm>
\plot  8.289 23.779  9.559 23.779 /
\setdots < 0.0508cm>
\plot  8.289 22.509  9.559 22.509 /
\setsolid
\ellipticalarc axes ratio  0.019:0.019  360 degrees
	from  4.676 22.479 center at  4.657 22.479
\ellipticalarc axes ratio  0.019:0.019  360 degrees
	from  3.152 22.246 center at  3.133 22.246
\plot  4.417 25.343  4.246 24.202 /
\plot  1.319 23.529  2.413 23.137 /
\plot  1.829 24.572  3.004 24.194 /
\plot  3.023 24.242  3.082 25.406 /
\plot  1.926 22.098  2.413 23.129 /
\plot  3.035 24.191  4.238 24.191 /
\plot  2.417 23.154  3.023 24.200 /
\plot  4.837 23.152  4.233 24.194 /
\plot  3.122 22.225  2.417 23.137 /
\plot  3.133 22.225  4.197 22.081 /
\plot  4.870 23.144  6.121 22.904 /
\plot  4.238 22.073  4.329 21.273 /
\plot  4.297 24.191  5.501 24.191 /
\plot  2.542 25.506  3.073 25.385 /
\plot  3.073 25.385  4.426 25.343 /
\plot  6.121 22.911  6.536 22.172 /
\plot  5.478 24.185  5.478 24.185 /
\plot  5.493 24.202  5.906 24.784 /
\plot  4.439 25.343  5.271 25.366 /
\plot  5.391 24.735  5.493 24.194 /
\plot  1.839 24.562  1.810 25.106 /
\plot  1.858 24.572  1.319 23.529 /
\plot  1.319 23.529  0.864 23.167 /
\plot  1.319 23.529  0.936 23.717 /
\plot  1.858 24.572  1.458 24.634 /
\plot  3.073 25.396  3.082 25.675 /
\plot  4.426 25.343  4.449 25.648 /
\plot  3.702 24.826  3.702 24.826 /
\plot  5.209 23.611  5.209 23.611 /
\setdots < 0.0508cm>
\plot  3.143 22.246  2.910 21.431 /
\plot  3.143 22.225  3.634 23.059 /
\plot  4.238 22.045  2.868 21.410 /
\plot  3.634 23.050  4.246 22.073 /
\plot  4.246 22.073  4.259 22.087 /
\plot  4.246 22.073  4.667 22.479 /
\plot  4.667 22.479  4.837 23.137 /
\plot  6.121 22.888  6.202 22.130 /
\plot  6.113 22.896  6.784 22.504 /
\plot  5.220 23.601  6.113 22.896 /
\plot  4.860 23.148  3.628 23.050 /
\plot  5.220 23.590  4.860 23.148 /
\plot  5.510 24.185  5.209 23.611 /
\plot  5.520 24.194  6.179 24.033 /
\plot  5.209 23.611  4.265 24.170 /
\plot  5.520 24.208  5.622 24.672 /
\plot  4.928 24.725  5.510 24.185 /
\plot  4.917 24.735  5.264 25.095 /
\plot  4.909 24.735  4.259 24.185 /
\plot  4.246 24.185  3.645 23.059 /
\plot  4.468 25.337  4.909 24.735 /
\plot  3.702 24.814  4.426 25.337 /
\plot  3.717 24.803  4.246 24.185 /
\plot  3.073 25.377  3.702 24.826 /
\plot  3.082 25.377  3.385 25.626 /
\plot  3.092 25.377  2.883 25.576 /
\plot  2.428 24.966  3.073 25.377 /
\plot  3.702 24.814  3.023 24.170 /
\plot  2.417 24.966  3.023 24.170 /
\plot  3.035 24.185  2.148 23.914 /
\plot  3.634 23.050  3.035 24.185 /
\plot  2.417 24.966  2.231 25.205 /
\plot  1.869 24.555  2.417 24.966 /
\plot  2.148 23.914  1.869 24.555 /
\plot  1.848 24.572  1.659 24.793 /
\plot  1.858 24.555  1.547 24.454 /
\plot  2.148 23.901  2.417 23.129 /
\plot  2.138 23.901  1.319 23.510 /
\plot  2.417 23.129  3.634 23.050 /
\plot  1.528 22.936  2.417 23.129 /
\plot  2.441 23.118  2.889 21.400 /
\plot  1.336 23.544  1.187 23.880 /
\plot  1.336 23.518  0.995 23.451 /
\plot  1.327 23.510  1.528 22.936 /
\plot  1.528 22.936  1.285 22.911 /
\plot  1.528 22.926  1.558 22.737 /
\plot  4.449 25.343  4.729 25.527 /
\plot  4.449 25.343  4.168 25.599 /
\setsolid
\plot  6.121 22.911  6.536 23.690 /
\plot  6.121 22.904  7.046 22.729 /
\setdots < 0.0508cm>
\plot  6.121 22.911  6.179 24.033 /
\plot  6.121 22.904  6.775 23.245 /
\plot  6.202 24.041  6.401 24.208 /
\plot  6.179 24.033  6.164 24.448 /
\setdashes < 0.0677cm>
\plot  3.768 25.379  3.998 25.593 /
\plot  3.768 25.370  3.584 25.578 /
\plot  4.906 25.353  5.057 25.521 /
\plot  4.906 25.353  4.841 25.512 /
\plot  5.853 24.687  6.052 24.670 /
\plot  5.853 24.687  5.812 24.845 /
\plot  5.844 24.678  5.749 24.693 /
\plot  6.464 23.516  6.617 23.539 /
\plot  6.464 23.523  6.433 23.721 /
\plot  6.775 22.792  6.960 22.911 /
\plot  6.767 22.767  6.951 22.600 /
\plot  6.449 22.346  6.680 22.236 /
\plot  6.441 22.322  6.394 22.077 /
\plot  2.106 22.479  2.191 22.204 /
\plot  2.106 22.490  1.820 22.511 /
\plot  1.579 24.041  1.437 24.304 /
\plot  1.587 24.041  1.300 23.937 /
\plot  4.286 21.474  4.710 21.262 /
\plot  4.286 21.463  4.032 21.315 /
\setsolid
\plot  5.501 24.208  6.104 22.911 /
\plot  6.104 22.911  5.948 21.992 /
\setdashes < 0.0677cm>
\plot  5.990 22.299  5.800 22.098 /
\plot  6.011 22.267  6.159 22.109 /
\setdots < 0.0508cm>
\plot  4.868 23.135  5.387 22.045 /
\plot  4.255 22.045  5.376 22.045 /
\plot  6.107 22.881  5.366 22.045 /
\plot  5.376 22.045  5.154 21.749 /
\plot  5.387 22.024  5.503 21.717 /
\plot  5.408 22.035  5.652 21.950 /
\plot  2.879 21.410  2.455 21.400 /
\plot  2.455 21.400  2.455 21.389 /
\plot  2.879 21.400  2.783 21.029 /
\plot  2.900 21.378  3.186 21.220 /
\setsolid
\plot  4.238 22.092 	 4.231 22.162
	 4.226 22.227
	 4.219 22.344
	 4.217 22.445
	 4.220 22.531
	 4.228 22.606
	 4.242 22.670
	 4.286 22.775
	 4.352 22.862
	 4.457 22.947
	 4.527 22.991
	 4.612 23.038
	 4.712 23.088
	 4.830 23.144
	/
\plot  4.238 22.073 	 4.309 22.083
	 4.376 22.093
	 4.494 22.116
	 4.594 22.142
	 4.677 22.171
	 4.804 22.244
	 4.889 22.341
	 4.923 22.406
	 4.946 22.476
	 4.957 22.554
	 4.956 22.641
	 4.944 22.740
	 4.920 22.854
	 4.904 22.918
	 4.884 22.985
	 4.862 23.058
	 4.837 23.135
	/
\setdashes < 0.0677cm>
\plot  5.495 23.015 	 5.537 22.896
	 5.571 22.809
	 5.622 22.703
	 5.670 22.635
	 5.741 22.550
	 5.845 22.438
	 5.912 22.368
	 5.990 22.289
	/
\plot  5.486 23.023 	 5.413 22.970
	 5.350 22.924
	 5.252 22.849
	 5.135 22.744
	 5.091 22.689
	 5.037 22.612
	 4.969 22.503
	 4.927 22.433
	 4.879 22.352
	/
\plot  4.879 22.352 	 4.875 22.280
	 4.871 22.219
	 4.861 22.120
	 4.826 21.996
	 4.762 21.896
	 4.658 21.782
	 4.589 21.716
	 4.504 21.642
	 4.404 21.558
	 4.286 21.463
	/
\plot  5.495 23.023 	 5.549 23.115
	 5.589 23.183
	 5.637 23.269
	 5.694 23.379
	 5.738 23.467
	 5.798 23.588
	/
\plot  5.486 23.023 	 5.408 23.106
	 5.339 23.177
	 5.227 23.287
	 5.141 23.362
	 5.072 23.412
	 5.002 23.450
	 4.905 23.492
	 4.842 23.515
	 4.768 23.541
	 4.681 23.570
	 4.580 23.603
	/
\plot  4.953 24.185 	 5.033 24.121
	 5.102 24.066
	 5.211 23.979
	 5.291 23.918
	 5.351 23.874
	 5.415 23.830
	 5.504 23.772
	 5.628 23.694
	 5.706 23.645
	 5.798 23.588
	/
\plot  4.587 23.611 	 4.678 23.712
	 4.744 23.788
	 4.820 23.891
	 4.864 23.993
	 4.893 24.077
	 4.930 24.191
	/
\plot  5.812 23.571 	 5.883 23.558
	 5.944 23.547
	 6.041 23.531
	 6.164 23.516
	 6.265 23.515
	 6.347 23.518
	 6.458 23.523
	/
\plot  6.458 23.508 	 6.526 23.384
	 6.575 23.292
	 6.634 23.173
	 6.655 23.116
	 6.681 23.037
	 6.716 22.926
	 6.737 22.856
	 6.761 22.775
	/
\plot  6.752 22.775 	 6.669 22.656
	 6.623 22.593
	 6.564 22.507
	 6.515 22.439
	 6.449 22.346
	/
\plot  6.449 22.339 	 6.337 22.344
	 6.254 22.346
	 6.149 22.341
	 6.093 22.329
	 5.990 22.299
	/
\plot  5.798 23.580 	 5.812 23.687
	 5.824 23.780
	 5.835 23.859
	 5.843 23.927
	 5.855 24.034
	 5.861 24.113
	 5.863 24.200
	 5.863 24.317
	 5.861 24.392
	 5.859 24.480
	 5.856 24.583
	 5.853 24.701
	/
\plot  4.938 24.191 	 4.981 24.260
	 5.017 24.319
	 5.076 24.412
	 5.152 24.526
	 5.198 24.588
	 5.287 24.701
	/
\plot  4.945 24.200 	 4.827 24.307
	 4.741 24.386
	 4.633 24.488
	 4.531 24.596
	 4.450 24.684
	 4.339 24.805
	/
\plot  4.350 24.788 	 4.422 24.832
	 4.485 24.871
	 4.583 24.934
	 4.699 25.027
	 4.777 25.134
	 4.830 25.223
	 4.898 25.347
	/
\plot  4.898 25.347 	 4.939 25.266
	 4.975 25.196
	 5.036 25.086
	 5.085 25.010
	 5.129 24.958
	 5.247 24.877
	/
\plot  4.570 23.597 	 4.449 23.611
	 4.345 23.625
	 4.256 23.640
	 4.180 23.656
	 4.061 23.693
	 3.975 23.738
	 3.907 23.797
	 3.836 23.887
	 3.797 23.946
	 3.755 24.017
	 3.708 24.101
	 3.655 24.200
	/
\plot  4.570 23.588 	 4.514 23.493
	 4.466 23.411
	 4.426 23.340
	 4.393 23.278
	 4.346 23.179
	 4.318 23.101
	 4.298 22.991
	 4.291 22.903
	 4.286 22.784
	/
\plot  4.269 22.761 	 4.159 22.663
	 4.079 22.591
	 3.981 22.498
	 3.871 22.376
	 3.787 22.277
	 3.734 22.214
	 3.672 22.140
	/
\plot  4.269 22.792 	 4.289 22.708
	 4.309 22.635
	 4.349 22.523
	 4.396 22.445
	 4.451 22.392
	 4.520 22.358
	 4.609 22.343
	 4.729 22.344
	 4.803 22.351
	 4.889 22.363
	/
\plot  4.276 22.752 	 4.352 22.762
	 4.418 22.769
	 4.524 22.773
	 4.604 22.761
	 4.665 22.735
	 4.779 22.610
	 4.832 22.498
	 4.860 22.426
	 4.889 22.341
	/
\plot  3.662 22.130 	 3.605 22.221
	 3.555 22.300
	 3.468 22.420
	 3.393 22.501
	 3.323 22.553
	 3.247 22.586
	 3.145 22.608
	 3.081 22.615
	 3.005 22.621
	 2.917 22.625
	 2.815 22.627
	/
\plot  2.805 22.617 	 2.825 22.715
	 2.842 22.799
	 2.855 22.872
	 2.865 22.935
	 2.876 23.034
	 2.877 23.110
	 2.865 23.199
	 2.837 23.315
	 2.816 23.388
	 2.789 23.472
	 2.757 23.570
	 2.718 23.683
	/
\plot  2.718 23.683 	 2.825 23.690
	 2.918 23.697
	 2.997 23.706
	 3.065 23.715
	 3.172 23.739
	 3.251 23.770
	 3.327 23.821
	 3.414 23.902
	 3.520 24.022
	 3.584 24.100
	 3.655 24.191
	/
\plot  3.662 22.140 	 3.699 22.050
	 3.731 21.973
	 3.788 21.851
	 3.837 21.766
	 3.884 21.706
	 3.938 21.659
	 4.017 21.608
	 4.130 21.548
	 4.202 21.513
	 4.286 21.474
	/
\plot  2.815 22.627 	 2.739 22.599
	 2.673 22.575
	 2.569 22.539
	 2.492 22.515
	 2.434 22.500
	 2.324 22.485
	 2.236 22.477
	 2.117 22.468
	/
\plot  2.106 22.479 	 2.071 22.589
	 2.040 22.684
	 2.014 22.766
	 1.992 22.835
	 1.956 22.943
	 1.928 23.023
	 1.885 23.139
	 1.850 23.232
	 1.801 23.357
	/
\plot  1.801 23.357 	 1.731 23.320
	 1.670 23.288
	 1.571 23.241
	 1.496 23.213
	 1.437 23.199
	 1.363 23.193
	 1.266 23.198
	 1.206 23.205
	 1.135 23.215
	 1.053 23.228
	 0.959 23.245
	/
\plot  1.818 23.357 	 1.911 23.393
	 1.991 23.424
	 2.060 23.450
	 2.119 23.473
	 2.211 23.508
	 2.280 23.533
	 2.345 23.554
	 2.433 23.583
	 2.557 23.622
	 2.634 23.647
	 2.724 23.675
	/
\plot  1.818 23.357 	 1.812 23.433
	 1.806 23.498
	 1.793 23.601
	 1.779 23.676
	 1.763 23.730
	 1.701 23.840
	 1.645 23.919
	 1.564 24.024
	/
\plot  1.564 24.024 	 1.658 24.056
	 1.739 24.084
	 1.808 24.108
	 1.866 24.129
	 1.958 24.164
	 2.026 24.191
	 2.082 24.220
	 2.157 24.260
	 2.259 24.318
	 2.324 24.355
	 2.398 24.399
	/
\plot  2.398 24.399 	 2.420 24.321
	 2.440 24.254
	 2.472 24.149
	 2.497 24.073
	 2.519 24.018
	 2.580 23.896
	 2.634 23.802
	 2.709 23.675
	/
\plot  2.398 24.382 	 2.494 24.479
	 2.566 24.548
	 2.661 24.630
	 2.716 24.666
	 2.794 24.711
	 2.903 24.769
	 2.972 24.805
	 3.052 24.845
	/
\plot  3.052 24.845 	 3.122 24.782
	 3.183 24.728
	 3.278 24.641
	 3.395 24.526
	 3.488 24.413
	 3.560 24.320
	 3.604 24.260
	 3.655 24.191
	/
\plot  3.664 24.185 	 3.779 24.306
	 3.864 24.395
	 3.975 24.503
	 4.099 24.605
	 4.200 24.683
	 4.265 24.732
	 4.339 24.788
	/
\plot  4.339 24.788 	 4.218 24.888
	 4.129 24.962
	 4.022 25.059
	 3.930 25.165
	 3.861 25.251
	 3.768 25.370
	/
\plot  3.768 25.370 	 3.692 25.316
	 3.627 25.269
	 3.523 25.194
	 3.449 25.139
	 3.395 25.099
	 3.275 25.011
	 3.180 24.941
	 3.121 24.897
	 3.052 24.845
	/
\plot  3.052 24.845 	 2.928 24.954
	 2.838 25.035
	 2.733 25.146
	 2.672 25.244
	 2.629 25.324
	 2.574 25.434
	/
\plot  2.407 24.399 	 2.356 24.475
	 2.313 24.541
	 2.243 24.645
	 2.192 24.720
	 2.153 24.773
	 2.076 24.866
	 2.014 24.938
	 1.928 25.036
	/
\plot  2.794 22.638 	 2.781 22.547
	 2.772 22.468
	 2.768 22.399
	 2.768 22.340
	 2.783 22.246
	 2.815 22.172
	 2.911 22.065
	 2.973 22.017
	 3.042 21.975
	 3.116 21.940
	 3.190 21.915
	 3.264 21.900
	 3.334 21.897
	 3.408 21.912
	 3.487 21.954
	 3.580 22.028
	 3.694 22.140
	/
\put{Model I.} [lB] at 10.353 24.890
\put{Model II.} [lB] at 10.353 23.620
\put{Model III.} [lB] at 10.353 22.413
\linethickness=0pt
\putrectangle corners at  0.838 25.701 and 11.980 21.004
\endpicture
\vskip 20pt
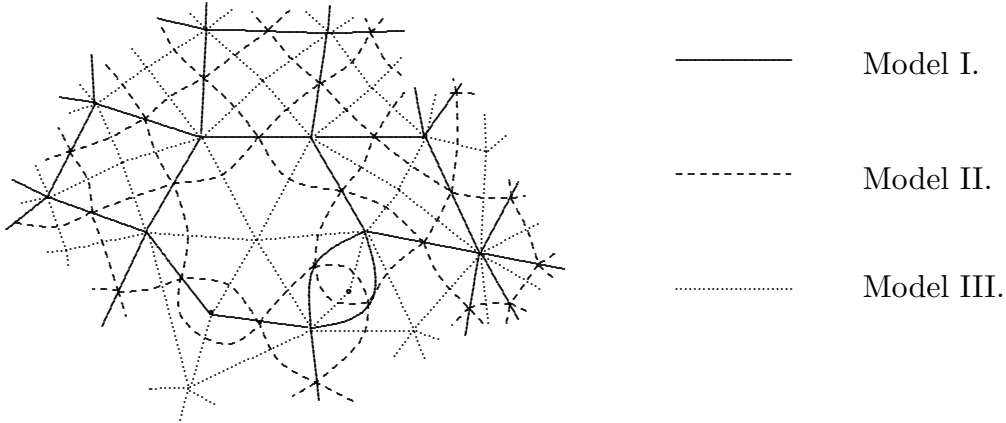
\centerline{{\bf Fig. 3:} Graphical relationship between models I, II
and III}
\vskip 20pt
\hskip -19pt {}From this line map one can see that the
expectation values in models I and III are also the same. More
specifically
\eqn\expsam{
\langle{1\over N}\Tr[(MA)^{2 k}]\rangle_{I}=
\langle{1\over N}\Tr[(MA_4)^{2 k}]\rangle_{III}}
Notice, however, that they are not equivalent to
$\langle{1\over N}\Tr[(MA)^k]\rangle_{II}$ in model II.

We can now return to the discussion of the large $N$ limit and write
the saddlepoint equation for these three models. Looking for the
stationary point in \IzDiFr, one finds from \KSW , in all three cases,
the following equation, valid on an interval $[b,a]$ with $0 \leq b
\leq 1 \leq a$:
\eqn\sdpt{
2F(h)+\barint_0^a\ dh'\ {\rho(h') \over h-h'}= -\ln h.}
The solution requires, evidently, the knowledge of the large $N$ limit of
the variation of the characters in eq.\Aa:
\eqn\defF{
F(h_k)=2{\partial \over \partial h^e_k}~\ln\ {\chi_{\{{h^e\over 2}\}}(a)
\over \Delta(h^e)}.}
The determination of $F(h)$ is the subject
of the next section. Let us also recall here the definition
of the resolvent $H(h)$:
\eqn\defH{
H(h)=\int_0^a dh'\ {\rho(h') \over h-h'}.}
In \KSW\  we demonstrated, via a simple functional inversion, how to
relate the results of the weight formalism to the resolvent
$W(P)=\langle {1\over N} \Tr {1\over P-M}\rangle$ of the matrix model
\DWGmatrix . In the model investigated in this paper, however, it is more
natural to study the correlators $\langle {1\over N} \Tr ((MA)^{2q})
\rangle$ . The results of the following section will provide a simple way
to calculate such moments.

\newsec{Large $N$ limit of the character}
In the saddle point equation \sdpt\ we
introduced the function $F(h)$ defined in eq. \defF\ as the
derivative of the logarithm of a character.
This function $F(h)$ depends upon the moments of the matrix
$A$, i.e.~it contains all the information
on the weights that one assigns to the faces of our discrete
surfaces. In order to proceed with the solution of the saddle point
equation, one would like to take the large $N$ limit of
\defF\ and express $F(h)$ in terms of $H(h)$
(which specifies the Young tableau) and the set of moments $t_{q}$ of
the matrix $A$ (the weights assigned to the faces).

In \KSW\ a contour integral formula
relating $H(h)$, $F(h)$ and the set of moments $t_{q}$ was derived. We
recall  here a single essential step of the derivation, which we will
make use of shortly. We observed that
\eqn\trachar{
\Tr[a^q]=\sum_{k=1}^{N/2}{\chi_{\{{\tilde h^e\over 2}\}}(a)\over
                          \chi_{\{{h^e\over 2}\}}(a)}\quad {\rm where}
\quad \tilde h^e_i=h^e_i+2q\delta_{i,k},}
and the matrix $a$ is the ${N\over 2}\times {N\over 2}$ matrix
introduced in \Aa .  For notational simplicity we omit an index $k$ on
$\tilde h$.  In the large $N$ limit \trachar\ was then reduced to a simple
contour integral
\eqn\tqHF{
t_{2q}={1\over q}\oint\,{dh\over 2\pi i} e^{q(H(h)+F(h))}\quad {\rm where}
                 \quad t_{2q}={2\over N}\Tr[a^q].}
Note that the definition of $F(h)$ \defF\ differs from that in the
derivation in \KSW\ since we are now restricting our attention to the
case where only the even moments of the matrix $A$ are non-zero\foot{
Indeed, it might be asked why we do not directly use formula (3.5)
derived in section 3.~of
\KSW. There, the contour integration relation was
derived for the general case where both even and odd moments are
non-zero. However, in the special case where we then set all odd
moments to zero, $e^{F(h)}$ contains a cut overlapping with the
cut of $e^{H(h)}$. In this case defining the
contour encircling the cut of $e^{H(h)}$ is ambiguous.
We have therefore
rederived the result for the reduced case of only even non-zero
moments. The same note of caution applies to formula (3.8) of \KSW.}.

As it stands, formula \tqHF\ is of little direct use. It can however be
dramatically simplified as we sketch out below. We introduce a function $G(h)$
defined as
\eqn\defG{
G(h) = e^{H(h) +F(h)},}
in terms of which \tqHF\ becomes
\eqn\tqG{
t_{2q}={1\over q}\oint\,{dh\over 2\pi i}~G(h)^q.}
Changing integration variables from $h$ to $G$ we arrive at
\eqn\tqGcont{
t_{2q}=\oint\,{dG\over 2\pi i G}~h(G)~G^q,}
where $h(G)$ is the inverse of the equation for $G(h)$ given in
\defG, and the contour in the complex $G$ plane encircles the
origin. We now assume that there are only a finite number of non-zero
couplings $t_q$. We obtain immediately the solution:
\eqn\hG{
h-1 = \sum_{q=1}^Q{t_{2q}\over G^q}~+~\psi(G).}
Here $\psi(G)$ is an as yet unknown function, analytic in the vicinity
of the origin, with $\psi(0)=0$.  It is trivial to see that this
satisfies \tqGcont. Note that, strictly speaking, we can not solve
equation \tqGcont\ for $q=0$ since \tqG\ is not defined there. The $1$
on the l.h.s. of \hG\ comes from the
normalization of the density $\rho(h)$ (See appendix A).

The unknown function $\psi(G)$ is not fixed by \tqG\ and depends on
the specific model being studied. We now give a very simple physical
interpretation to this function. Let us return to the Schur polynomial
definition of the character \schuchar\ . Differentiating eq. \schupol\
with respect to $\theta_i$ we see that
\eqn\difpoly{
{\partial\over\partial\theta_q}P_n(\theta)=P_{n-q}(\theta)\quad
{\rm with}\quad \theta_q={N\over 2q}t_{2q}.}
This implies immediately that
\eqn\difchar{
{2q\over N}{\partial\over \partial t_{2q}}\ln\Bigl(
\chi_{\{{h^e\over 2}\}}(a)\Bigr)=
\sum_{k=1}^{N/2}{\chi_{\{{\tilde h^e\over 2}\}}(a)\over
                          \chi_{\{{h^e\over 2}\}}(a)}\quad {\rm where}
\quad \tilde h^e_i=h^e_i-2q\delta_{i,k}.}
{}From \DWGmatrix, \IzDiFr\ and \Aa ,  we see that the left hand side
of this equation is equivalent to differentiating the
logarithm of the original matrix integral \DWGmatrix\ with respect to
$t_{2q}$. In terms of the dual to this matrix integral (in which the
weights $t_{2q}$ assigned to the faces are now the weights of the
vertices) this is equivalent to differentiating the coupling
constants of the dual potential. So, denoting the dual matrix by
$\tilde M$, the left hand side of eq. \difchar\ is equivalent to
the expectation value $\langle\Tr (\tilde MB)^{2q} \rangle$. Now,
comparing the right hand side
of \difchar\ to equations \trachar\ and \tqHF , we
see that we have the following relation in the large $N$ limit,
\eqn\expcont{
\langle {1\over N}\Tr (\tilde MB)^{2q} \rangle=
\oint\,{dG\over 2\pi i G}~h(G)~G^{-q},}
$G(h)$ being defined by \defG . It is now simple to follow
identical arguments to those used to simplify \tqG\ to \hG\ to arrive
at
\eqn\hGexp{
h-1 = \sum_{q=1}^Q{t_q\over G^q} +
\sum_{q=1}^{\infty}\langle {1\over N}\Tr~(\tilde MB)^{2q} \rangle
{}~G^q.}
Given $G(h)$, we have,
after a functional inversion, the correlators of the dual model.

To find $G(h)$ we have to connect eq. \hGexp\ with the saddle point
equation \sdpt .  From
\defH\ we obtain
\eqn\Hdef{
H(h)=\ln {h\over h-b}+ \tilde H(h)
\quad{\rm with}\quad \tilde H(h)=\int_b^a\,dh' {\rho(h')\over h-h'},}
where the first term on the right is the contribution from the flat
part of the density, i.e.~the empty part of the Young tableau.
The integral from $b$ to $a$ is the contribution from the
``excited'' part of the density, i.e.~the non-empty part of the Young
tableau. Noting, from the definition of $G(h)$ \defG ,
that $\ln G(h)=H(h) + F(h)$, we replace the
integral of \Hdef\ by the contour integral
\eqn\Hinva{
\tilde H(h)=\oint\,{dh' \over 2\pi i}~{\ln G(h')\over h-h'}}
where the contour encircles the $[b,a]$ part of the cut of $H(h)$. The
discontinuity across this cut is precisely $\pm i\pi\rho(h)$. Note
also that $F(h)$ has -- at least for some range of the couplings --
no cut on the interval $[b,a]$. If we
now change the variables of integration from $h$ to $G$, as previously,
and shrink the contour in the complex $G$ plane catching poles on the
way (see appendix A), we arrive at the following simple relationship
between equation \hG\ and $H(h)$:
\eqn\HGprod{
e^{H(h)}={(-1)^{(Q-1)}h\over t_Q}\prod_{q=1}^Q G_q(h).}
Some words of explanation are in order to clarify the meaning of this
equation. Inverting eq. \hG\ leads to a multi-sheeted function
$G(h)$. The general picture is illustrated in Fig. 4. One of the
sheets is the physical sheet and has two cuts, one corresponding to
$e^{H(h)}$, the other to $e^{F(h)}$; we label this sheet $G_1(h)$. The
sheets $G_2(h),\,\dots\,,G_Q(h)$ are all the sheets connected to
$G_1(h)$ by the cut of $e^{F(h)}$; there are exactly $Q$ of these
sheets, where $Q$ is the maximum inverse power of $G$ in \hGexp .
\vskip 25pt
\beginpicture
\setcoordinatesystem units <1.00000cm,1.00000cm>
\linethickness=1pt
\plot  5.080 19.571 10.207 18.860 /
\plot 10.221 18.860 11.345 19.571 /
\plot  5.080 19.571  6.505 20.142 /
\plot  5.080 20.426 10.207 19.814 /
\plot  5.080 21.281 10.207 20.726 /
\plot 10.207 19.799 11.345 20.426 /
\plot 10.207 20.726 11.345 21.281 /
\plot  7.074 21.380   7.074 21.139 /
\plot  7.074 20.968   7.074 20.540 /
\plot  9.495 20.269   9.495 19.986 /
\plot  9.495 19.799   9.495 19.372 /
\plot  5.806 21.510  7.074 21.380 /
\plot  5.791 20.682  7.074 20.540 /
\plot  9.495 20.269 10.791 20.127 /
\plot  9.495 19.387 10.791 19.215 /
\plot  6.932 20.983   6.932 20.555 /
\plot  6.788 20.995   6.788 20.570 /
\plot  6.646 20.995   6.646 20.582 /
\plot  6.505 21.010   6.505 20.597 /
\plot  6.363 21.025   6.363 20.625 /
\plot  6.219 21.040   6.219 20.640 /
\plot  6.077 21.052   6.077 20.654 /
\plot  5.935 21.067   5.935 20.712 /
\plot  5.791 21.040   5.791 20.769 /
\plot  5.649 20.995   5.649 20.826 /
\plot  9.637 19.757   9.637 19.359 /
\plot  9.779 19.757   9.779 19.344 /
\plot  9.923 19.742   9.923 19.329 /
\plot 10.065 19.715  10.065 19.315 /
\plot 10.207 19.715  10.207 19.287 /
\plot 10.348 19.729  10.348 19.272 /
\plot 10.492 19.772  10.492 19.257 /
\plot 10.634 19.829  10.634 19.245 /
\plot 10.776 19.886  10.776 19.287 /
\plot 10.918 19.842  10.918 19.344 /
\plot 11.062 19.772  11.062 19.486 /
\plot  6.505 20.142  9.409 19.799 /
\plot 11.345 19.571  11.189 19.571 /
\plot 11.345 20.426  7.231 20.868 /
\plot  6.505 21.736  7.643 21.637 /
\plot  9.565 19.784  9.565 19.784 /
\plot  9.722 19.757  9.722 19.757 /
\plot  9.851 19.742  9.851 19.742 /
\plot  9.993 19.729  9.993 19.729 /
\plot 10.135 19.715 10.135 19.715 /
\plot 10.279 19.685 10.279 19.685 /
\plot 10.420 19.685 10.420 19.685 /
\plot 10.562 19.657 10.562 19.657 /
\plot 10.706 19.643 10.706 19.643 /
\plot 10.990 19.615 10.990 19.615 /
\plot  5.863 20.712  5.863 20.712 /
\plot  6.005 20.769  6.005 20.769 /
\plot  6.149 20.811  6.149 20.811 /
\plot  6.291 20.868  6.291 20.868 /
\plot  6.433 20.925  6.433 20.925 /
\plot  6.574 20.940  6.574 20.940 /
\plot  6.718 20.925  6.718 20.925 /
\plot  6.860 20.896  6.860 20.896 /
\plot  7.002 20.883  7.002 20.883 /
\plot  5.080 20.426  5.791 20.682 /
\plot 11.153 21.050 10.899 20.987 11.153 20.923 /
\plot 10.899 20.987  12.495 20.987 /
\plot  7.785 21.311  8.911 21.196 /
\plot  7.857 21.609  7.857 21.609 /
\plot  7.999 21.579  7.999 21.579 /
\plot  8.143 21.567  8.143 21.567 /
\plot  8.285 21.567  8.285 21.567 /
\plot  8.426 21.552  8.426 21.552 /
\plot  8.568 21.552  8.568 21.552 /
\plot  8.712 21.537  8.712 21.537 /
\plot  8.854 21.510  8.854 21.510 /
\plot  7.789 22.193   7.789 21.315 /
\plot  8.498 22.130   8.498 21.241 /
\plot  8.636 22.109   8.636 21.230 /
\plot  8.784 22.098   8.784 21.209 /
\plot  8.922 22.087   8.922 21.198 /
\plot  7.933 22.193   7.933 21.294 /
\plot  8.075 22.172   8.075 21.283 /
\plot  8.219 22.151   8.219 21.273 /
\plot  8.361 22.140   8.361 21.262 /
\plot  9.239 21.406  9.028 21.251  9.287 21.288 /
\plot  9.028 21.251 11.091 22.092 /
\plot  9.790 21.425 11.331 21.283 /
\plot  9.076 21.474  9.267 21.457 /
\plot  6.134 21.692  6.346 21.537  6.232 21.773 /
\plot  6.346 21.537  5.884 22.092 /
\plot  6.075 21.601  5.108 21.283 /
\plot  6.392 21.696  6.505 21.742 /
\setdashes < 0.0677cm>
\plot  9.779 20.170  9.779 19.941 /
\plot 10.065 20.127 10.065 19.928 /
\plot 10.336 20.085 10.336 19.971 /
\plot  6.788 21.311  6.788 21.196 /
\plot  6.505 21.338  6.505 21.224 /
\plot  6.219 21.366  6.219 21.253 /
\plot 10.848 19.628 10.848 19.628 /
\plot  5.935 21.380  5.935 21.296 /
\setsolid
\put{$\biggr\lmoustache$} [lB] at  4.255 20.733
\put{$\biggr\rmoustache$} [lB] at  4.255 19.907
\put{$G_q$} [lB] at  3.524 20.288
\put{physical sheet} [lB] at 12.810 20.906
\put{cut of $e^{H(h)}$} [lB] at 11.383 22.210
\put{other sheets} [lB] at  7.601 22.536
\put{cut of $e^{F(h)}$} [lB] at  4.748 22.274
\linethickness=0pt
\putrectangle corners at  3.524 22.765 and 15.153 18.834
\endpicture
\vskip 20pt
\centerline{{\bf Fig. 4:} Analytic structure of $G(h)$}
\vskip 20pt

\hskip -19pt In appendix B several examples are presented to illustrate
explicitly this general analytic structure.

Equation \HGprod\ together with \hGexp\ contains sufficient
information to find the logarithmic derivative of the character. These
two equations represent a well-defined Cauchy-Riemann problem for
$F(h)$ which can be explicitly solved. We will present the solution
elsewhere.

\newsec{Almost flat planar diagrams}

We now have all the tools necessary to reduce our model of dually
weighted graphs to a well defined Cauchy-Riemann problem. In this
section we will analyse the case in which only positive curvature
defects are allowed on the surface, arbitrary amounts of negative
curvature being introduced at a single point. This is done by studying
the particular case $t_q=t_2 \delta_{q,2}+ t_4 \delta_{q,4}$, which
generates the flat patches (see Fig.~1(a)) and the positive defects
(see Fig.~1(b)). The correlators \expsam\ then correspond to the
insertion of a single defect of curvature $(2-k)~\pi$ (see also
Fig.~2). They will be extracted using eq.\hGexp, after explicit
calculation of the function $h(G)$.

{}From the analysis of the large $N$ limit of the character in the
previous section, we know that the product in \HGprod\ contains only
two sheets $G_q(h)$ (see Fig.~4). We label the physical sheet $G_1(h)$
and the sheet connected to it by the cut of $e^{F(h)}$, $G_2(h)$.
Taking the logarithm of equation \HGprod, we summarize the information
extracted from the large $N$ limit of the character by
\eqn\fpfph{
F_1(h) + F_2(h) + H(h) = -\ln (-{h \over t_4}),}
where $\ln G_i(h) = F_i(h) + H(h)$. The two sheets $G_1(h)$ and
$G_2(h)$ are glued together by the square root cut coming from
$F(h)$. The combination $F_1(h) + F_2(h)$, evaluated on the cut of
$F(h)$, is twice the constant part of $F(h)$ on the cut (the
discontinuous part of $F(h)$ is of opposite sign on $F_1(h)$ and
$F_2(h)$ and is therefore canceled). We thus have the two equations
\eqn\twofph{\eqalign{
2\cut F(h)+H(h)=&-\ln (-{h \over t_4})\cr
2F(h)+\cut H(h)=&-\ln h,}}
the first coming from the large $N$ limit of the character \HGprod\
and the second being the saddlepoint equation \sdpt. These two
equations tell us about the behaviour of the function $2F(h)+H(h)$ on
the cuts of $F(h)$ and $H(h)$ respectively. We have introduced the
notation $\cut F(h)$ to denote the real part on the cut of $F(h)$, and
similarly for $\cut H(h)$. The principal part integral in \sdpt\ is
thus denoted in \twofph\ by $\cut H(h)$.

Our object now is to reconstruct the analytic function $2F(h)+H(h)$
from its behaviour on its cuts. To do this we have to understand the
complete structure of cuts. First we notice from \hG\ that $G(h)$ is
non zero everywhere in the complex $h$ plane except at infinity. The
combination $F(h)+H(h)$ thus has no logarithmic cut point except for
the one which starts from $h=b$. This corresponds to the end of the
flat part of the density $\rho(h)$. We introduce two functions
$\tilde F(h)$ and $\tilde H(h)$ defined by
\eqn\tildefh{
F(h)=\tilde F(h)-\ln h\quad{\rm and}\quad H(h) = \tilde H(h) + \ln
{h\over h-b},}
in terms of which \twofph\ becomes
\eqn\twoftpht{\eqalign{
2\cut \tilde F(h)+\tilde H(h)=&\ln \bigl(-t_4(h-b)\bigr)\cr
2\tilde F(h)+\cut \tilde H(h)=&\ln (h-b).}}
These two equations define the behaviour of $2\tilde F(h) + \tilde
H(h)$ on all of its cuts. By standard methods we now generate the
full analytic function $2\tilde F(h) + \tilde H(h)$. We introduce
three cut points, $a$, $b$ and $c$ whose values are fixed by
boundary conditions (the points $a$ and $b$ define the cut of $\tilde
H(h)$ and $c$
defines the starting point of the cut of $\tilde F(h)$ which
goes from $c$ to $-\infty$) and generate the full
analytic function by performing the contour integral
\eqn\contint{\eqalign{
2\tilde F(h)+\tilde H(h)=\sqrt{(h-c)(h-b)(h-a)}\biggl[&
\oint_{C_H}\,{ds\over 2\pi i}
          {\ln (s-b) \over (h-s)\sqrt{(s-c)(s-b)(s-a)}}\cr+&
\oint_{C_F}\,{ds\over 2\pi i}
     {\ln \bigl(-t_4(s-b)\bigr) \over
(h-s)\sqrt{(s-c)(s-b)(s-a)}}\biggr].}}
The contours $C_H$ and $C_F$ are illustrated in Fig.~5(a).
The slanted zigzag line corresponds to the cut of $\ln(h-b)$.
Expanding the contours, catching poles on the way and using the fact that
logarithmic cuts have a discontinuity of $\pm i\pi$, we arrive at
\eqn\cbbaint{\eqalign{
2F(h)+H(h)=\ln {t_4\over h}+\sqrt{(h-c)(h-b)(h-a)}\biggl[&
\int_c^b\,{ds\over (h-s)}{1\over\sqrt{(s-c)(s-b)(s-a)}}\cr+&
\int_b^a\,{ds\over (h-s)}
              {{1\over \pi i}\ln
t_4\over\sqrt{(s-c)(s-b)(s-a)}}\biggr].}}
Fig.~5(b) clarifies the sign convention for
$\sqrt{(h-c)(h-b)(h-a)}$ on the real axis above and below the cuts.
Note that, for the cuts of $1/\sqrt{(h-c)(h-b)(h-a)}$  the signs on the
cuts are inverted compared to Fig.~5(b),
i.e. $+i\leftrightarrow -i$. The integrals in \cbbaint\ are
defined to be along the upper side.
\vskip 25pt
\beginpicture
\setcoordinatesystem units <1.00000cm,1.00000cm>
\linethickness=1pt
\plot  6.780 24.799  6.780 24.585 /
\plot  9.055 24.799  9.055 24.585 /
\plot 13.919 24.788 13.919 24.634 /
\plot 14.800 24.773 14.800 24.617 /
\plot 12.526 24.773 12.526 24.617 /
\plot  5.999 23.920  6.058 23.876 /
\plot  6.058 23.876  6.092 24.016 /
\plot  6.092 24.016  6.206 23.931 /
\plot  6.206 23.931  6.240 24.071 /
\plot  6.240 24.071  6.358 23.986 /
\plot  6.358 23.986  6.390 24.124 /
\plot  6.390 24.124  6.507 24.041 /
\plot  6.507 24.041  6.538 24.179 /
\plot  6.538 24.179  6.655 24.094 /
\plot  6.655 24.094  6.689 24.234 /
\plot  6.689 24.234  6.748 24.191 /
\plot  6.746 24.191  6.803 24.149 /
\plot  6.803 24.149  6.837 24.289 /
\plot  6.837 24.289  6.951 24.202 /
\plot  6.951 24.202  6.987 24.342 /
\plot  6.987 24.342  7.101 24.257 /
\plot  7.101 24.257  7.135 24.395 /
\plot  7.135 24.395  7.250 24.312 /
\plot  7.250 24.312  7.286 24.450 /
\plot  7.286 24.450  7.400 24.365 /
\plot  7.400 24.365  7.436 24.505 /
\plot  7.436 24.505  7.491 24.462 /
\plot  5.258 23.650  5.313 23.605 /
\plot  5.313 23.605  5.349 23.747 /
\plot  5.349 23.747  5.463 23.660 /
\plot  5.463 23.660  5.499 23.800 /
\plot  5.499 23.800  5.613 23.715 /
\plot  5.613 23.715  5.647 23.855 /
\plot  5.647 23.855  5.762 23.770 /
\plot  5.762 23.770  5.795 23.910 /
\plot  5.795 23.910  5.914 23.823 /
\plot  5.914 23.823  5.946 23.963 /
\plot  5.946 23.963  6.003 23.920 /
\plot  7.495 24.462  7.552 24.422 /
\plot  7.552 24.422  7.588 24.560 /
\plot  7.588 24.560  7.700 24.475 /
\plot  7.700 24.475  7.734 24.613 /
\plot  7.734 24.613  7.851 24.530 /
\plot  7.851 24.530  7.885 24.668 /
\plot  7.885 24.668  8.001 24.583 /
\plot  8.001 24.583  8.035 24.723 /
\plot  8.035 24.723  8.151 24.638 /
\plot  8.151 24.638  8.170 24.714 /
\plot  4.974 24.699  9.430 24.699 /
\plot  8.668 24.363  8.755 24.450 /
\plot  8.666 24.539  8.748 24.454 /
\plot  5.654 24.395  5.740 24.481 /
\plot  5.652 24.572  5.736 24.486 /
\plot 10.725 24.706 15.181 24.706 /
\plot  5.078 24.486 	 5.159 24.485
	 5.237 24.485
	 5.312 24.484
	 5.384 24.484
	 5.454 24.484
	 5.522 24.484
	 5.586 24.484
	 5.649 24.484
	 5.766 24.484
	 5.875 24.485
	 5.976 24.486
	 6.068 24.487
	 6.154 24.488
	 6.232 24.490
	 6.305 24.493
	 6.372 24.495
	 6.492 24.501
	 6.596 24.509
	 6.656 24.513
	 6.733 24.519
	 6.809 24.532
	 6.871 24.555
	 6.928 24.638
	 6.941 24.704
	 6.928 24.784
	 6.883 24.843
	 6.827 24.864
	 6.755 24.877
	 6.684 24.883
	 6.627 24.888
	 6.523 24.896
	 6.402 24.903
	 6.335 24.906
	 6.262 24.909
	 6.182 24.911
	 6.096 24.913
	 6.003 24.915
	 5.902 24.916
	 5.792 24.917
	 5.674 24.918
	 5.611 24.918
	 5.546 24.918
	 5.478 24.918
	 5.408 24.918
	 5.335 24.918
	 5.259 24.918
	 5.180 24.918
	 5.099 24.917
	/
\plot  8.164 24.765 	 8.242 24.850
	 8.341 24.895
	 8.446 24.926
	 8.563 24.936
	 8.680 24.939
	 8.781 24.938
	 8.882 24.932
	 8.975 24.914
	 9.068 24.888
	 9.188 24.809
	 9.239 24.740
	 9.239 24.704
	 9.233 24.619
	 9.182 24.549
	 9.119 24.496
	 9.004 24.471
	 8.888 24.460
	 8.824 24.455
	 8.744 24.452
	 8.663 24.451
	 8.600 24.452
	 8.491 24.461
	 8.382 24.479
	 8.293 24.506
	 8.208 24.549
	 8.164 24.613
	/
\plot	 8.172 24.699  8.194 24.732
 	 8.215 24.749
	 8.237 24.732
	 8.259 24.699
	 8.280 24.667
	 8.302 24.650
	 8.324 24.667
	 8.346 24.699
	 8.368 24.732
	 8.390 24.749
	 8.412 24.732
	 8.433 24.699
	 8.455 24.667
	 8.477 24.650
	 8.498 24.667
	 8.520 24.699
	 8.543 24.732
	 8.566 24.749
	 8.587 24.732
	 8.608 24.699
	 8.630 24.667
	 8.651 24.650
	 8.672 24.667
	 8.694 24.699
	 8.716 24.732
	 8.739 24.749
	 8.761 24.732
	 8.782 24.699
	 8.804 24.667
	 8.826 24.650
	 8.848 24.667
	 8.869 24.699
	 8.890 24.732
	 8.912 24.749
	 8.934 24.732
	 8.957 24.699
	 8.979 24.667
	 9.000 24.650
	 9.022 24.667
	 /
\plot  9.022 24.667  9.045 24.699 /
\plot	 5.897 24.699  5.920 24.732
 	 5.943 24.749
	 5.965 24.732
	 5.986 24.699
	 6.007 24.667
	 6.028 24.650
	 6.049 24.667
	 6.071 24.699
	 6.094 24.732
	 6.117 24.749
	 6.138 24.732
	 6.160 24.699
	 6.182 24.667
	 6.204 24.650
	 6.225 24.667
	 6.247 24.699
	 6.269 24.732
	 6.291 24.749
	 6.313 24.732
	 6.335 24.699
	 6.356 24.667
	 6.378 24.650
	 6.399 24.667
	 6.420 24.699
	 6.442 24.732
	 6.464 24.749
	 6.487 24.732
	 6.509 24.699
	 6.531 24.667
	 6.553 24.650
	 6.574 24.667
	 6.596 24.699
	 6.617 24.732
	 6.638 24.749
	 6.660 24.732
	 6.682 24.699
	 6.705 24.667
	 6.727 24.650
	 6.749 24.667
	 /
\plot  6.749 24.667  6.771 24.699 /
\plot	 5.025 24.699  5.047 24.732
 	 5.069 24.749
	 5.091 24.732
	 5.112 24.699
	 5.135 24.667
	 5.158 24.650
	 5.179 24.667
	 5.201 24.699
	 5.222 24.732
	 5.243 24.749
	 5.265 24.732
	 5.287 24.699
	 5.310 24.667
	 5.332 24.650
	 5.354 24.667
	 5.376 24.699
	 5.397 24.732
	 5.419 24.749
	 5.441 24.732
	 5.463 24.699
	 5.483 24.667
	 5.504 24.650
	 5.527 24.667
	 5.549 24.699
	 5.571 24.732
	 5.592 24.749
	 5.614 24.732
	 5.636 24.699
	 5.658 24.667
	 5.680 24.650
	 5.702 24.667
	 5.725 24.699
	 5.747 24.732
	 5.768 24.749
	 5.789 24.732
	 5.811 24.699
	 5.832 24.667
	 5.854 24.650
	 5.876 24.667
	 /
\plot  5.876 24.667  5.897 24.699 /
\plot	13.923 24.706 13.945 24.739
 	13.966 24.755
	13.988 24.739
	14.010 24.705
	14.031 24.672
	14.053 24.655
	14.075 24.672
	14.097 24.705
	14.119 24.739
	14.141 24.755
	14.163 24.739
	14.184 24.705
	14.206 24.672
	14.228 24.655
	14.249 24.672
	14.271 24.705
	14.294 24.739
	14.317 24.755
	14.338 24.739
	14.359 24.705
	14.381 24.672
	14.402 24.655
	14.423 24.672
	14.445 24.705
	14.467 24.739
	14.490 24.755
	14.512 24.739
	14.533 24.705
	14.555 24.672
	14.577 24.655
	14.599 24.672
	14.620 24.705
	14.641 24.739
	14.663 24.755
	14.685 24.739
	14.708 24.705
	14.730 24.672
	14.751 24.655
	14.773 24.672
	 /
\plot 14.773 24.672 14.796 24.706 /
\plot	11.648 24.706 11.670 24.739
 	11.692 24.755
	11.715 24.739
	11.737 24.705
	11.758 24.672
	11.779 24.655
	11.800 24.672
	11.822 24.705
	11.845 24.739
	11.868 24.755
	11.889 24.739
	11.911 24.705
	11.933 24.672
	11.955 24.655
	11.976 24.672
	11.998 24.705
	12.019 24.739
	12.042 24.755
	12.064 24.739
	12.086 24.705
	12.107 24.672
	12.129 24.655
	12.150 24.672
	12.171 24.705
	12.193 24.739
	12.215 24.755
	12.238 24.739
	12.260 24.705
	12.282 24.672
	12.304 24.655
	12.325 24.672
	12.347 24.705
	12.368 24.739
	12.389 24.755
	12.411 24.739
	12.433 24.705
	12.456 24.672
	12.478 24.655
	12.500 24.672
	 /
\plot 12.500 24.672 12.522 24.706 /
\plot	10.776 24.706 10.798 24.739
 	10.820 24.755
	10.842 24.739
	10.863 24.705
	10.886 24.672
	10.909 24.655
	10.930 24.672
	10.952 24.705
	10.973 24.739
	10.994 24.755
	11.016 24.739
	11.038 24.705
	11.061 24.672
	11.083 24.655
	11.105 24.672
	11.127 24.705
	11.148 24.739
	11.170 24.755
	11.191 24.739
	11.212 24.705
	11.233 24.672
	11.255 24.655
	11.278 24.672
	11.300 24.705
	11.322 24.739
	11.343 24.755
	11.365 24.739
	11.387 24.705
	11.409 24.672
	11.430 24.655
	11.452 24.672
	11.474 24.705
	11.497 24.739
	11.519 24.755
	11.540 24.739
	11.561 24.705
	11.583 24.672
	11.605 24.655
	11.627 24.672
	 /
\plot 11.627 24.672 11.648 24.706 /
\put{$c$} [lB] at  6.731 25.015
\put{$b$} [lB] at  8.098 25.030
\put{$a$} [lB] at  8.994 25.030
\put{$C_F$} [lB] at  5.414 24.107
\put{$C_H$} [lB] at  8.424 24.011
\put{$+i$} [lB] at 11.430 24.826
\put{$c$} [lB] at 12.446 24.862
\put{$b$} [lB] at 13.805 24.831
\put{$+$ve} [lB] at 15.456 24.687
\put{$+i$} [lB] at 14.228 24.331
\put{$-i$} [lB] at 11.436 24.352
\put{$a$} [lB] at 14.734 24.831
\put{$-$ve} [lB] at 12.958 24.718
\put{$-i$} [lB] at 14.222 24.816
\put{(a) Contours for \contint\ } [1B] at 6.620 23.067
\put{(b) Sign convention for $\sqrt{(h-c)(h-b)(h-a)}$} [1B] at 14.000 23.067
\linethickness=0pt
\putrectangle corners at  4.949 25.309 and 15.456 22.991
\endpicture
\vskip 20pt
\centerline{{\bf Fig. 5:} Contours and sign conventions for \contint\
and \cbbaint\ }
\vskip 20pt

To fix the constants $a$, $b$ and $c$, we expand \cbbaint\ for large
$h$ and compare the resulting power series expansion to that obtained
from inverting \hG :
\eqn\bcexpn{
2F(h)+H(h)=\ln {t_4\over h}+{1\over\sqrt{h}}{t_2\over\sqrt{t_4}}\quad+\quad
        {\cal O}\bigl({1\over h\sqrt{h}}).}
The terms of ${\cal O}\bigl({1\over h\sqrt{h}})$ depend on the as yet
unknown function $\psi(G)$. Expanding \cbbaint\ for large $h$ and
comparing to \bcexpn\ we find the two boundary conditions
\eqn\twobc{
t_4=q=e^{-\pi {K'\over K}}\quad{\rm and}\quad
                         {t_2\over \sqrt{t_4}}={\pi\over K}\sqrt{a-c},}
with $K$ and $K'$ complete elliptic integrals of the first kind, defined
in terms of their respective moduli $k$ and $k'=\sqrt{1-k^2}$ through
\eqn\kdef{
k=\sqrt{{a-b\over a-c}}.}
The first condition fixes $k$ and hence the ratio of the distances
separating the cut points, and the second condition fixes $\,\,a-c\,\,$,
i.e.~the scale. The condition needed to fix the position of
the cut points along the real axis is provided by the
condition that the density must be normalized to one.

We now perform the integrals in \cbbaint\ and, after using the first
boundary condition and an identity between elliptic functions\foot{
For this and many other relations between Jacobi's elliptic
functions and theta functions useful for performing the
calculations of this section see e.g.~\BYRD,\LAWD.},
we obtain
\eqn\twoFH{
2F(h)+H(h)=-\ln h -{i\pi\over K}
             \sn^{-1}\bigl(\sqrt{{a-h\over a-b}},k\bigr),}
where $\sn^{-1}(z,k)$ is the inverse Jacobi elliptic function.
Using the saddle point equation $2F(h)+\cut H(h)=-\ln h$ and the fact
that the resolvent for the Young tableau can be written as $H(h)=\cut
H(h) \mp i\pi\rho(h)$, we can immediately write down the expression for
the density of Young tableau boxes as
\eqn\dens{
\rho(h)={1\over K}\sn^{-1}\bigl(\sqrt{{a-h\over a-b}},k\bigr).}

The Jacobi elliptic function $\sn(z,k)$ is a generalisation of
$\sin(z)$ with quarter period $K$. In fact, in the limit $k\rightarrow
0$, which corresponds to $t_4\rightarrow 0$, the expression for the
density becomes precisely $(2/\pi) \sin^{-1}(\sqrt{(a-h)/(a-b)})$.

Integrating $\rho(h)$ from $b$ to $a$ and equating the answer to $1-b$
to ensure that the density is normalized to $1$ (the flat portion from $0$
to $b$ gives a contribution $b$), gives the final boundary condition
\eqn\abc{
a=1+{t_2^2\over \pi^2 t_4}(K^2-EK),}
where $E$ is the complete elliptic integral of the second kind.

{}From the expression for the density we now generate the full Young
tableau resolvent, $H(h)$, in the standard way and obtain
following expression:
\eqn\Hsoln{\eqalign{
H(h)=&{h\over h-b}+\int_b^a dh'\ {\rho(h') \over h-h'}\cr
    =&\ln h -{i\pi\over K}\sn^{-1}\sqrt{{a-h\over a-b}}+
2\ln\Biggl({
     \theta_4\Bigl({\pi\over 2K}\sn^{-1}\sqrt{{a-h\over a-b}}\Bigr)
      \over q^{1/4}\bigl[(a-c)(a-b)\bigr]^{1/4}\theta_4(0)}\Biggr).}}
Using the above expression for $H(h)$, the eq. \twoFH\ for $2F(h)+H(h)$
and the quasi-periodicity of theta functions, we can
write the expression for $G(h)$,
\eqn\Gsoln{
G(h)=-{1\over D}
  \theta_4\Bigl({\pi\over 2K}\sn^{-1}\sqrt{{a-h\over a-b}}+
     {i\pi K'\over K}\Bigr)}
and its inverse
\eqn\hofG{
h=a-{q^{3/2}\over G^2}\Bigl[{\theta_1\bigl[\theta_4^{-1}(-GD)\bigr]\over
                            \theta_4(0)}\Bigr]^2,}
where the constant $D$ is given by
\eqn\Ddef{
D={t_2K\sqrt{k}~\theta_4(0)\over\pi q^{5/4}}=
                        {t_2\theta_1'(0)\over 2q^{5/4}}.}
To simplify \hofG\ we have used the definition of the Jacobi elliptic
function in terms of theta functions.  In view of eq.\hGexp , we see
that we have now explicitly calculated the generating function for
the correlators for models I and III.

We will now expand eq.\hofG\ and read off the correlators as the
coefficients of the positive powers of $G$.  Notice that
\hofG\ is a multivalued function since the function $\theta_4(z)$ is
periodic as $z$ is varied in the real direction and quasiperiodic in
the imaginary direction. We must thus choose the correct zero of
the $\theta_4(z)$ function about which to expand.
The physical sheet
corresponds to expanding about the zero $z={i\pi K'\over K}$. Using
the definition of the Jacobi elliptic function $\sn(u)$ in terms of
theta functions and shifting the arguments of the theta functions
using their quasi-periodicity, we rewrite \hofG\ as the pair
of equations
\eqn\hofGph{
h=a+{q\over G^2}\Bigl(e^{iz}{\theta_4(z)\over \theta_4(0)}\Bigr)^2\quad
{\rm with\,}\,z\,{\rm\,the\,\,solution\,\,of}\quad
{t_2G\over 2q}=ie^{iz}{\theta_1(z)\over \theta_1'(0)}.}
Expanding this for small $G$ we find that the first three terms
give (as expected from \hGexp ) $h={q\over G^2}+{t_2\over G}+1+{\cal
O}(G)$. Expanding three orders further, permits us
(using \hGexp ) to read off the first three moments of model III
(which are also the moments of model I):
\eqn\moms{\eqalign{
\langle~{1\over N}~\Tr~[(MA_4)^2]~\rangle_{III}=&
       {t_2^3\over 24 q^2}(1+f_3)\cr
\langle~{1\over N}~\Tr~[(MA_4)^4]~\rangle_{III}=&
     {t_2^4\over 192 q^3}\bigl(-8(1+f_3)+3f_2^2-4f_2f_3+f_4\bigr)\cr
\langle~{1\over N}~\Tr~[(MA_4)^6]~\rangle_{III}=&
   {t_2^5\over 1920 q^4}
        \bigl(81+90f_3-30f_2^2+40f_2f_3+10f_3^2-10f_4-f_5\bigr).}}
where for convenience we have defined
\eqn\fdef{\eqalign{
f_2=&{\theta_4''(0)\over \theta_4(0)}\,\,\,=
{4\over\pi^2}(K^2-EK)\cr
f_4=&{\theta_4''''(0)\over \theta_4(0)}\,=
{16K^2\over\pi^4}\bigl((3-2k^2)K^2-6EK+3E^2\bigr)\cr
f_3=&{\theta_1'''(0)\over \theta_1'(0)}\,\,=
{4\over\pi^2}\bigl((2-k^2)K^2-3EK)\cr
f_5=&{\theta_1'''''(0)\over \theta_1'(0)}=
{16K^2\over\pi^4}\bigl((6-6k^2+k^4)K^2-10(2-k^2)EK+15E^2\bigr).}}
and have then expressed these derivatives as combinations of the
complete elliptic integrals $K$, $E$ and their modulus $k$.

We can now give a simple physical interpretation of these moments. The first
two are directly related to the free energy ${\cal F}(t_2,t_4)$. The latter is
defined as the sum over all possible surfaces with the
topology of a sphere that can be constructed out of flat space and
positive curvature defects. It is impossible to put a flat surface
onto the sphere, so positive curvature defects are needed to close the
surface. Since the defects in this model have a deficit angle of $\pi$
it takes precisely four of them to close the surface into a
sphere. The surfaces are in the form of a cylinder with both ends
flattened. The four $t_2$ defects sit at the corners. Below we
illustrate the free energy for model III:
\eqn\fdiag{
{\cal F}(t_2,t_4)=\sum
\beginpicture
\setcoordinatesystem units <1.00000cm,1.00000cm>
\linethickness=1pt
\plot  1.822 24.280  2.644 23.556 /
\plot  5.609 25.383  5.908 24.392 /
\plot  2.225 24.392 	 2.337 24.361
	 2.414 24.331
	 2.489 24.295
	 2.590 24.228
	 2.684 24.151
	 2.747 24.078
	 2.805 23.999
	 2.855 23.906
	 2.893 23.806
	 2.885 23.692
	/
\plot  2.707 24.545 	 2.784 24.522
	 2.840 24.503
	 2.908 24.473
	 2.997 24.410
	 3.078 24.337
	 3.143 24.255
	 3.200 24.168
	 3.232 24.085
	 3.255 23.999
	 3.266 23.929
	 3.264 23.857
	 3.217 23.777
	/
\plot  3.152 24.668 	 3.273 24.631
	 3.336 24.602
	 3.426 24.523
	 3.505 24.433
	 3.554 24.364
	 3.594 24.289
	 3.613 24.180
	 3.620 24.071
	 3.614 24.003
	 3.603 23.935
	 3.554 23.855
	/
\plot  3.505 24.748 	 3.582 24.720
	 3.638 24.698
	 3.708 24.668
	 3.783 24.620
	 3.852 24.562
	 3.898 24.485
	 3.931 24.401
	 3.949 24.294
	 3.956 24.185
	 3.948 24.104
	 3.931 24.024
	 3.882 23.944
	/
\plot  3.893 24.829 	 4.004 24.796
	 4.062 24.771
	 4.172 24.678
	 4.215 24.588
	 4.244 24.494
	 4.262 24.385
	 4.269 24.276
	 4.262 24.195
	 4.244 24.115
	 4.197 24.035
	/
\plot  4.316 24.934 	 4.411 24.900
	 4.467 24.828
	 4.508 24.748
	 4.548 24.657
	 4.580 24.564
	 4.598 24.455
	 4.606 24.346
	 4.598 24.265
	 4.580 24.185
	 4.532 24.105
	/
\plot  4.646 25.038 	 4.743 25.006
	 4.798 24.933
	 4.839 24.854
	 4.879 24.762
	 4.911 24.668
	 4.929 24.560
	 4.936 24.450
	 4.928 24.369
	 4.911 24.289
	 4.862 24.210
	/
\plot  4.976 25.118 	 5.074 25.087
	 5.128 25.013
	 5.169 24.934
	 5.209 24.843
	 5.241 24.748
	 5.259 24.640
	 5.266 24.530
	 5.259 24.450
	 5.241 24.371
	 5.194 24.289
	/
\plot  5.283 25.199 	 5.378 25.167
	 5.434 25.094
	 5.476 25.015
	 5.515 24.923
	 5.548 24.829
	 5.565 24.721
	 5.573 24.613
	 5.565 24.531
	 5.548 24.450
	 5.501 24.371
	/
\plot  2.153 24.007 	 2.263 24.048
	 2.322 24.071
	 2.403 24.110
	 2.483 24.151
	 2.595 24.232
	 2.665 24.273
	 2.755 24.323
	 2.845 24.372
	 2.917 24.409
	 2.973 24.437
	 3.045 24.471
	 3.118 24.505
	 3.175 24.530
	 3.235 24.554
	 3.311 24.584
	 3.388 24.612
	 3.448 24.634
	 3.571 24.673
	 3.647 24.697
	 3.726 24.721
	 3.806 24.745
	 3.882 24.767
	 4.005 24.803
	 4.101 24.832
	 4.223 24.868
	 4.345 24.904
	 4.441 24.932
	 4.547 24.964
	 4.613 24.984
	 4.682 25.005
	 4.752 25.025
	 4.817 25.045
	 4.923 25.078
	 4.979 25.095
	 5.055 25.119
	 5.160 25.153
	 5.226 25.174
	 5.302 25.199
	/
\plot  2.337 23.846 	 2.420 23.878
	 2.480 23.901
	 2.555 23.933
	 2.631 23.978
	 2.725 24.038
	 2.818 24.098
	 2.893 24.143
	 2.951 24.174
	 3.022 24.210
	 3.103 24.250
	 3.188 24.292
	 3.274 24.332
	 3.355 24.370
	 3.429 24.402
	 3.490 24.426
	 3.552 24.447
	 3.629 24.470
	 3.716 24.493
	 3.807 24.517
	 3.899 24.541
	 3.986 24.563
	 4.063 24.584
	 4.125 24.602
	 4.195 24.624
	 4.279 24.652
	 4.373 24.684
	 4.473 24.718
	 4.572 24.753
	 4.666 24.785
	 4.751 24.813
	 4.820 24.835
	 4.938 24.871
	 5.012 24.892
	 5.090 24.915
	 5.167 24.937
	 5.241 24.959
	 5.359 24.998
	 5.475 25.042
	 5.566 25.080
	 5.690 25.133
	/
\plot  2.506 23.692 	 2.579 23.722
	 2.633 23.744
	 2.701 23.774
	 2.788 23.818
	 2.898 23.877
	 3.008 23.936
	 3.095 23.982
	 3.172 24.023
	 3.270 24.074
	 3.369 24.124
	 3.448 24.160
	 3.530 24.186
	 3.634 24.215
	 3.738 24.243
	 3.821 24.265
	 3.891 24.284
	 3.981 24.309
	 4.071 24.333
	 4.142 24.354
	 4.215 24.376
	 4.308 24.404
	 4.400 24.434
	 4.473 24.458
	 4.551 24.485
	 4.647 24.517
	 4.755 24.555
	 4.869 24.594
	 4.982 24.635
	 5.089 24.674
	 5.185 24.709
	 5.262 24.740
	 5.333 24.769
	 5.429 24.812
	 5.489 24.840
	 5.560 24.873
	 5.643 24.912
	 5.738 24.958
	/
\plot  1.998 24.151 	 2.084 24.177
	 2.147 24.196
	 2.225 24.223
	 2.313 24.267
	 2.423 24.326
	 2.532 24.387
	 2.618 24.433
	 2.700 24.474
	 2.803 24.525
	 2.907 24.575
	 2.991 24.610
	 3.104 24.648
	 3.196 24.674
	 3.321 24.708
	/
\plot  2.813 23.643 	 2.924 23.703
	 3.006 23.746
	 3.112 23.798
	 3.216 23.843
	 3.282 23.870
	 3.351 23.898
	 3.421 23.925
	 3.487 23.951
	 3.594 23.990
	 3.719 24.032
	 3.794 24.055
	 3.874 24.079
	 3.959 24.105
	 4.047 24.131
	 4.139 24.157
	 4.231 24.184
	 4.323 24.211
	 4.414 24.238
	 4.503 24.264
	 4.588 24.289
	 4.669 24.313
	 4.743 24.336
	 4.868 24.378
	 4.995 24.424
	 5.073 24.454
	 5.155 24.487
	 5.237 24.519
	 5.315 24.550
	 5.440 24.602
	 5.563 24.657
	 5.661 24.702
	 5.722 24.730
	 5.793 24.763
	/
\plot  5.850 24.579 	 5.769 24.544
	 5.700 24.513
	 5.588 24.466
	 5.508 24.433
	 5.448 24.409
	 5.363 24.379
	 5.255 24.343
	 5.146 24.308
	 5.061 24.280
	 4.972 24.252
	 4.900 24.230
	 4.803 24.200
	/
\plot  2.652 23.565 	 2.772 23.622
	 2.836 23.652
	 2.937 23.698
	 3.040 23.741
	 3.166 23.781
	 3.242 23.803
	 3.323 23.825
	 3.409 23.849
	 3.499 23.872
	 3.591 23.896
	 3.685 23.920
	 3.779 23.943
	 3.872 23.966
	 3.962 23.988
	 4.049 24.009
	 4.130 24.029
	 4.206 24.047
	 4.275 24.064
	 4.335 24.079
	 4.458 24.111
	 4.530 24.130
	 4.607 24.151
	 4.690 24.172
	 4.776 24.195
	 4.864 24.218
	 4.953 24.241
	 5.043 24.264
	 5.131 24.286
	 5.218 24.307
	 5.301 24.327
	 5.379 24.345
	 5.452 24.361
	 5.577 24.384
	 5.689 24.396
	 5.802 24.401
	 5.908 24.392
	/
\plot  1.854 24.280 	 1.975 24.312
	 2.096 24.351
	 2.216 24.392
	 2.305 24.430
	 2.394 24.467
	 2.457 24.487
	 2.524 24.507
	 2.597 24.528
	 2.674 24.550
	 2.755 24.571
	 2.840 24.593
	 2.928 24.616
	 3.020 24.638
	 3.114 24.661
	 3.210 24.684
	 3.309 24.707
	 3.409 24.731
	 3.511 24.754
	 3.614 24.778
	 3.717 24.801
	 3.821 24.825
	 3.924 24.848
	 4.028 24.872
	 4.130 24.895
	 4.232 24.919
	 4.332 24.942
	 4.431 24.965
	 4.527 24.988
	 4.621 25.011
	 4.713 25.034
	 4.801 25.056
	 4.886 25.078
	 4.967 25.099
	 5.044 25.121
	 5.117 25.142
	 5.184 25.162
	 5.247 25.182
	 5.356 25.223
	 5.463 25.271
	 5.518 25.303
	 5.618 25.368
	/
\linethickness=0pt
\putrectangle corners at  1.776 25.430 and  5.954 24.250
\endpicture
}
\vskip 20pt
\hskip -20pt
Note that the flattened ends can have an angle of twist between them.
The four $t_2$ defects correspond to vertices ${t_2\over
2}~\Tr~[(MA_4)^2]$, and all other vertices (with four legs) correspond
to the vertex ${t_4\over 4}~\Tr~[(MA_4)^4]$. We see that the
first two moments can be written in terms of the free energy ${\cal
F}(t_2,t_4)$ as
\eqn\Fmom{\eqalign{
\langle~{1\over N}~\Tr~[(MA_4)^2]~\rangle_{III}=&
                      2{\partial\over\partial t_2}{\cal F}(t_2,t_4)\cr
\langle~{1\over N}~\Tr~[(MA_4)^4]~\rangle_{III}=&
                      4{\partial\over\partial t_4}{\cal F}(t_2,t_4).}}
We thus read off the free energy
\eqn\fe{
{\cal F}(t_2,t_4)={t_2^4\over 192 q^2}(1+f_3).}
Using \fdef , the identity
${\partial\over\partial q}=
                 {2K^2kk'^2\over\pi^2q}{\partial\over\partial k}$,
along with standard identies for differentiating complete elliptic
integrals with respect to the modulus $k$, it is trivial to verify that
the moment $\langle~{1\over N}~\Tr~[(MA_4)^4]~\rangle_{III}$ given in
\moms\ is indeed four times the derivative of the free energy
with respect to $t_4=q$.

Using the definition of $f_3$ in terms of derivatives of the first
theta function, $\theta_1(z)$, along with the standard definition of
the theta function as an infinite product, allows us to write the free
energy as
\eqn\feq{
{\cal F}(t_2,t_4)={-t_2^4\over 8}{\partial\over\partial q^2}
\ln\Bigl[\prod_{n=1}^{\infty}(1-q^{2n})\Bigr]=
\sum \hskip 15pt \raise-20pt\hbox{
\beginpicture
\setcoordinatesystem units <1.00000cm,1.00000cm>
\linethickness=1pt
\ellipticalarc axes ratio  1.611:0.961  360 degrees
	from  4.367 24.678 center at  2.756 24.678
\put{$\bullet$} [1B] at 3.203 23.943
\plot  2.805 24.532  2.796 24.437 /
\plot  2.796 24.437  2.790 24.240 /
\plot  2.790 24.240  2.800 24.043 /
\plot  2.800 24.043  2.822 23.829 /
\plot  2.822 23.829  2.845 23.717 /
\plot  3.488 24.767 	 3.380 24.860
	 3.319 24.907
	 3.220 24.966
	 3.118 25.019
	 3.056 25.043
	 2.976 25.072
	 2.896 25.098
	 2.832 25.116
	 2.733 25.136
	 2.635 25.152
	 2.527 25.169
	 2.419 25.184
	 2.296 25.190
	 2.174 25.193
	 2.059 25.189
	 1.945 25.180
	 1.838 25.155
	 1.734 25.121
	 1.650 25.079
	 1.573 25.023
	 1.529 24.943
	 1.501 24.858
	 1.501 24.772
	 1.516 24.687
	 1.561 24.596
	 1.617 24.513
	 1.695 24.438
	 1.780 24.369
	 1.869 24.315
	 1.962 24.268
	 2.023 24.244
	 2.100 24.217
	 2.178 24.192
	 2.239 24.172
	 2.305 24.154
	 2.387 24.131
	 2.470 24.108
	 2.536 24.092
	 2.597 24.080
	 2.674 24.066
	 2.752 24.053
	 2.813 24.043
	 2.877 24.035
	 2.959 24.026
	 3.040 24.018
	 3.105 24.014
	 3.170 24.011
	 3.252 24.010
	 3.334 24.011
	 3.399 24.014
	 3.473 24.019
	 3.566 24.028
	 3.659 24.039
	 3.732 24.052
	 3.849 24.087
	 3.965 24.128
	 4.046 24.164
	 4.125 24.204
	 4.238 24.291
	/
\plot  3.427 24.678 	 3.507 24.731
	 3.552 24.795
	 3.575 24.867
	 3.560 24.928
	 3.512 24.991
	 3.418 25.089
	 3.322 25.154
	 3.222 25.212
	 3.114 25.254
	 3.006 25.290
	 2.921 25.310
	 2.836 25.328
	 2.758 25.341
	 2.680 25.353
	 2.574 25.362
	 2.470 25.366
	 2.370 25.371
	 2.271 25.372
	 2.191 25.366
	 2.110 25.358
	 1.985 25.349
	 1.891 25.329
	 1.797 25.305
	 1.720 25.280
	 1.645 25.252
	 1.576 25.211
	 1.511 25.165
	 1.420 25.076
	 1.372 24.996
	 1.331 24.903
	 1.308 24.778
	 1.317 24.687
	 1.336 24.598
	 1.371 24.511
	 1.416 24.428
	 1.498 24.339
	 1.587 24.257
	 1.710 24.173
	 1.780 24.129
	 1.837 24.096
	 1.964 24.037
	 2.035 24.007
	 2.093 23.984
	 2.166 23.962
	 2.260 23.937
	 2.354 23.913
	 2.428 23.895
	 2.501 23.880
	 2.593 23.862
	 2.686 23.845
	 2.760 23.834
	 2.833 23.827
	 2.925 23.820
	 3.017 23.816
	 3.090 23.815
	 3.173 23.819
	 3.277 23.827
	 3.381 23.837
	 3.463 23.846
	 3.537 23.859
	 3.596 23.871
	 3.677 23.887
	/
\plot  2.764 24.541 	 2.863 24.530
	 2.917 24.526
	 2.974 24.527
	 3.046 24.531
	 3.119 24.535
	 3.177 24.541
	 3.252 24.549
	 3.347 24.561
	 3.442 24.575
	 3.518 24.589
	 3.627 24.619
	 3.736 24.657
	 3.829 24.710
	 3.920 24.767
	 4.022 24.839
	 4.068 24.903
	 4.102 24.972
	 4.111 25.040
	 4.111 25.076
	 4.108 25.144
	 4.085 25.188
	 4.034 25.265
	 4.030 25.269
	/
\plot  3.283 24.602 	 3.388 24.630
	 3.465 24.651
	 3.560 24.682
	 3.681 24.746
	 3.797 24.835
	 3.876 24.951
	 3.901 25.053
	 3.893 25.129
	 3.869 25.188
	 3.844 25.241
	 3.776 25.313
	 3.721 25.358
	 3.655 25.413
	 3.575 25.461
	 3.512 25.489
	 3.448 25.514
	 3.373 25.543
	 3.296 25.569
	 3.207 25.587
	 3.118 25.603
	 2.993 25.626
	/
\plot  3.355 24.678 	 3.247 24.759
	 3.144 24.818
	 3.037 24.871
	 2.919 24.912
	 2.800 24.947
	 2.686 24.973
	 2.572 24.996
	 2.458 25.009
	 2.343 25.019
	 2.247 25.020
	 2.151 25.019
	 2.036 25.010
	 1.922 24.991
	 1.835 24.953
	 1.757 24.892
	 1.734 24.771
	 1.767 24.691
	 1.814 24.619
	 1.911 24.538
	 2.015 24.469
	 2.073 24.441
	 2.147 24.410
	 2.221 24.382
	 2.280 24.361
	 2.356 24.338
	 2.454 24.313
	 2.552 24.289
	 2.629 24.272
	 2.688 24.261
	 2.764 24.249
	 2.839 24.239
	 2.898 24.232
	 2.973 24.224
	 3.068 24.217
	 3.163 24.211
	 3.238 24.208
	 3.311 24.209
	 3.402 24.212
	 3.493 24.217
	 3.564 24.223
	 3.671 24.240
	 3.776 24.263
	 3.873 24.292
	 3.969 24.325
	 4.062 24.369
	 4.153 24.418
	 4.224 24.477
	 4.290 24.541
	 4.343 24.606
	 4.367 24.701
	/
\plot  1.985 24.731 	 2.053 24.657
	 2.161 24.577
	 2.253 24.543
	 2.347 24.517
	 2.468 24.493
	 2.589 24.473
	 2.681 24.459
	 2.797 24.442
	 2.913 24.427
	 3.006 24.418
	 3.085 24.416
	 3.185 24.417
	 3.285 24.420
	 3.363 24.424
	 3.444 24.430
	 3.546 24.440
	 3.647 24.453
	 3.727 24.469
	 3.785 24.487
	 3.856 24.515
	 3.981 24.570
	 4.046 24.610
	 4.108 24.653
	 4.202 24.727
	 4.250 24.799
	 4.286 24.858
	 4.290 24.911
	 4.290 24.979
	/
\plot  1.649 25.381 	 1.756 25.426
	 1.814 25.447
	 1.940 25.477
	 2.066 25.502
	 2.127 25.510
	 2.204 25.520
	 2.282 25.527
	 2.343 25.531
	 2.428 25.532
	 2.536 25.529
	 2.644 25.524
	 2.728 25.519
	 2.800 25.512
	 2.890 25.502
	 2.980 25.490
	 3.050 25.478
	 3.168 25.449
	 3.283 25.413
	 3.375 25.366
	 3.463 25.313
	 3.551 25.247
	 3.632 25.169
	 3.681 25.059
	 3.696 24.979
	 3.677 24.911
	 3.643 24.826
	 3.596 24.767
	 3.552 24.731
	 3.507 24.701
	 3.427 24.661
	 3.363 24.634
	/
\plot  1.327 25.127 	 1.251 25.004
	 1.223 24.907
	 1.206 24.807
	 1.210 24.700
	 1.223 24.594
	 1.240 24.530
	 1.265 24.452
	 1.294 24.376
	 1.323 24.316
	 1.414 24.196
	 1.520 24.088
	 1.580 24.045
	 1.659 23.996
	 1.740 23.951
	 1.806 23.918
	 1.884 23.887
	 1.985 23.852
	 2.088 23.820
	 2.170 23.798
	 2.266 23.778
	 2.390 23.759
	 2.514 23.742
	 2.612 23.730
	 2.688 23.724
	 2.749 23.721
	 2.832 23.717
	/
\plot  2.572 24.545 	 2.500 24.481
	 2.465 24.409
	 2.441 24.333
	 2.429 24.231
	 2.424 24.128
	 2.427 24.016
	 2.436 23.906
	 2.449 23.826
	 2.479 23.749
	 2.523 23.726
	/
\plot  2.324 24.589 	 2.214 24.549
	 2.166 24.482
	 2.134 24.409
	 2.106 24.317
	 2.087 24.223
	 2.081 24.106
	 2.083 23.990
	 2.099 23.895
	 2.134 23.802
	 2.170 23.781
	/
\plot  2.178 24.630 	 2.070 24.634
	 1.979 24.598
	 1.897 24.545
	 1.833 24.455
	 1.789 24.356
	 1.772 24.247
	 1.770 24.138
	 1.788 24.020
	 1.825 23.906
	 1.854 23.874
	/
\plot  2.057 24.701 	 2.015 24.742
	 1.899 24.759
	 1.823 24.754
	 1.748 24.737
	 1.653 24.673
	 1.573 24.589
	 1.547 24.532
	 1.526 24.460
	 1.512 24.386
	 1.505 24.325
	 1.508 24.264
	 1.519 24.189
	 1.536 24.115
	 1.560 24.056
	 1.649 23.982
	/
\plot  2.034 24.759 	 2.002 24.820
	 1.939 24.871
	 1.869 24.907
	 1.782 24.922
	 1.695 24.926
	 1.616 24.903
	 1.541 24.871
	 1.420 24.771
	 1.369 24.677
	 1.331 24.577
	 1.318 24.480
	 1.317 24.384
	 1.343 24.297
	 1.376 24.215
	 1.412 24.143
	/
\plot  2.087 24.687 	 2.102 24.723
	 2.110 24.782
	 2.098 24.848
	 2.030 24.955
	 1.930 25.027
	 1.812 25.063
	 1.753 25.068
	 1.679 25.071
	 1.606 25.068
	 1.547 25.059
	 1.438 25.007
	 1.340 24.939
	 1.275 24.853
	 1.223 24.759
	 1.202 24.646
	 1.198 24.532
	 1.216 24.449
	 1.238 24.387
	 1.270 24.304
	/
\plot  2.123 24.674 	 2.214 24.701
	 2.303 24.718
	/
\plot  2.214 24.661 	 2.239 24.727
	 2.244 24.792
	 2.239 24.858
	 2.215 24.942
	 2.178 25.023
	 2.097 25.110
	 2.002 25.184
	 1.913 25.218
	 1.820 25.241
	 1.719 25.253
	 1.617 25.254
	 1.541 25.242
	 1.465 25.220
	 1.411 25.189
	 1.317 25.121
	/
\plot  3.061 24.562 	 3.109 24.517
	 3.142 24.454
	 3.169 24.388
	 3.184 24.294
	 3.194 24.200
	 3.202 24.093
	 3.207 23.984
	 3.202 23.893
	 3.194 23.802
	 3.190 23.745
	/
\plot  1.985 24.731 	 1.985 24.782
	 2.030 24.820
	 2.110 24.840
	 2.191 24.848
	 2.292 24.855
	 2.394 24.858
	 2.473 24.854
	 2.553 24.848
	 2.597 24.845
	 2.680 24.839
	/
\plot  2.275 24.695 	 2.339 24.786
	 2.362 24.848
	 2.371 24.960
	 2.365 25.034
	 2.352 25.108
	 2.313 25.180
	 2.267 25.245
	 2.163 25.349
	 2.047 25.390
	 1.926 25.415
	 1.860 25.424
	 1.793 25.425
	 1.739 25.414
	 1.640 25.385
	/
\plot  2.083 25.555 	 2.189 25.544
	 2.246 25.531
	 2.317 25.498
	 2.383 25.453
	 2.446 25.369
	 2.496 25.277
	 2.522 25.174
	 2.536 25.068
	 2.529 24.980
	 2.512 24.892
	 2.470 24.799
	 2.396 24.763
	/
\plot  2.548 25.631 	 2.603 25.582
	 2.648 25.491
	 2.684 25.394
	 2.695 25.316
	 2.701 25.237
	 2.706 25.144
	 2.705 25.051
	 2.696 24.975
	 2.680 24.903
	 2.648 24.839
	 2.629 24.826
	/
\plot  2.877 24.826 	 2.862 24.934
	 2.864 25.025
	 2.868 25.116
	 2.869 25.222
	 2.872 25.328
	 2.878 25.408
	 2.885 25.489
	 2.898 25.590
	 2.925 25.639
	/
\plot  3.086 24.790 	 3.046 24.852
	 3.037 24.979
	 3.043 25.079
	 3.054 25.180
	 3.075 25.283
	 3.105 25.385
	 3.150 25.465
	 3.207 25.538
	 3.296 25.586
	/
\plot  3.247 24.723 	 3.198 24.795
	 3.181 24.920
	 3.187 25.007
	 3.203 25.093
	 3.252 25.199
	 3.315 25.296
	 3.398 25.372
	 3.495 25.434
	 3.590 25.452
	 3.685 25.453
	 3.772 25.425
	/
\plot  3.372 24.657 	 3.334 24.714
	 3.315 24.807
	 3.331 24.913
	 3.363 25.013
	 3.428 25.097
	 3.503 25.169
	 3.597 25.227
	 3.696 25.273
	 3.786 25.301
	 3.880 25.317
	 4.005 25.296
	/
\plot  3.399 24.666 	 3.399 24.746
	 3.412 24.822
	 3.435 24.896
	 3.498 24.968
	 3.571 25.027
	 3.655 25.069
	 3.744 25.099
	 3.860 25.114
	 3.977 25.112
	 4.074 25.081
	 4.166 25.036
	 4.259 24.943
	 4.335 24.826
	 4.367 24.704
	/
\plot  3.497 24.795 	 3.539 24.843
	 3.628 24.892
	 3.699 24.911
	 3.772 24.920
	 3.858 24.909
	 3.941 24.884
	 4.018 24.831
	 4.085 24.767
	 4.140 24.682
	 4.183 24.589
	 4.199 24.477
	 4.202 24.365
	 4.192 24.308
	 4.166 24.204
	/
\plot  3.306 24.610 	 3.418 24.570
	 3.471 24.503
	 3.512 24.428
	 3.552 24.304
	 3.562 24.200
	 3.564 24.096
	 3.562 24.010
	 3.556 23.923
	 3.552 23.846
	/
\plot  3.503 24.737 	 3.587 24.737
	 3.632 24.731
	 3.719 24.695
	 3.829 24.589
	 3.874 24.497
	 3.905 24.401
	 3.917 24.292
	 3.920 24.183
	 3.915 24.104
	 3.901 24.026
	 3.861 23.963
	/
\plot  2.030 24.790 	 2.066 24.691
	 2.151 24.638
	 2.218 24.618
	 2.286 24.602
	 2.359 24.585
	 2.432 24.570
	 2.508 24.558
	 2.584 24.549
	 2.670 24.546
	 2.756 24.545
	 2.870 24.547
	 2.985 24.553
	 3.058 24.565
	 3.131 24.579
	 3.219 24.596
	 3.306 24.617
	 3.363 24.638
	 3.448 24.704
	 3.503 24.799
	/
\plot  2.187 24.638 	 2.288 24.714
	 2.392 24.761
	 2.500 24.799
	 2.576 24.817
	 2.652 24.831
	 2.769 24.835
	 2.857 24.829
	 2.944 24.820
	 3.063 24.795
	 3.135 24.772
	 3.207 24.746
	 3.302 24.687
	 3.342 24.646
	/
\linethickness=0pt
\putrectangle corners at  1.113 25.667 and  4.398 25.000
\endpicture
%
%
}
}
\vskip 20pt
\hskip -20pt
In this form we recognize the argument of the logarithm to be the
partition function for the torus. The derivative operator acts to mark
a single point. We have thus found, as illustrated in
 equation \feq , that the free energy can be written
as the free energy for a marked torus. Below we illustrate the
connection between a marked torus and the flattened cylinder
diagrammatically.
\vskip 20pt
\beginpicture
\setcoordinatesystem units <1.00000cm,1.00000cm>
\linethickness=1pt
\setlinear
\ellipticalarc axes ratio  1.611:0.961  360 degrees
	from  7.383 24.045 center at  5.772 24.045
\put {$\bullet$} [1B] at 6.219 23.300
\plot  5.821 23.899  5.812 23.804 /
\plot  5.812 23.804  5.806 23.607 /
\plot  5.806 23.607  5.817 23.410 /
\plot  5.817 23.410  5.838 23.197 /
\plot  5.838 23.197  5.861 23.084 /
\plot 10.590 23.832 10.560 23.760 /
\plot 10.560 23.760 10.577 23.594 /
\plot 10.577 23.594 10.645 23.351 /
\plot 10.645 23.351 10.702 23.150 /
\plot 10.702 23.150 10.774 23.055 /
\plot 11.242 23.061 10.850 23.724 /
\plot 13.568 23.806 14.389 23.082 /
\plot 17.355 24.909 17.653 23.918 /
\plot  6.505 24.134 	 6.396 24.227
	 6.335 24.274
	 6.236 24.333
	 6.134 24.386
	 6.072 24.411
	 5.992 24.439
	 5.912 24.465
	 5.848 24.483
	 5.749 24.503
	 5.652 24.519
	 5.543 24.537
	 5.436 24.551
	 5.313 24.557
	 5.190 24.560
	 5.075 24.556
	 4.961 24.547
	 4.855 24.522
	 4.750 24.488
	 4.666 24.446
	 4.589 24.390
	 4.545 24.311
	 4.517 24.225
	 4.517 24.139
	 4.532 24.054
	 4.577 23.963
	 4.633 23.880
	 4.712 23.805
	 4.796 23.736
	 4.885 23.682
	 4.978 23.635
	 5.039 23.611
	 5.117 23.584
	 5.194 23.559
	 5.256 23.539
	 5.321 23.521
	 5.404 23.498
	 5.486 23.476
	 5.552 23.459
	 5.613 23.447
	 5.690 23.433
	 5.768 23.420
	 5.829 23.410
	 5.893 23.402
	 5.975 23.393
	 6.057 23.385
	 6.121 23.381
	 6.186 23.378
	 6.268 23.377
	 6.351 23.378
	 6.416 23.381
	 6.489 23.386
	 6.582 23.395
	 6.675 23.407
	 6.748 23.419
	 6.865 23.454
	 6.981 23.495
	 7.062 23.531
	 7.142 23.571
	 7.254 23.658
	/
\plot  6.443 24.045 	 6.524 24.098
	 6.568 24.162
	 6.591 24.234
	 6.576 24.295
	 6.528 24.359
	 6.435 24.456
	 6.338 24.521
	 6.238 24.579
	 6.131 24.621
	 6.022 24.657
	 5.937 24.677
	 5.853 24.695
	 5.774 24.708
	 5.696 24.721
	 5.590 24.729
	 5.486 24.733
	 5.387 24.738
	 5.287 24.740
	 5.207 24.733
	 5.127 24.725
	 5.002 24.716
	 4.907 24.696
	 4.813 24.672
	 4.736 24.647
	 4.661 24.619
	 4.592 24.578
	 4.528 24.532
	 4.437 24.443
	 4.388 24.363
	 4.348 24.270
	 4.324 24.145
	 4.334 24.054
	 4.352 23.965
	 4.388 23.878
	 4.432 23.796
	 4.514 23.706
	 4.604 23.624
	 4.726 23.540
	 4.796 23.496
	 4.854 23.463
	 4.980 23.405
	 5.052 23.374
	 5.110 23.351
	 5.182 23.329
	 5.276 23.304
	 5.370 23.280
	 5.444 23.262
	 5.517 23.247
	 5.610 23.229
	 5.702 23.212
	 5.776 23.201
	 5.849 23.194
	 5.941 23.187
	 6.034 23.183
	 6.107 23.182
	 6.189 23.186
	 6.293 23.194
	 6.397 23.204
	 6.479 23.213
	 6.554 23.226
	 6.613 23.238
	 6.693 23.254
	/
\plot  5.781 23.908 	 5.880 23.897
	 5.933 23.893
	 5.990 23.894
	 6.063 23.898
	 6.136 23.903
	 6.193 23.908
	 6.268 23.916
	 6.363 23.928
	 6.459 23.942
	 6.534 23.956
	 6.643 23.986
	 6.752 24.024
	 6.845 24.077
	 6.936 24.134
	 7.038 24.206
	 7.084 24.270
	 7.118 24.340
	 7.127 24.407
	 7.127 24.443
	 7.125 24.511
	 7.101 24.555
	 7.051 24.632
	 7.046 24.636
	/
\plot  6.299 23.969 	 6.405 23.997
	 6.482 24.018
	 6.576 24.050
	 6.697 24.113
	 6.814 24.202
	 6.892 24.318
	 6.917 24.420
	 6.909 24.496
	 6.886 24.555
	 6.860 24.608
	 6.792 24.680
	 6.737 24.725
	 6.672 24.780
	 6.591 24.829
	 6.528 24.856
	 6.464 24.881
	 6.389 24.910
	 6.312 24.936
	 6.224 24.955
	 6.134 24.970
	 6.009 24.994
	/
\plot  6.371 24.045 	 6.263 24.126
	 6.160 24.185
	 6.054 24.238
	 5.936 24.279
	 5.817 24.314
	 5.703 24.340
	 5.588 24.363
	 5.474 24.376
	 5.359 24.386
	 5.263 24.387
	 5.167 24.386
	 5.053 24.378
	 4.938 24.359
	 4.851 24.320
	 4.773 24.259
	 4.750 24.138
	 4.783 24.058
	 4.830 23.986
	 4.927 23.905
	 5.031 23.836
	 5.089 23.808
	 5.163 23.777
	 5.237 23.749
	 5.296 23.728
	 5.373 23.705
	 5.470 23.680
	 5.568 23.656
	 5.645 23.639
	 5.705 23.628
	 5.780 23.616
	 5.855 23.606
	 5.914 23.599
	 5.989 23.592
	 6.084 23.584
	 6.180 23.578
	 6.255 23.575
	 6.327 23.576
	 6.418 23.579
	 6.509 23.584
	 6.581 23.590
	 6.687 23.607
	 6.792 23.630
	 6.889 23.659
	 6.985 23.692
	 7.078 23.736
	 7.169 23.785
	 7.240 23.844
	 7.307 23.908
	 7.360 23.973
	 7.383 24.069
	/
\plot  5.002 24.098 	 5.069 24.024
	 5.177 23.944
	 5.269 23.910
	 5.364 23.884
	 5.484 23.860
	 5.605 23.840
	 5.697 23.826
	 5.813 23.809
	 5.930 23.794
	 6.022 23.785
	 6.101 23.784
	 6.201 23.784
	 6.301 23.787
	 6.380 23.791
	 6.460 23.797
	 6.562 23.807
	 6.664 23.820
	 6.744 23.836
	 6.801 23.854
	 6.873 23.882
	 6.998 23.937
	 7.062 23.977
	 7.125 24.020
	 7.218 24.094
	 7.267 24.166
	 7.303 24.225
	 7.307 24.278
	 7.307 24.346
	/
\plot  4.665 24.748 	 4.773 24.793
	 4.830 24.814
	 4.956 24.844
	 5.082 24.869
	 5.143 24.877
	 5.220 24.887
	 5.298 24.895
	 5.359 24.898
	 5.445 24.899
	 5.552 24.897
	 5.660 24.891
	 5.745 24.886
	 5.816 24.879
	 5.906 24.869
	 5.996 24.858
	 6.066 24.845
	 6.184 24.816
	 6.299 24.780
	 6.391 24.733
	 6.479 24.680
	 6.568 24.614
	 6.648 24.536
	 6.697 24.426
	 6.712 24.346
	 6.693 24.278
	 6.659 24.194
	 6.612 24.134
	 6.568 24.098
	 6.524 24.069
	 6.443 24.028
	 6.380 24.001
	/
\plot  4.343 24.494 	 4.267 24.371
	 4.239 24.275
	 4.223 24.174
	 4.226 24.067
	 4.240 23.961
	 4.257 23.898
	 4.282 23.820
	 4.311 23.743
	 4.339 23.683
	 4.430 23.563
	 4.536 23.455
	 4.596 23.412
	 4.675 23.363
	 4.756 23.318
	 4.822 23.285
	 4.901 23.254
	 5.002 23.219
	 5.104 23.187
	 5.186 23.165
	 5.282 23.146
	 5.406 23.126
	 5.530 23.109
	 5.628 23.097
	 5.704 23.092
	 5.765 23.088
	 5.848 23.084
	/
\plot  5.588 23.912 	 5.516 23.848
	 5.482 23.777
	 5.457 23.700
	 5.445 23.598
	 5.440 23.495
	 5.443 23.384
	 5.453 23.273
	 5.465 23.193
	 5.495 23.116
	 5.539 23.093
	/
\plot  5.340 23.956 	 5.230 23.916
	 5.183 23.849
	 5.150 23.777
	 5.122 23.684
	 5.103 23.590
	 5.097 23.473
	 5.099 23.357
	 5.115 23.262
	 5.150 23.169
	 5.186 23.148
	/
\plot  5.194 23.997 	 5.086 24.001
	 4.995 23.965
	 4.913 23.912
	 4.849 23.822
	 4.805 23.724
	 4.788 23.614
	 4.786 23.506
	 4.804 23.387
	 4.841 23.273
	 4.870 23.241
	/
\plot  5.074 24.069 	 5.031 24.109
	 4.915 24.126
	 4.840 24.121
	 4.765 24.105
	 4.669 24.040
	 4.589 23.956
	 4.563 23.899
	 4.543 23.827
	 4.528 23.753
	 4.521 23.692
	 4.524 23.631
	 4.535 23.556
	 4.552 23.482
	 4.576 23.423
	 4.665 23.349
	/
\plot  5.050 24.126 	 5.019 24.187
	 4.955 24.238
	 4.885 24.274
	 4.799 24.289
	 4.712 24.293
	 4.632 24.270
	 4.557 24.238
	 4.437 24.138
	 4.385 24.044
	 4.348 23.944
	 4.334 23.847
	 4.333 23.751
	 4.359 23.664
	 4.392 23.582
	 4.428 23.510
	/
\plot  5.103 24.054 	 5.118 24.090
	 5.127 24.149
	 5.114 24.215
	 5.046 24.323
	 4.947 24.395
	 4.828 24.431
	 4.769 24.436
	 4.695 24.438
	 4.622 24.435
	 4.564 24.426
	 4.454 24.374
	 4.356 24.306
	 4.291 24.220
	 4.240 24.126
	 4.218 24.013
	 4.214 23.899
	 4.232 23.816
	 4.254 23.754
	 4.286 23.671
	/
\plot  5.139 24.041 	 5.230 24.069
	 5.319 24.086
	/
\plot  5.230 24.028 	 5.256 24.094
	 5.261 24.160
	 5.256 24.225
	 5.231 24.309
	 5.194 24.390
	 5.113 24.478
	 5.019 24.551
	 4.929 24.585
	 4.837 24.608
	 4.735 24.620
	 4.633 24.621
	 4.557 24.609
	 4.481 24.587
	 4.428 24.556
	 4.333 24.488
	/
\plot  6.077 23.929 	 6.126 23.884
	 6.158 23.821
	 6.185 23.755
	 6.200 23.661
	 6.210 23.567
	 6.219 23.460
	 6.223 23.351
	 6.218 23.260
	 6.210 23.169
	 6.206 23.112
	/
\plot  5.002 24.098 	 5.002 24.149
	 5.046 24.187
	 5.126 24.207
	 5.207 24.215
	 5.308 24.222
	 5.410 24.225
	 5.489 24.221
	 5.569 24.215
	 5.613 24.212
	 5.696 24.206
	/
\plot  5.292 24.062 	 5.355 24.153
	 5.378 24.215
	 5.387 24.327
	 5.382 24.401
	 5.368 24.475
	 5.330 24.547
	 5.283 24.613
	 5.179 24.716
	 5.063 24.757
	 4.942 24.782
	 4.876 24.791
	 4.809 24.793
	 4.755 24.782
	 4.657 24.752
	/
\plot  5.099 24.922 	 5.205 24.911
	 5.262 24.898
	 5.333 24.865
	 5.400 24.820
	 5.463 24.736
	 5.512 24.644
	 5.538 24.541
	 5.552 24.435
	 5.546 24.347
	 5.529 24.259
	 5.486 24.166
	 5.412 24.130
	/
\plot  5.565 24.998 	 5.620 24.949
	 5.664 24.858
	 5.700 24.761
	 5.711 24.683
	 5.717 24.604
	 5.722 24.511
	 5.721 24.418
	 5.712 24.343
	 5.696 24.270
	 5.664 24.206
	 5.645 24.194
	/
\plot  5.893 24.194 	 5.878 24.301
	 5.880 24.392
	 5.884 24.483
	 5.885 24.589
	 5.889 24.695
	 5.894 24.775
	 5.901 24.856
	 5.914 24.958
	 5.941 25.006
	/
\plot  6.102 24.158 	 6.062 24.219
	 6.054 24.283
	 6.054 24.346
	 6.060 24.446
	 6.071 24.547
	 6.091 24.650
	 6.121 24.752
	 6.166 24.832
	 6.223 24.905
	 6.312 24.953
	/
\plot  6.263 24.090 	 6.215 24.162
	 6.198 24.287
	 6.203 24.374
	 6.219 24.460
	 6.268 24.566
	 6.331 24.663
	 6.414 24.740
	 6.511 24.801
	 6.606 24.819
	 6.701 24.820
	 6.788 24.793
	/
\plot  6.388 24.024 	 6.350 24.081
	 6.331 24.174
	 6.347 24.280
	 6.380 24.380
	 6.444 24.464
	 6.519 24.536
	 6.613 24.594
	 6.712 24.640
	 6.802 24.668
	 6.896 24.685
	 7.021 24.663
	/
\plot  6.416 24.033 	 6.416 24.113
	 6.428 24.189
	 6.452 24.263
	 6.514 24.335
	 6.587 24.395
	 6.671 24.436
	 6.761 24.467
	 6.876 24.481
	 6.993 24.479
	 7.091 24.448
	 7.182 24.403
	 7.275 24.310
	 7.351 24.194
	 7.383 24.071
	/
\plot  6.513 24.162 	 6.555 24.210
	 6.644 24.259
	 6.715 24.278
	 6.788 24.287
	 6.874 24.276
	 6.957 24.251
	 7.034 24.198
	 7.101 24.134
	 7.157 24.049
	 7.199 23.956
	 7.216 23.845
	 7.218 23.732
	 7.208 23.676
	 7.182 23.571
	/
\plot  6.322 23.978 	 6.435 23.937
	 6.487 23.870
	 6.528 23.796
	 6.568 23.671
	 6.578 23.567
	 6.581 23.463
	 6.578 23.377
	 6.572 23.290
	 6.568 23.213
	/
\plot  6.519 24.105 	 6.604 24.104
	 6.648 24.098
	 6.735 24.062
	 6.845 23.956
	 6.890 23.864
	 6.921 23.768
	 6.934 23.659
	 6.936 23.550
	 6.931 23.471
	 6.917 23.393
	 6.877 23.330
	/
\plot 11.108 23.940 	11.213 23.967
	11.290 23.989
	11.386 24.020
	11.506 24.083
	11.625 24.172
	11.701 24.289
	11.728 24.392
	11.718 24.469
	11.697 24.528
	11.669 24.581
	11.601 24.653
	11.549 24.697
	11.481 24.752
	11.398 24.801
	11.273 24.856
	11.198 24.884
	11.121 24.909
	11.031 24.928
	10.941 24.945
	10.816 24.968
	/
\plot  9.468 24.721 	 9.575 24.766
	 9.633 24.788
	 9.758 24.818
	 9.885 24.841
	 9.946 24.850
	10.024 24.860
	10.102 24.869
	10.164 24.873
	10.249 24.873
	10.357 24.870
	10.465 24.865
	10.549 24.860
	10.621 24.853
	10.712 24.843
	10.802 24.831
	10.873 24.820
	10.992 24.790
	11.108 24.752
	11.199 24.706
	11.288 24.653
	11.376 24.587
	11.458 24.511
	11.506 24.397
	11.521 24.316
	11.504 24.249
	11.466 24.164
	11.422 24.107
	11.377 24.069
	11.333 24.039
	11.252 23.999
	11.189 23.971
	/
\plot  9.142 24.464 	 9.066 24.344
	 9.038 24.246
	 9.021 24.145
	 9.025 24.038
	 9.040 23.931
	 9.057 23.867
	 9.082 23.789
	 9.111 23.711
	 9.140 23.652
	 9.231 23.531
	 9.337 23.423
	 9.397 23.380
	 9.477 23.331
	 9.558 23.285
	 9.624 23.252
	 9.703 23.221
	 9.804 23.186
	 9.907 23.154
	 9.989 23.131
	10.086 23.112
	10.210 23.092
	10.335 23.075
	10.433 23.063
	10.510 23.058
	10.570 23.055
	10.653 23.050
	/
\plot 10.393 23.880 	10.321 23.819
	10.286 23.746
	10.262 23.671
	10.249 23.567
	10.245 23.463
	10.248 23.351
	10.257 23.239
	10.270 23.160
	10.298 23.082
	10.342 23.061
	/
\plot 10.145 23.925 	10.033 23.887
	 9.986 23.819
	 9.953 23.747
	 9.925 23.653
	 9.906 23.559
	 9.900 23.441
	 9.904 23.324
	 9.918 23.229
	 9.953 23.137
	 9.989 23.114
	/
\plot  9.997 23.965 	 9.889 23.971
	 9.797 23.935
	 9.713 23.880
	 9.651 23.791
	 9.605 23.692
	 9.590 23.583
	 9.588 23.472
	 9.606 23.354
	 9.641 23.239
	 9.673 23.207
	/
\plot  9.876 24.039 	 9.836 24.079
	 9.718 24.096
	 9.642 24.091
	 9.565 24.075
	 9.470 24.010
	 9.390 23.925
	 9.364 23.868
	 9.344 23.796
	 9.329 23.722
	 9.324 23.662
	 9.326 23.600
	 9.336 23.525
	 9.353 23.450
	 9.377 23.391
	 9.468 23.315
	/
\plot  9.853 24.096 	 9.821 24.160
	 9.760 24.208
	 9.688 24.244
	 9.600 24.260
	 9.512 24.263
	 9.433 24.241
	 9.360 24.208
	 9.237 24.109
	 9.185 24.014
	 9.148 23.912
	 9.134 23.816
	 9.133 23.719
	 9.159 23.633
	 9.193 23.550
	 9.229 23.478
	/
\plot  9.906 24.024 	 9.921 24.060
	 9.929 24.119
	 9.917 24.185
	 9.849 24.293
	 9.751 24.367
	 9.629 24.403
	 9.570 24.408
	 9.497 24.409
	 9.423 24.406
	 9.364 24.397
	 9.255 24.346
	 9.157 24.276
	 9.091 24.191
	 9.040 24.096
	 9.017 23.983
	 9.013 23.868
	 9.032 23.785
	 9.053 23.723
	 9.085 23.639
	/
\plot  9.944 24.011 	10.033 24.039
	10.122 24.056
	/
\plot 10.033 23.999 	10.061 24.066
	10.064 24.130
	10.061 24.196
	10.035 24.280
	 9.997 24.361
	 9.916 24.449
	 9.821 24.524
	 9.731 24.558
	 9.637 24.581
	 9.537 24.593
	 9.436 24.596
	 9.358 24.581
	 9.284 24.558
	 9.228 24.528
	 9.133 24.460
	/
\plot  9.804 24.069 	 9.804 24.119
	 9.849 24.160
	 9.930 24.178
	10.012 24.185
	10.112 24.193
	10.213 24.196
	10.294 24.191
	10.374 24.185
	10.501 24.177
	/
\plot 10.097 24.033 	10.158 24.124
	10.181 24.185
	10.190 24.299
	10.186 24.373
	10.173 24.448
	10.134 24.519
	10.088 24.587
	 9.984 24.689
	 9.866 24.731
	 9.745 24.757
	 9.678 24.764
	 9.610 24.765
	 9.556 24.754
	 9.457 24.725
	/
\plot  9.904 24.894 	10.009 24.885
	10.065 24.873
	10.137 24.838
	10.204 24.793
	10.267 24.709
	10.317 24.617
	10.343 24.513
	10.357 24.405
	10.350 24.318
	10.334 24.232
	10.289 24.136
	10.217 24.100
	/
\plot 10.370 24.970 	10.425 24.924
	10.469 24.833
	10.505 24.733
	10.516 24.655
	10.524 24.577
	10.528 24.483
	10.526 24.388
	10.517 24.314
	10.501 24.240
	10.469 24.177
	10.450 24.164
	/
\plot 10.698 24.164 	10.685 24.272
	10.686 24.364
	10.689 24.456
	10.691 24.561
	10.693 24.666
	10.700 24.747
	10.708 24.829
	10.721 24.932
	10.748 24.981
	/
\plot 10.909 24.128 	10.869 24.191
	10.861 24.316
	10.866 24.418
	10.878 24.519
	10.898 24.623
	10.928 24.725
	10.973 24.805
	11.030 24.877
	11.121 24.928
	/
\plot 11.070 24.060 	11.021 24.132
	11.009 24.195
	11.005 24.259
	11.010 24.346
	11.026 24.433
	11.075 24.538
	11.138 24.636
	11.222 24.712
	11.320 24.773
	11.415 24.792
	11.513 24.793
	11.597 24.765
	/
\plot 11.197 23.992 	11.157 24.052
	11.138 24.145
	11.155 24.251
	11.189 24.352
	11.252 24.437
	11.328 24.511
	11.422 24.567
	11.521 24.613
	11.611 24.640
	11.705 24.657
	11.830 24.636
	/
\plot 11.225 24.001 	11.225 24.083
	11.237 24.160
	11.261 24.236
	11.321 24.308
	11.396 24.367
	11.495 24.398
	11.597 24.416
	11.705 24.425
	11.813 24.424
	11.904 24.406
	11.993 24.376
	12.086 24.280
	12.164 24.164
	12.194 24.043
	/
\plot 11.322 24.132 	11.354 24.204
	11.458 24.263
	11.529 24.276
	11.601 24.280
	11.694 24.274
	11.786 24.253
	11.864 24.199
	11.930 24.132
	11.984 24.040
	12.025 23.940
	12.034 23.820
	12.029 23.702
	12.019 23.645
	11.993 23.539
	/
\plot 11.328 24.075 	11.413 24.074
	11.458 24.069
	11.544 24.033
	11.656 23.925
	11.700 23.834
	11.733 23.738
	11.744 23.628
	11.745 23.518
	11.741 23.439
	11.728 23.360
	11.688 23.298
	/
\plot 11.102 23.944 	11.201 23.872
	11.266 23.801
	11.322 23.724
	11.366 23.639
	11.405 23.552
	11.435 23.487
	11.462 23.419
	11.474 23.363
	11.489 23.258
	/
\plot 11.314 24.107 	11.204 24.198
	11.144 24.244
	11.044 24.304
	10.941 24.356
	10.879 24.382
	10.799 24.410
	10.718 24.437
	10.653 24.456
	10.555 24.475
	10.456 24.492
	10.348 24.509
	10.240 24.524
	10.116 24.530
	 9.993 24.532
	 9.878 24.528
	 9.764 24.519
	 9.657 24.494
	 9.553 24.460
	 9.468 24.418
	 9.390 24.361
	 9.346 24.282
	 9.318 24.196
	 9.318 24.109
	 9.332 24.024
	 9.378 23.933
	 9.436 23.848
	 9.513 23.774
	 9.597 23.705
	 9.687 23.651
	 9.781 23.603
	 9.842 23.579
	 9.920 23.553
	 9.999 23.528
	10.061 23.508
	10.148 23.482
	10.261 23.449
	10.373 23.416
	10.461 23.387
	10.536 23.355
	10.632 23.312
	10.726 23.268
	10.801 23.235
	10.885 23.201
	10.990 23.160
	11.096 23.119
	11.178 23.086
	11.242 23.061
	11.269 23.095
	11.351 23.161
	11.436 23.222
	11.514 23.278
	11.593 23.334
	11.696 23.390
	11.800 23.444
	11.877 23.491
	11.953 23.539
	12.065 23.626
	/
\plot 11.252 24.016 	11.333 24.069
	11.377 24.132
	11.398 24.204
	11.386 24.268
	11.337 24.329
	11.242 24.428
	11.146 24.493
	11.045 24.549
	10.938 24.594
	10.829 24.632
	10.744 24.650
	10.657 24.666
	10.579 24.681
	10.501 24.693
	10.395 24.702
	10.289 24.708
	10.191 24.711
	10.092 24.712
	10.011 24.705
	 9.929 24.697
	 9.804 24.689
	 9.709 24.669
	 9.616 24.644
	 9.538 24.619
	 9.462 24.589
	 9.393 24.550
	 9.326 24.505
	 9.237 24.416
	 9.188 24.335
	 9.148 24.240
	 9.125 24.115
	 9.134 24.024
	 9.152 23.935
	 9.188 23.847
	 9.233 23.764
	 9.314 23.675
	 9.402 23.594
	 9.527 23.509
	 9.599 23.465
	 9.656 23.431
	 9.782 23.372
	 9.854 23.342
	 9.912 23.319
	 9.986 23.297
	10.080 23.272
	10.174 23.248
	10.249 23.230
	10.322 23.215
	10.415 23.197
	10.508 23.180
	10.581 23.167
	10.667 23.152
	10.777 23.132
	10.886 23.113
	10.973 23.099
	11.043 23.091
	11.099 23.086
	11.176 23.078
	/
\plot 10.549 23.827 	10.672 23.732
	10.763 23.664
	10.886 23.588
	10.941 23.575
	10.973 23.616
	11.043 23.685
	11.117 23.751
	11.230 23.835
	11.350 23.912
	11.456 23.953
	11.561 23.992
	11.655 24.048
	11.745 24.107
	11.849 24.177
	11.894 24.240
	11.930 24.312
	11.940 24.380
	11.940 24.416
	11.934 24.481
	11.913 24.528
	11.864 24.604
	11.858 24.608
	/
\plot  9.804 24.069 	 9.872 23.992
	 9.978 23.912
	10.073 23.881
	10.168 23.855
	10.293 23.814
	10.416 23.768
	10.474 23.738
	10.543 23.698
	10.620 23.650
	10.700 23.599
	10.779 23.548
	10.855 23.501
	10.981 23.427
	11.034 23.406
	11.077 23.436
	11.178 23.534
	11.282 23.628
	11.341 23.670
	11.417 23.719
	11.495 23.765
	11.557 23.800
	11.683 23.854
	11.809 23.908
	11.934 23.988
	12.029 24.066
	12.078 24.136
	12.114 24.196
	12.118 24.249
	12.118 24.316
	/
\plot 11.178 24.016 	11.070 24.096
	10.967 24.156
	10.861 24.208
	10.742 24.250
	10.621 24.285
	10.508 24.312
	10.393 24.335
	10.278 24.348
	10.164 24.356
	10.067 24.359
	 9.970 24.356
	 9.855 24.349
	 9.741 24.329
	 9.653 24.291
	 9.574 24.232
	 9.553 24.109
	 9.585 24.029
	 9.633 23.956
	 9.730 23.874
	 9.836 23.804
	 9.893 23.777
	 9.967 23.747
	10.041 23.718
	10.101 23.696
	10.177 23.673
	10.274 23.645
	10.371 23.616
	10.446 23.588
	10.504 23.558
	10.576 23.514
	10.702 23.436
	10.785 23.395
	10.891 23.344
	10.997 23.295
	11.081 23.258
	11.136 23.235
	11.184 23.271
	11.241 23.316
	11.313 23.375
	11.385 23.434
	11.445 23.478
	11.538 23.528
	11.633 23.575
	11.714 23.619
	11.796 23.662
	11.888 23.706
	11.980 23.755
	12.051 23.813
	12.118 23.876
	12.173 23.944
	12.194 24.039
	/
\plot 13.970 23.918 	14.082 23.887
	14.160 23.857
	14.235 23.821
	14.335 23.754
	14.429 23.677
	14.493 23.604
	14.550 23.525
	14.600 23.432
	14.639 23.332
	14.630 23.218
	/
\plot 14.453 24.071 	14.530 24.048
	14.586 24.029
	14.654 23.999
	14.743 23.935
	14.823 23.863
	14.888 23.781
	14.946 23.694
	14.977 23.610
	15.001 23.525
	15.011 23.455
	15.009 23.383
	14.963 23.302
	/
\plot 14.897 24.194 	15.019 24.157
	15.081 24.128
	15.172 24.048
	15.251 23.959
	15.299 23.890
	15.339 23.815
	15.358 23.706
	15.365 23.597
	15.359 23.529
	15.348 23.461
	15.299 23.381
	/
\plot 15.251 24.274 	15.327 24.246
	15.384 24.224
	15.454 24.194
	15.528 24.145
	15.598 24.088
	15.643 24.011
	15.676 23.927
	15.695 23.820
	15.701 23.711
	15.694 23.630
	15.676 23.550
	15.627 23.470
	/
\plot 15.638 24.354 	15.749 24.321
	15.807 24.297
	15.917 24.204
	15.960 24.114
	15.989 24.020
	16.007 23.911
	16.015 23.802
	16.007 23.721
	15.989 23.641
	15.943 23.561
	/
\plot 16.061 24.460 	16.157 24.426
	16.212 24.354
	16.254 24.274
	16.294 24.183
	16.326 24.090
	16.344 23.981
	16.351 23.872
	16.343 23.791
	16.326 23.711
	16.277 23.630
	/
\plot 16.391 24.564 	16.489 24.532
	16.543 24.459
	16.584 24.380
	16.624 24.288
	16.656 24.194
	16.674 24.086
	16.681 23.975
	16.674 23.895
	16.656 23.815
	16.607 23.736
	/
\plot 16.722 24.644 	16.819 24.613
	16.873 24.539
	16.914 24.460
	16.954 24.368
	16.986 24.274
	17.004 24.166
	17.012 24.056
	17.004 23.976
	16.986 23.897
	16.940 23.815
	/
\plot 17.029 24.725 	17.124 24.693
	17.179 24.620
	17.221 24.541
	17.261 24.449
	17.293 24.354
	17.311 24.247
	17.319 24.138
	17.310 24.056
	17.293 23.975
	17.247 23.897
	/
\plot 13.898 23.533 	14.008 23.574
	14.067 23.597
	14.148 23.635
	14.228 23.677
	14.340 23.757
	14.410 23.799
	14.500 23.849
	14.591 23.898
	14.662 23.935
	14.719 23.963
	14.791 23.997
	14.863 24.030
	14.920 24.056
	14.980 24.080
	15.057 24.110
	15.133 24.138
	15.193 24.160
	15.316 24.199
	15.392 24.223
	15.472 24.247
	15.552 24.271
	15.628 24.293
	15.750 24.329
	15.846 24.358
	15.968 24.394
	16.090 24.430
	16.186 24.458
	16.293 24.490
	16.358 24.510
	16.428 24.530
	16.497 24.551
	16.563 24.571
	16.669 24.604
	16.725 24.621
	16.801 24.645
	16.905 24.679
	16.971 24.700
	17.048 24.725
	/
\plot 14.082 23.372 	14.165 23.403
	14.225 23.427
	14.300 23.459
	14.376 23.504
	14.470 23.564
	14.564 23.624
	14.639 23.669
	14.697 23.699
	14.768 23.736
	14.848 23.776
	14.933 23.818
	15.019 23.858
	15.101 23.895
	15.174 23.928
	15.236 23.952
	15.298 23.973
	15.374 23.995
	15.461 24.019
	15.553 24.043
	15.644 24.067
	15.731 24.089
	15.808 24.110
	15.871 24.128
	15.940 24.150
	16.024 24.178
	16.119 24.210
	16.218 24.244
	16.318 24.278
	16.412 24.310
	16.496 24.339
	16.565 24.361
	16.684 24.397
	16.757 24.418
	16.835 24.441
	16.913 24.463
	16.986 24.485
	17.105 24.524
	17.220 24.568
	17.311 24.606
	17.435 24.659
	/
\plot 14.252 23.218 	14.325 23.248
	14.379 23.270
	14.446 23.300
	14.534 23.344
	14.643 23.403
	14.753 23.462
	14.840 23.508
	14.917 23.549
	15.015 23.600
	15.114 23.650
	15.193 23.685
	15.275 23.712
	15.379 23.741
	15.484 23.769
	15.566 23.791
	15.637 23.810
	15.727 23.834
	15.817 23.859
	15.888 23.880
	15.961 23.902
	16.053 23.930
	16.145 23.960
	16.218 23.984
	16.297 24.010
	16.393 24.043
	16.501 24.081
	16.614 24.120
	16.728 24.161
	16.835 24.199
	16.930 24.235
	17.007 24.265
	17.078 24.295
	17.174 24.338
	17.234 24.366
	17.305 24.399
	17.388 24.438
	17.484 24.483
	/
\plot 13.744 23.677 	13.829 23.703
	13.892 23.722
	13.970 23.749
	14.059 23.793
	14.168 23.852
	14.277 23.913
	14.364 23.959
	14.445 24.000
	14.548 24.051
	14.652 24.100
	14.736 24.136
	14.849 24.173
	14.941 24.200
	15.066 24.234
	/
\plot 14.558 23.169 	14.670 23.229
	14.752 23.272
	14.857 23.324
	14.962 23.369
	15.027 23.396
	15.096 23.424
	15.166 23.451
	15.232 23.477
	15.339 23.516
	15.465 23.558
	15.539 23.581
	15.619 23.605
	15.704 23.630
	15.793 23.657
	15.884 23.683
	15.976 23.710
	16.069 23.737
	16.160 23.763
	16.248 23.789
	16.334 23.815
	16.414 23.839
	16.488 23.862
	16.614 23.904
	16.740 23.950
	16.818 23.980
	16.900 24.012
	16.982 24.045
	17.060 24.076
	17.185 24.128
	17.309 24.183
	17.406 24.228
	17.468 24.256
	17.539 24.289
	/
\plot 17.596 24.105 	17.515 24.069
	17.445 24.039
	17.334 23.992
	17.253 23.959
	17.194 23.935
	17.109 23.905
	17.000 23.869
	16.892 23.834
	16.806 23.806
	16.717 23.778
	16.645 23.756
	16.548 23.726
	/
\plot  5.046 24.158 	 5.082 24.058
	 5.167 24.005
	 5.234 23.985
	 5.302 23.969
	 5.375 23.952
	 5.448 23.937
	 5.524 23.926
	 5.601 23.916
	 5.686 23.913
	 5.772 23.912
	 5.886 23.914
	 6.001 23.920
	 6.074 23.932
	 6.147 23.946
	 6.235 23.964
	 6.322 23.984
	 6.380 24.005
	 6.464 24.071
	 6.519 24.166
	/
\plot  5.203 24.005 	 5.304 24.081
	 5.409 24.128
	 5.516 24.166
	 5.592 24.184
	 5.668 24.198
	 5.785 24.202
	 5.873 24.196
	 5.961 24.187
	 6.079 24.162
	 6.152 24.140
	 6.223 24.113
	 6.318 24.054
	 6.358 24.014
	/
\plot 10.374 24.970 	10.450 24.973
	10.516 24.975
	10.621 24.978
	10.697 24.978
	10.753 24.977
	10.833 24.970
	10.935 24.957
	11.037 24.943
	11.117 24.928
	11.213 24.903
	11.333 24.867
	11.452 24.828
	11.544 24.793
	11.609 24.761
	11.688 24.718
	11.766 24.672
	11.826 24.632
	11.899 24.569
	11.985 24.484
	12.066 24.395
	12.124 24.316
	12.161 24.222
	12.186 24.124
	12.194 24.032
	12.192 23.940
	12.157 23.818
	12.109 23.702
	12.068 23.637
	12.021 23.575
	11.907 23.467
	11.786 23.368
	11.684 23.313
	11.580 23.262
	11.516 23.231
	11.435 23.193
	11.354 23.152
	11.292 23.114
	11.246 23.063
	11.178 23.082
	11.086 23.088
	10.994 23.086
	10.880 23.071
	10.765 23.055
	10.666 23.052
	10.541 23.053
	10.416 23.057
	10.317 23.063
	10.236 23.074
	10.134 23.091
	10.032 23.110
	 9.953 23.127
	 9.886 23.144
	 9.801 23.168
	 9.717 23.194
	 9.652 23.218
	 9.540 23.273
	 9.432 23.334
	 9.322 23.409
	 9.216 23.491
	 9.143 23.566
	 9.076 23.647
	 9.026 23.735
	 8.985 23.827
	 8.964 23.934
	 8.956 24.043
	 8.974 24.147
	 9.004 24.249
	 9.049 24.337
	 9.102 24.420
	 9.178 24.504
	 9.260 24.581
	 9.378 24.664
	 9.500 24.740
	 9.621 24.798
	 9.745 24.850
	 9.811 24.871
	 9.894 24.895
	 9.979 24.917
	10.046 24.932
	10.166 24.951
	10.262 24.962
	10.323 24.969
	10.393 24.977
	/
\plot  9.974 23.948 	10.060 24.018
	10.109 24.052
	10.214 24.098
	10.321 24.136
	10.397 24.154
	10.473 24.168
	10.590 24.172
	10.678 24.168
	10.765 24.160
	10.886 24.132
	10.959 24.110
	11.030 24.083
	11.117 24.031
	11.214 23.963
	11.214 23.956
	/
\plot  9.849 24.128 	 9.885 24.031
	 9.970 23.975
	10.037 23.955
	10.105 23.940
	10.179 23.923
	10.253 23.908
	10.329 23.897
	10.406 23.887
	10.474 23.872
	10.541 23.855
	10.601 23.832
	10.675 23.800
	10.749 23.767
	10.808 23.743
	10.837 23.732
	10.869 23.764
	10.936 23.821
	11.005 23.876
	11.129 23.952
	11.189 23.975
	11.273 24.043
	11.328 24.136
	/
\plot 14.398 23.091 	14.518 23.148
	14.582 23.178
	14.683 23.224
	14.785 23.266
	14.912 23.307
	14.987 23.329
	15.068 23.351
	15.154 23.375
	15.244 23.398
	15.337 23.422
	15.431 23.445
	15.524 23.469
	15.617 23.492
	15.707 23.514
	15.794 23.535
	15.876 23.555
	15.952 23.573
	16.020 23.590
	16.080 23.605
	16.203 23.637
	16.275 23.656
	16.353 23.676
	16.435 23.698
	16.521 23.721
	16.609 23.744
	16.699 23.767
	16.788 23.790
	16.877 23.812
	16.963 23.833
	17.046 23.853
	17.125 23.871
	17.198 23.886
	17.323 23.910
	17.435 23.921
	17.547 23.927
	17.653 23.918
	/
\linethickness=2pt
\plot 13.600 23.806 	13.720 23.838
	13.841 23.877
	13.962 23.918
	14.050 23.956
	14.139 23.992
	14.202 24.013
	14.270 24.033
	14.342 24.054
	14.420 24.075
	14.501 24.097
	14.586 24.119
	14.674 24.142
	14.765 24.164
	14.859 24.187
	14.956 24.210
	15.054 24.233
	15.155 24.256
	15.256 24.280
	15.359 24.303
	15.462 24.327
	15.566 24.351
	15.670 24.374
	15.773 24.398
	15.876 24.421
	15.977 24.445
	16.077 24.468
	16.176 24.491
	16.272 24.514
	16.367 24.537
	16.458 24.559
	16.546 24.582
	16.631 24.604
	16.712 24.625
	16.789 24.647
	16.862 24.667
	16.930 24.688
	16.993 24.708
	17.102 24.749
	17.209 24.797
	17.263 24.829
	17.363 24.894
	/
\linethickness=0pt
\putrectangle corners at  4.130 25.034 and 17.700 22.991
\endpicture
\vskip 20pt
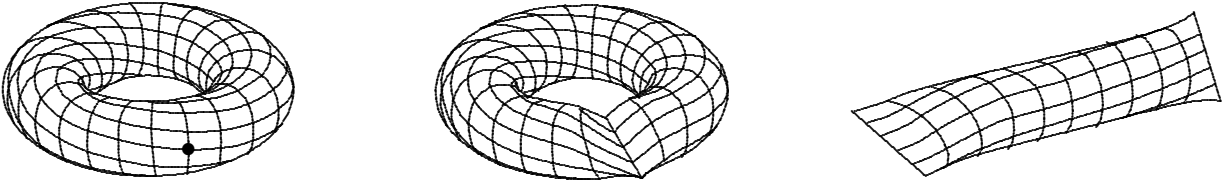
\centerline{{\bf Fig. 6} Diagrammatic connection between marked torus
and flattened cylinder.}
\vskip 20pt
\hskip -19pt
Starting from the mark on the twisted torus,
flatten the torus
across its width (this defines a point on the opposite side), then cut
along the flattening and open out the crimped torus into a cylinder
with flattened ends. The two points at either end of the flattening
on the torus become the four $t_2$ defects of the flattened cylinder.

Higher order moments correspond to inserting a single negative
curvature defect. The lowest order defect of this type is the
insertion of negative curvature of deficit angle $-\pi$ introduced by
the vertex $\Tr[(MA_4)^6]$ (see Fig. 1(c) and eq. \moms).

As a final check of our solution, we expand
$\langle~{1\over N}~\Tr~[(MA_4)^6]~\rangle_{III}$ in powers of $q$
(this can be done directly from the expression for the moment
in terms of theta functions \moms ):
\eqn\pert{
\langle~{1\over N}~\Tr~[(MA_4)^6]~\rangle_{III}=
t_2^5~(~9q^2~+~27q^4~+~81q^6~+~\dots ).}
It is easy to verify that this correctly counts the number of diagrams.

Further moments can be calculated by expanding \hofGph\ to
higher order. They can always be written as sums of products of
complete elliptic integrals.

We now look for a continuum limit in which the
size of graphs tends to infinity. One can see that the critical
point, at which the size of the graphs diverges, is at $q=1$.
Note, however, that (since the critical $q$ is $1$),
in stark contrast to two-dimensional quantum gravity \DAVID\ \VOL,
the leading behaviour for the growth of diagrams
is {\it not} exponential but merely power-like. To extract the power,
notice that the product
$\prod_{n=1}^{\infty}(1-q^{2n})$ in \feq\ can be written in terms of
the $\eta$ function as
$q^{-1/12}\eta(i\tau)$ where
$q=e^{-\pi\tau}$. Making use of the modular invariance of the
$\eta(\tau)$ function under the the modular transformation
$\tau\rightarrow 1/\tau$, we extract
\eqn\freelim{
{\cal F}(t_2,1-\mu)\sim {\pi^2t_2^4\over 192}{1\over\mu^2},}
where we have defined a ``continuum cosmological constant'' $\mu$ through
$q=1-\mu$ i.e. $\mu=\pi\tau+{\cal O}(\tau^2)$.
We thus see, changing from fixed cosmological constant to fixed area
by Laplace transform, that the number of graphs grows as a linear power of the
area. Employing the conventions of quantum gravity \DAVID\ \VOL,
this would formally correspond to a ``string susceptibility''
$\gamma_{str}=4$.

We can easily understand this
result by performing the calculation directly in the continuum limit. We
thus integrate over cylinders of all possible lengths $t$ and
circumferences $s$ weighted with a factor $s$ (corresponding to the
modular twist between the two flattened ends) and
a delta function for the area so as to count the number of surfaces of a
given area $A$. We thus perform the following integral:
\eqn\contfe{
{\cal F}\sim\int_0^{\infty}~dt~ds~s~\delta(ts-A)=
            A\int_0^{\infty}~{dt\over t^2}.}
We see immediately a linear dependence on the area $A$, but also a
divergently large contribution coming from small $t$. The most
important contribution comes from the cylinders which are infinitely
short and thus have the maximum amount of entropy coming from the
modular twist.

It is interesting now to investigate the behaviour of the correlators
in the large area ($q\rightarrow 1$) limit. Quite generally, for matrix
models the correlators correspond to surfaces with a boundary of
length proportional to the power of the correlator, and one seeks a
continuum scaling limit for very long boundaries. The correlators
$\langle~{1\over N}~\Tr~[(MA_4)^{2n}]~\rangle$ in the present model,
however, introduce point-like negative curvature
and we cannot look for a scaling
limit involving long boundaries\foot{In principle, it is possible to
study boundaries of arbitrary length by taking correlators of
$\langle~{1\over N}~\Tr~[(MA_4^2)^{2n}]~\rangle$, which correspond to
a boundary in the form of the end of a cylinder. Technically, however,
we do not at present have the means to calculate such quantities.}.
Nevertheless we can find the limiting behaviour of these negative
curvature insertions in the limit of large area.

Using the modular transformation $\tau\rightarrow 1/\tau$ (with
$q=e^{-\pi\tau}$) for the formula \hofG, we can also extract
a scaling limit for the generating function for the
moments. Specifically, for the theta functions $\theta_1(z)$
and $\theta_4(z)$, we find that the dominant contributions are
\eqn\thlim{\eqalign{
\theta_4(z)=&{1\over\sqrt{\tau}}e^{-{\tau\over\pi}y^2}2q'^{1/4}
\bigl(\cosh y + {\cal O}(q'^2)\bigr)\cr
\theta_1(z)=&{1\over\sqrt{\tau}}e^{-{\tau\over\pi}y^2}2q'^{1/4}
\bigl(\sinh y + {\cal O}(q'^2)\bigr),}}
where $y=z/\tau$ and $q'=e^{-{\pi\over\tau}}$. Holding $y$ fixed as a
parameter of order 1, we take the limit as
$\tau\rightarrow 0$ (corresponding to $q\rightarrow 1$) and
work to the first two orders in $\tau$. Remembering that
$q=1-\pi\tau+\dots$, the constants $D$ and $a$
to the first two orders in $\tau$ are given by
\eqn\Da{
D=t_2{1\over\tau^{3/2}}q'^{1/4}\bigl(1+{5\over 4}\pi\tau\bigr)
\quad {\rm and} \quad
a=1+t_2^2\Bigl[{1\over 4\tau^2}-
  \bigl({1\over 2\pi}-{\pi\over
4}\bigr){1\over\tau}\Bigr],}
We can now define a natural rescaled parameter $x={Gt_2\over 2\tau}$
and perform the inversion of the theta function to the first two
orders in $\tau$ to find the generating function for the
correlators. The lowest order term gives the contribution ${q\over G^2}$
in \hGexp\ and also a part that cancels with $a-1$. The next order
gives the contribution ${t_2\over G}$ along with the generating
function which we read off as:
\eqn\scalmom{
\sum_{n=1}^{\infty}\langle~{1\over N}~\Tr~[(MA_4)^{2n}]~\rangle G^n=
        {t_2\over\tau}
   \Bigl[{\sin^{-1}x-x\over 2x^2}+{(\sin^{-1}x)^2-x^2\over 2\pi
                                                      x^2}\Bigr]
\quad {\rm with} \quad x={Gt_2\over 2\tau}.}
This has a simple square root singularity at the point $x=1$. The
series expansions for $\sin^{-1}x$ and $(\sin^{-1}(x))^2$ then give us
the dominant contribution to the correlators in the large area limit:
\eqn\momscal{
\langle~{1\over N}~\Tr~[(MA_4)^{2n}]~\rangle \simeq
                            {C_nt_2^{n+2}\over\mu^{n+1}}
\quad {\rm with} \quad \cases{
C_{2n}={\pi^{2n}(n!)^2\over 2~(2n+1)!~(n+1)}\cr
C_{2n+1}={\pi^{2n+2}(2n+2)!\over 2^{4n+4}((n+1)!)^2(2n+3)},}}
where we have again introduced the parameter $\mu$ defined by
$q=1-\mu$ i.e. $\mu=\pi\tau+{\cal O}(\tau^2)$. The number of surfaces
of fixed area $A$ for a correlator $\langle~{1\over
N}~\Tr~[(MA_4)^{2n}]~\rangle$ is thus seen to be of the order of
$A^n$, with entropy coming from modular twists analogous to those for
the free energy. The rather curious structure of the considered
surfaces is thus evident - they consist of cylindrical ``fingers''
growing out from the negative curvature defect (see Fig.~2(b)).  The
square root singularity at $x=1$ means that there is a tree-like
growth of the number of ways to attach the fingers to their base at
the negative defect.  As in the case of the free energy, modular
integrations cause a filamentary structure of very long cylinders to
dominate in the large area limit.

\newsec{The onset of quantum gravity: Adding negative curvature defects}
The introduction of arbitrary numbers of negative defects,
specifically $t_6$, alongside
the components $t_4$, $t_4^*$ and positive
curvature defects, $t_2$,  gives us a model in which we can tune away
the curvature fluctuations of two-dimensional quantum gravity. The large
$N$ limit of the character in section 3 allows us to understand the analytic
structure of the solution and thus reduce the model to
a well defined Cauchy-Riemann problem. The function
$G(h)$ now consists of two sheets below the physical sheet (see Fig. 4).
An extra
sheet which we label $G_3(h)$ is now attached to the sheet $G_2(h)$ of
the previous section by a square root cut. We thus have the following
two equations
\eqn\grav{\eqalign{
2F(h) + \cut H(h) =& -\ln h\cr
F_1(h)+F_2(h)+F_3(h)+2H(h)=&-\ln({h\over t_6})}.}
The first is the saddle point equation \sdpt. The second comes from
the logarithm of eq. \HGprod, where we define, as before,
$F_i(h)$ by $\ln G_i(h)=H(h)+F_i(h)$.  Along with the boundary
conditions provided by the coefficients of the negative powers of $G$
in \hGexp, the system of equations \grav\ completely determines the solution
to this problem.

\newsec{Conclusions and outlook}

In the present work we have demonstrated that our technique of
character expansions for large $N$ matrix models may be successfully
applied to the study of a novel, up to now inaccessible phase of almost
regular planar diagrams. This required determining -- quite generally --
the large $N$ limit of Weyl characters through the functional equation
\HGprod. Specializing to almost flat graphs, we have then
found the exact generating function \hGexp, \hofG\
of planar square lattices endowed with a single negative curvature insertion
balanced by a number of positive defects.

We feel that our observations
could trigger the investigation of many new phenomena in two-dimensional
physics and the combinatorial theory of planar graphs.
However, most urgent is the understanding of the crossover phenomenon from
the phase of almost flat two-dimensional space to the phase of
two-dimensional quantum gravity. It requires the careful analysis
of the well-posed Cauchy-Riemann problem of the last section.
This investigation is pending. Aside its obvious mathematical interest,
the solution of this problem could help to solve
the hitherto inaccessible problem of $R^2$ quantum gravity in two dimensions.

In addition to the even lattices considered in this paper, our methods allow
the study of the ``melting'' of more general regular, or almost
regular, lattices; e.g.~triangular lattices.

It is well known that there are many intriguing relations between integrable
two-dimensional models on regular lattices and dynamical planar
random lattices. It is tempting to try to unify the two classes of
models, a project one might term $GUT_2$. Our work should be considered
a first attempt into this direction, even though it must be noted
that further methods will have to be developed in order to successfully
treat matter coupled to dually weighted graphs.

Some of the results presented above could be interpreted as insights into
the structure of the group $SU(\infty)$ (see section 3 on
the large $N$ limit of Weyl characters.). Further insights into this
direction might prove very useful for the treatment of higher
dimensional matrix models, e.g.~the principal chiral field, discrete
string theories in physical dimensions and, one hopes, QCD.
\vskip 30pt
\hskip -20pt {\bf Acknowledgements}

We would like to thank E.~Br\'ezin and I. Kostov for
useful discussions.

\appendix{A}{Derivation of the inversion formula}

We start by proving that the constant coefficient in \hG\ is equal to
1 (the normalization of the density).  To correctly normalize the
density, $\rho(h)$, we have to ensure that
\eqn\rhonorm{
1-b=\int_b^a\,dh\rho(h).}
Using the fact that $\ln G(h)=H(h)+F(h)$, we replace the integral
by the contour integral
\eqn\rhocont{
1-b=\oint_{C_H}\,{dh\over 2\pi i}\ln G(h),}
with the contour $C_h$ encircling the $[b,a]$ part of the cut of
$H(h)$, as shown in Fig. 7.  The zig-zag line corresponds to the
logarithmic cut starting at $h=b$. Note that this is not a closed contour
since at $b$ there is a discontinuity across the cut of $\pm i\pi$.
Evaluating $G(h)$ around this contour we see that its argument goes
from $+i\pi$ at $h=b$ (below the cut) all the way around to $-i\pi$ at
$h=b$ (above the cut). We now change integration variables from $h$ to
$G$, with (in light of the comment above) the contour $C_G$ in the complex
$G$ plane encircling the origin (see Fig. 7):
\eqn\rhocond{
1-b=-\oint_{C_G}\,{dG\over 2\pi i}h'(G)\ln G(h)=
         -\oint_{C_G}\,{dG\over 2\pi i}
    \bigl[{\partial\over\partial G}(h(G)\ln G)-{h(G)\over G}\bigr],}
where $h(G)$ is defined through \hG . The contour starts and finishes
on either side of the cut generated by $\ln G$ illustrated in Fig.7 by
a zig-zag line.  The total derivative term picks up the discontinuity
across the cut giving $-b$. The final term, which picks up the
constant coefficient of $h(G)$, is thus equal to $1$.
\vskip 15pt
\hskip 10pt
\beginpicture
\setcoordinatesystem units <1.00000cm,1.00000cm>
\linethickness=1pt
\plot	 1.916 25.478  1.946 25.523
 	 1.977 25.545
	 2.008 25.523
	 2.038 25.478
	 2.069 25.433
	 2.099 25.410
	 2.129 25.433
	 2.159 25.478
	 2.190 25.523
	 2.220 25.545
	 2.251 25.523
	 2.282 25.478
	 2.311 25.433
	 2.341 25.410
	 2.371 25.433
	 2.401 25.478
	 2.432 25.523
	 2.463 25.545
	 2.493 25.523
	 2.523 25.478
	 2.554 25.433
	 2.584 25.410
	 2.615 25.433
	 2.646 25.478
	 2.675 25.523
	 2.705 25.545
	 2.736 25.523
	 2.766 25.478
	 2.796 25.433
	 2.826 25.410
	 2.857 25.433
	 2.889 25.478
	 2.919 25.523
	 2.949 25.545
	 2.979 25.523
	 3.010 25.478
	 3.040 25.433
	 3.069 25.410
	 3.100 25.433
	 /
\plot  3.100 25.433  3.131 25.478 /
\plot  0.572 25.470  0.637 25.569 /
\plot  0.637 25.569  0.773 25.368 /
\plot  0.773 25.368  0.906 25.569 /
\plot  0.906 25.569  1.039 25.368 /
\plot  1.039 25.368  1.175 25.569 /
\plot  1.175 25.569  1.308 25.368 /
\plot  1.308 25.368  1.441 25.569 /
\plot  1.441 25.569  1.575 25.368 /
\plot  1.575 25.368  1.708 25.569 /
\plot  1.708 25.569  1.844 25.368 /
\plot  1.844 25.368  1.909 25.470 /
\plot  2.534 25.262  2.623 25.174 /
\plot  2.534 25.078  2.623 25.167 /
\plot  5.558 25.470  5.624 25.569 /
\plot  5.624 25.569  5.759 25.368 /
\plot  5.759 25.368  5.893 25.569 /
\plot  5.893 25.569  6.026 25.368 /
\plot  6.026 25.368  6.159 25.569 /
\plot  6.159 25.569  6.293 25.368 /
\plot  6.293 25.368  6.428 25.569 /
\plot  6.428 25.569  6.562 25.368 /
\plot  6.562 25.368  6.695 25.569 /
\plot  6.695 25.569  6.828 25.368 /
\plot  6.828 25.368  6.896 25.470 /
\plot  5.656 25.463  8.149 25.463 /
\plot  6.902 25.567  6.902 25.360 /
\plot  9.610 25.470  9.675 25.569 /
\plot  9.675 25.569  9.811 25.368 /
\plot  9.811 25.368  9.944 25.569 /
\plot  9.944 25.569 10.077 25.368 /
\plot 10.077 25.368 10.211 25.569 /
\plot 10.211 25.569 10.344 25.368 /
\plot 10.344 25.368 10.480 25.569 /
\plot 10.480 25.569 10.613 25.368 /
\plot 10.613 25.368 10.746 25.569 /
\plot 10.746 25.569 10.880 25.368 /
\plot 10.880 25.368 10.947 25.470 /
\plot  9.707 25.463 12.200 25.463 /
\plot 10.954 25.567 10.954 25.360 /
\plot  6.856 24.746  6.945 24.657 /
\plot  6.856 24.562  6.945 24.649 /
\plot 10.977 25.358 11.098 25.328 /
\plot 11.070 25.199 11.102 25.320 /
\plot 10.207 25.614 10.861 25.614 /
\plot 10.207 25.320 10.861 25.320 /
\plot 10.867 25.324 	10.884 25.307
	10.933 25.286
	10.983 25.284
	11.032 25.288
	11.074 25.303
	11.113 25.334
	11.138 25.364
	11.159 25.400
	11.172 25.434
	11.176 25.476
	11.172 25.514
	11.163 25.548
	11.142 25.582
	11.115 25.620
	11.074 25.646
	11.032 25.662
	10.983 25.669
	10.937 25.665
	10.888 25.646
	10.854 25.624
	/
\plot  0.671 25.463  3.994 25.463 /
\plot  1.916 25.567  1.916 25.360 /
\plot  3.143 25.567  3.143 25.360 /
\plot  1.922 25.339 	 2.026 25.250
	 2.142 25.216
	 2.208 25.202
	 2.279 25.190
	 2.352 25.180
	 2.429 25.173
	 2.507 25.168
	 2.587 25.165
	 2.666 25.165
	 2.744 25.168
	 2.821 25.174
	 2.896 25.182
	 2.967 25.193
	 3.034 25.208
	 3.152 25.245
	 3.266 25.366
	 3.283 25.470
	 3.262 25.588
	 3.162 25.682
	 3.044 25.719
	 2.977 25.735
	 2.906 25.748
	 2.831 25.758
	 2.753 25.766
	 2.674 25.771
	 2.594 25.773
	 2.513 25.773
	 2.434 25.770
	 2.356 25.765
	 2.281 25.756
	 2.209 25.744
	 2.141 25.730
	 2.021 25.692
	 1.922 25.595
	/
\plot  6.104 25.351 	 6.124 25.242
	 6.140 25.184
	 6.181 25.097
	 6.227 25.013
	 6.279 24.938
	 6.335 24.867
	 6.417 24.806
	 6.502 24.752
	 6.599 24.706
	 6.701 24.670
	 6.807 24.657
	 6.913 24.653
	 7.018 24.660
	 7.120 24.674
	 7.208 24.704
	 7.292 24.742
	 7.376 24.801
	 7.453 24.867
	 7.508 24.931
	 7.559 24.998
	 7.604 25.069
	 7.645 25.142
	 7.674 25.216
	 7.698 25.292
	 7.711 25.378
	 7.717 25.463
	 7.710 25.542
	 7.698 25.620
	 7.684 25.690
	 7.667 25.760
	 7.627 25.832
	 7.582 25.900
	 7.523 25.981
	 7.457 26.056
	 7.380 26.114
	 7.298 26.164
	 7.211 26.208
	 7.120 26.245
	 7.018 26.261
	 6.913 26.268
	 6.817 26.262
	 6.720 26.249
	 6.620 26.213
	 6.524 26.170
	 6.449 26.127
	 6.378 26.077
	 6.259 25.952
	 6.209 25.874
	 6.166 25.792
	 6.142 25.727
	 6.104 25.603
	/
\put{$b$} [lB] at  1.839 25.796
\put{$a$} [lB] at  3.040 25.796
\put{0} [lB] at  6.841 25.095
\put{$G(b)$} [lB] at  9.967 25.745
\put{$C_G$} [lB] at  6.475 24.130
\put{$C_h$} [lB] at  2.214 24.672
\put{$G(b)$} [lB] at  5.425 25.944
\put{$C_0$} [lB] at 10.651 24.801
\linethickness=0pt
\putrectangle corners at  0.546 26.285 and 12.226 24.000
\endpicture
\vskip 15pt
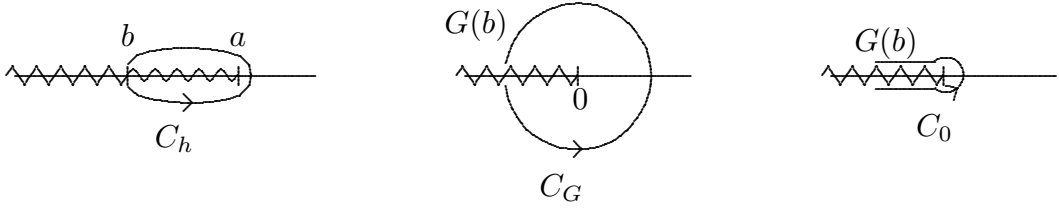
\centerline{{\bf Fig. 7:} Definition of contours $C_h$ in the complex
$h$ plane and}
\centerline{ $C_G$ and $C_0$ in the complex $G$ plane.}
\vskip 20pt

We now complete the derivation of \HGprod .
As discussed in section 3. we start by generating $\tilde H(h)$
(related to the full resolvent by $H(h)= \tilde H(h)+\ln{h\over h-b}$ )
from the contour integral
\eqn\Hinva{
\tilde H(h)=\oint_{C_h}\,{dh_1\over 2\pi i}{\ln G(h_1)\over h-h_1}.}
Changing integration variables from $h$ to $G$, as above, this can be
written as
\eqn\Hinvb{
\tilde H(h)=-\oint_{C_G}\,{dG\over 2\pi i}\ln G {h'(G) \over h-h(G)},}
where $h(G)$ is defined through \hG\ and $h'(G)$ is the derivative
with respect to $G$.

We now simplify this contour integral by evaluating it for large
$h$. Knowing the solution in any neighbourhood of $h$ means that, by
analytic continuation, we know it everywhere. For large enough $h$, we
see from \hG\ that the contour in \Hinvb\ will encircle precisely $Q$
zeros of $h(G)$, the zeros corresponding to the inverse powers of
$G$. If we shrink the contour in \Hinvb\ so that the contour hugs
either side of the cut (see Fig. 7, contour $C_0$) we pick up these
$Q$ poles:
\eqn\Hinvc{
\tilde H(h)=\sum_{q=1}^Q\ln G_q(h)
-\oint_{C_0}\,{dG\over 2\pi i}\ln G {h'(G) \over h-h(G)}.}
The remaining contour integral is relatively easy to evaluate provided
careful attention is paid to the
contribution coming from encircling the origin. The net result is that
the contour $C_0$ contributes $\ln\bigl((-1)^{Q-1}t_{2Q}\bigr)$ from
encircling the origin and $\ln (h-b)$ from the discontinuity across
the end points of the contour. Putting these results together and
making use of the relationship between $H(h)$ and $\tilde H(h)$ we arrive
at \HGprod.

\appendix{B}{Analytic structure of $G(h)$}

As is discussed in Appendix A, the sheets $G_q(h)$ in the product of
\HGprod\ are the physical sheet and all the sheets attached to the
physical sheet by the cut of $e^{H(h)}$.
To clarify this we provide some simple examples.

\subsec{Example 1. $V_B(MA)=0$}

In this simplest case it is immediate from eq.\hGexp\ (since
$B=0$ and thus $\psi(G)=0$) that
\eqn\hGzero{
h-1=\sum_{q=1}^Q{t_{2q}\over G^q}.}
This is a polynomial equation of degree $Q$. $G(h)$ will thus be a
multivalued analytic function with $Q$ sheets. The different sheets
are connected by square root cuts, represented in Fig. 8
below by the vertical walls.
\vskip 20pt
\beginpicture
\setcoordinatesystem units <1.00000cm,1.00000cm>
\linethickness=1pt
\plot  5.080 19.571 10.207 18.860 /
\plot 10.221 18.860 11.345 19.571 /
\plot  5.080 19.571  6.505 20.142 /
\plot  5.080 20.426 10.207 19.814 /
\plot  5.080 21.281 10.207 20.726 /
\plot 10.207 19.799 11.345 20.426 /
\plot 10.207 20.726 11.345 21.281 /
\plot  5.080 21.281  6.505 21.736 /
\plot  7.074 21.380  7.074 21.139 /
\plot  7.074 20.968  7.074 20.540 /
\plot  9.495 20.269  9.495 19.986 /
\plot  9.495 19.799  9.495 19.372 /
\plot  5.806 21.510  7.074 21.380 /
\plot  5.791 20.682  7.074 20.540 /
\plot  9.495 20.269 10.791 20.127 /
\plot  9.495 19.387 10.791 19.215 /
\plot  6.932 20.983  6.932 20.555 /
\plot  6.788 20.995  6.788 20.570 /
\plot  6.646 20.995  6.646 20.582 /
\plot  6.505 21.010  6.505 20.597 /
\plot  6.363 21.025  6.363 20.625 /
\plot  6.219 21.040  6.219 20.640 /
\plot  6.077 21.052  6.077 20.654 /
\plot  5.935 21.067  5.935 20.712 /
\plot  5.791 21.040  5.791 20.769 /
\plot  5.649 20.995  5.649 20.826 /
\plot  9.637 19.757  9.637 19.359 /
\plot  9.779 19.757  9.779 19.344 /
\plot  9.923 19.742  9.923 19.329 /
\plot 10.065 19.715 10.065 19.315 /
\plot 10.207 19.715 10.207 19.287 /
\plot 10.348 19.729 10.348 19.272 /
\plot 10.492 19.772 10.492 19.257 /
\plot 10.634 19.829 10.634 19.245 /
\plot 10.776 19.886 10.776 19.287 /
\plot 10.918 19.842 10.918 19.344 /
\plot 11.062 19.772 11.062 19.486 /
\plot  6.505 20.142  9.409 19.799 /
\plot 11.345 19.571 11.189 19.571 /
\plot 11.345 20.426  7.231 20.868 /
\plot  9.565 19.784  9.565 19.784 /
\plot  9.722 19.757  9.722 19.757 /
\plot  9.851 19.742  9.851 19.742 /
\plot  9.993 19.729  9.993 19.729 /
\plot 10.135 19.715 10.135 19.715 /
\plot 10.279 19.685 10.279 19.685 /
\plot 10.420 19.685 10.420 19.685 /
\plot 10.562 19.657 10.562 19.657 /
\plot 10.706 19.643 10.706 19.643 /
\plot 10.990 19.615 10.990 19.615 /
\plot  5.863 20.712  5.863 20.712 /
\plot  6.005 20.769  6.005 20.769 /
\plot  6.149 20.811  6.149 20.811 /
\plot  6.291 20.868  6.291 20.868 /
\plot  6.433 20.925  6.433 20.925 /
\plot  6.574 20.940  6.574 20.940 /
\plot  6.718 20.925  6.718 20.925 /
\plot  6.860 20.896  6.860 20.896 /
\plot  7.002 20.883  7.002 20.883 /
\plot  5.080 20.426  5.791 20.682 /
\plot 11.015 21.046 10.761 20.983 11.015 20.919 /
\plot 10.761 20.983 12.357 20.983 /
\plot  9.017 21.512 11.335 21.289 /
\plot  6.477 21.717  8.668 21.526 /
\plot  8.802 21.567  8.651 21.353  8.885 21.471 /
\plot  8.651 21.353  9.938 22.464 /
\setdashes < 0.0677cm>
\plot  9.779 20.170  9.779 19.941 /
\plot 10.065 20.127 10.065 19.928 /
\plot 10.336 20.085 10.336 19.971 /
\plot  6.788 21.311  6.788 21.196 /
\plot  6.505 21.338  6.505 21.224 /
\plot  6.219 21.366  6.219 21.253 /
\plot 10.848 19.628 10.848 19.628 /
\plot  5.935 21.380  5.935 21.296 /
\setsolid
\put{$\bullet$} [1B] at 8.492 21.200
\put{physical sheet} [lB] at 12.700 20.892
\put{$\biggr\lmoustache$} [lB] at  4.255 20.733
\put{$\biggr\rmoustache$} [lB] at  4.255 19.907
\put{$G_q$} [lB] at  3.524 20.288
\put{pole at $h=1$} [lB] at  9.224 22.782
\linethickness=0pt
\putrectangle corners at  3.524 23.112 and 15.043 18.834
\endpicture
\vskip 20pt
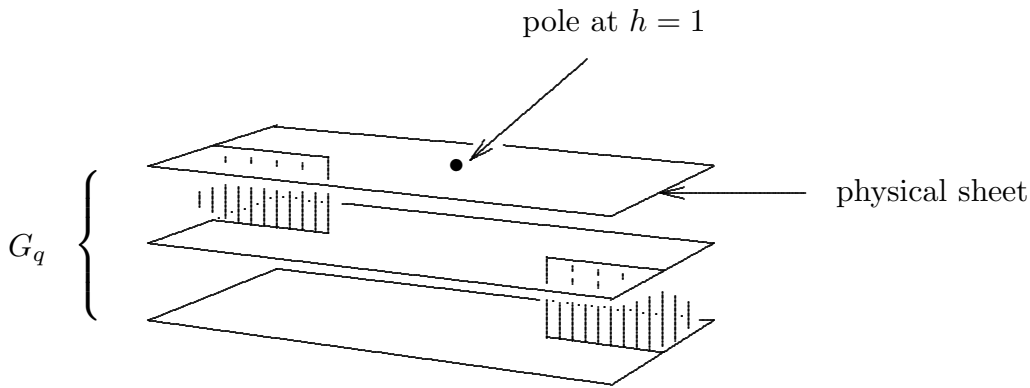
\centerline{{\bf Fig. 8:} Analytic structure of $G(h)$ for
$V_B(MA)=0$}
\vskip 20pt
\hskip -20pt
Note that cube roots and higher order roots are just special cases of
the above structure. For example, a cube root in the diagram above is
generated when the two square root cut points touch.

The $G_q(h)$ that enter the
product in \HGprod\ are precisely all the solutions, i.e.~all the
sheets. It then follows that
\eqn\eHzero{
e^{H(h)}={h\over h-1},}
which corresponds to a completely flat density $\rho(h)=1$ with
support $[0,1]$.

It is seen from equation \hGzero\ that at $h=1$, $G(h)$ becomes infinite
on one of its
sheets, so there is a pole at $h=1$ on what we call the physical
sheet.  For $\psi(G)$ non zero, the positive powers of $G$ ``soften'' this
pole and stretch it into a cut. The cut corresponds to exciting
boxes in the Young tableau. The next example illustrates this.

\subsec{Example 2. $V_B(MA)=MA$}

Here $B=A_1$.  It follows from eq.\hGexp\ and a simple
diagrammatic inspection that
\eqn\hGzero{
h-1=\sum_{q=1}^Q{t_{2q}\over G^q}\quad + \quad G}
This increases the degree of the polynomial by one from the
previous example, introducing an extra sheet. The pole that was at
$h=1$ has now opened into a cut (the cut of $e^{H(h)}$) connected
to this extra sheet (see Fig. 9).
\vskip 20pt
\beginpicture
\setcoordinatesystem units <1.00000cm,1.00000cm>
\linethickness=1pt
\plot  9.260 21.307  9.017 21.209  9.277 21.182 /
\plot  9.017 21.209 12.509 21.702 /
\plot  5.080 19.571 10.207 18.860 /
\plot 10.221 18.860 11.345 19.571 /
\plot  5.080 19.571  6.505 20.142 /
\plot  5.080 20.426 10.207 19.814 /
\plot  5.080 21.281 10.207 20.726 /
\plot  5.080 22.136 10.207 21.637 /
\plot 10.207 19.799 11.345 20.426 /
\plot 10.207 20.726 11.345 21.281 /
\plot 10.207 21.637 11.345 22.136 /
\plot  5.080 21.281  6.505 21.736 /
\plot  7.785 22.151  7.785 21.937 /
\plot  7.785 21.780  7.785 21.311 /
\plot  8.926 22.049  8.926 21.838 /
\plot  8.926 21.679  8.926 21.196 /
\plot  7.074 21.380  7.074 21.139 /
\plot  7.074 20.968  7.074 20.540 /
\plot  9.495 20.269  9.495 19.986 /
\plot  9.495 19.799  9.495 19.372 /
\plot  7.785 22.151  8.926 22.049 /
\plot  7.785 21.311  8.911 21.196 /
\plot  5.806 21.510  7.074 21.380 /
\plot  5.791 20.682  7.074 20.540 /
\plot  9.495 20.269 10.791 20.127 /
\plot  9.495 19.387 10.791 19.215 /
\plot  7.929 21.736  7.929 21.296 /
\plot  8.071 21.709  8.071 21.281 /
\plot  8.213 21.709  8.213 21.266 /
\plot  8.354 21.694  8.354 21.253 /
\plot  8.498 21.666  8.498 21.239 /
\plot  8.640 21.637  8.640 21.224 /
\plot  8.782 21.637  8.782 21.209 /
\plot  6.932 20.983  6.932 20.555 /
\plot  6.788 20.995  6.788 20.570 /
\plot  6.646 20.995  6.646 20.582 /
\plot  6.505 21.010  6.505 20.597 /
\plot  6.363 21.025  6.363 20.625 /
\plot  6.219 21.040  6.219 20.640 /
\plot  6.077 21.052  6.077 20.654 /
\plot  5.935 21.067  5.935 20.712 /
\plot  5.791 21.040  5.791 20.769 /
\plot  5.649 20.995  5.649 20.826 /
\plot  9.637 19.757  9.637 19.359 /
\plot  9.779 19.757  9.779 19.344 /
\plot  9.923 19.742  9.923 19.329 /
\plot 10.065 19.715 10.065 19.315 /
\plot 10.207 19.715 10.207 19.287 /
\plot 10.348 19.729 10.348 19.272 /
\plot 10.492 19.772 10.492 19.257 /
\plot 10.634 19.829 10.634 19.245 /
\plot 10.776 19.886 10.776 19.287 /
\plot 10.918 19.842 10.918 19.344 /
\plot 11.062 19.772 11.062 19.486 /
\plot  6.505 20.142  9.409 19.799 /
\plot 11.345 19.571 11.189 19.571 /
\plot 11.345 20.426  7.231 20.868 /
\plot  6.505 21.736  7.643 21.637 /
\plot  9.565 19.784  9.565 19.784 /
\plot  9.722 19.757  9.722 19.757 /
\plot  9.851 19.742  9.851 19.742 /
\plot  9.993 19.729  9.993 19.729 /
\plot 10.135 19.715 10.135 19.715 /
\plot 10.279 19.685 10.279 19.685 /
\plot 10.420 19.685 10.420 19.685 /
\plot 10.562 19.657 10.562 19.657 /
\plot 10.706 19.643 10.706 19.643 /
\plot 10.990 19.615 10.990 19.615 /
\plot  5.863 20.712  5.863 20.712 /
\plot  6.005 20.769  6.005 20.769 /
\plot  6.149 20.811  6.149 20.811 /
\plot  6.291 20.868  6.291 20.868 /
\plot  6.433 20.925  6.433 20.925 /
\plot  6.574 20.940  6.574 20.940 /
\plot  6.718 20.925  6.718 20.925 /
\plot  6.860 20.896  6.860 20.896 /
\plot  7.002 20.883  7.002 20.883 /
\plot  7.857 21.609  7.857 21.609 /
\plot  7.999 21.579  7.999 21.579 /
\plot  8.143 21.567  8.143 21.567 /
\plot  8.285 21.567  8.285 21.567 /
\plot  8.426 21.552  8.426 21.552 /
\plot  8.568 21.552  8.568 21.552 /
\plot  8.712 21.537  8.712 21.537 /
\plot  8.854 21.510  8.854 21.510 /
\plot  5.080 20.426  5.791 20.682 /
\plot  5.112 22.130  6.477 22.543 /
\plot  6.460 22.540 11.350 22.128 /
\plot 11.153 21.050 10.899 20.987 11.153 20.923 /
\plot 10.899 20.987 12.495 20.987 /
\plot 11.689 22.291 11.462 22.162 11.723 22.169 /
\plot 11.462 22.162 12.541 22.464 /
\plot  9.081 21.495  9.923 21.416 /
\plot 10.668 21.353 11.352 21.289 /
\setdashes < 0.0677cm>
\plot  9.779 20.170  9.779 19.941 /
\plot 10.065 20.127 10.065 19.928 /
\plot 10.336 20.085 10.336 19.971 /
\plot  8.071 22.037  8.071 21.937 /
\plot  8.354 22.022  8.354 21.937 /
\plot  8.640 21.979  8.640 21.865 /
\plot  6.788 21.311  6.788 21.196 /
\plot  6.505 21.338  6.505 21.224 /
\plot  6.219 21.366  6.219 21.253 /
\plot 10.848 19.628 10.848 19.628 /
\plot  5.935 21.380  5.935 21.296 /
\setsolid
\put{$\biggr\lmoustache$} [lB] at  4.255 20.733
\put{$\biggr\rmoustache$} [lB] at  4.255 19.907
\put{$G_q$} [lB] at  3.524 20.288
\put{physical sheet} [lB] at 12.810 20.906
\put{cut of $e^{H(h)}$} [lB] at 12.795 21.702
\put{sheet of $he^{-H^(h)}$} [lB] at 12.829 22.464
\linethickness=0pt
\putrectangle corners at  3.524 23.074 and 17.302 18.834
\endpicture
\vskip 20pt
\centerline{{\bf Fig. 9:} Analytic structure of $G(h)$ for
$V_B(MA)=MA$}
\vskip 20pt
The $G_q(h)$ that go into the product of eq.\HGprod\ are the physical
sheet and all the sheets below. We thus obtain
\eqn\eHone{
e^{H(h)}={h\over G^-(h)},}
where $G^-(h)$ is the topmost sheet.

\subsec{Example 3. $V_B(MA)=(MA)^2$}
By inspecting the moments of the dual model, we obtain
\eqn\hGtwo{
h-1=\sum_{q=1}^Q t_{2q}~\bigl({1\over G^q}+G^q\bigr).}
The sheet structure is still polynomial, but now, due to the symmetry
$G \rightarrow G^{-1}$ of
equation \hGtwo, the top sheets are the mirror image inverses of the
bottom sheets.
\vskip 20pt
\beginpicture
\setcoordinatesystem units <1.00000cm,1.00000cm>
\linethickness=1pt
\plot  5.080 19.571 10.207 18.860 /
\plot 10.221 18.860 11.345 19.571 /
\plot  5.080 23.844  6.505 24.130 /
\plot  6.505 24.130 11.345 23.844 /
\plot  5.080 23.844 10.221 23.489 /
\plot 10.207 23.489 11.345 23.844 /
\plot  5.080 20.426 10.207 19.814 /
\plot  5.080 21.281 10.207 20.726 /
\plot  5.080 22.136 10.207 21.637 /
\plot  5.080 22.991 10.207 22.606 /
\plot 10.207 19.799 11.345 20.426 /
\plot 10.207 20.726 11.345 21.281 /
\plot 10.207 21.637 11.345 22.136 /
\plot 10.207 22.606 11.345 22.991 /
\plot  5.080 21.281  6.505 21.736 /
\plot  5.080 22.991  6.505 23.332 /
\plot  7.785 22.151  7.785 21.937 /
\plot  7.785 21.780  7.785 21.311 /
\plot  8.926 22.049  8.926 21.838 /
\plot  8.926 21.679  8.926 21.196 /
\plot  7.074 21.380  7.074 21.139 /
\plot  7.074 23.061  7.074 22.904 /
\plot  7.074 22.748  7.074 22.221 /
\plot  9.495 20.269  9.495 19.986 /
\plot  9.495 19.799  9.495 19.372 /
\plot  9.495 23.760  9.495 23.618 /
\plot  9.495 23.419  9.495 22.904 /
\plot  9.495 23.760 10.776 23.675 /
\plot  9.495 22.904 10.776 22.805 /
\plot  5.806 23.161  7.074 23.076 /
\plot  5.806 22.335  7.074 22.221 /
\plot  7.785 22.151  8.926 22.049 /
\plot  7.785 21.311  8.911 21.196 /
\plot  5.806 21.510  7.074 21.380 /
\plot  9.495 20.269 10.791 20.127 /
\plot  9.495 19.387 10.791 19.215 /
\plot  7.929 21.736  7.929 21.296 /
\plot  8.071 21.709  8.071 21.281 /
\plot  8.213 21.709  8.213 21.266 /
\plot  8.354 21.694  8.354 21.253 /
\plot  8.498 21.666  8.498 21.239 /
\plot  8.640 21.637  8.640 21.224 /
\plot  8.782 21.637  8.782 21.209 /
\plot  9.637 23.404  9.637 22.892 /
\plot  9.779 23.404  9.779 22.892 /
\plot  9.923 23.389  9.923 22.877 /
\plot 10.065 23.389 10.065 22.862 /
\plot 10.207 23.374 10.207 22.847 /
\plot 10.348 23.419 10.348 22.835 /
\plot 10.492 23.431 10.492 22.835 /
\plot 10.634 23.461 10.634 22.820 /
\plot 10.776 23.503 10.776 22.847 /
\plot 10.918 23.489 10.918 22.877 /
\plot 11.062 23.419 11.062 23.004 /
\plot  6.932 22.748  6.932 22.236 /
\plot  6.788 22.763  6.788 22.250 /
\plot  6.646 22.763  6.646 22.263 /
\plot  6.505 22.763  6.505 22.278 /
\plot  5.935 22.790  5.935 22.350 /
\plot  5.791 22.805  5.791 22.449 /
\plot  5.649 22.805  5.649 22.507 /
\plot  9.637 19.757  9.637 19.359 /
\plot  9.779 19.757  9.779 19.344 /
\plot  9.923 19.742  9.923 19.329 /
\plot 10.065 19.715 10.065 19.315 /
\plot 10.207 19.715 10.207 19.287 /
\plot 10.348 19.729 10.348 19.272 /
\plot 10.492 19.772 10.492 19.257 /
\plot 10.634 19.829 10.634 19.245 /
\plot 10.776 19.886 10.776 19.287 /
\plot 10.918 19.842 10.918 19.344 /
\plot 11.062 19.772 11.062 19.486 /
\plot  6.505 20.142  9.409 19.799 /
\plot 11.345 19.571 11.189 19.571 /
\plot 11.345 20.426  7.231 20.868 /
\plot  6.505 21.736  7.643 21.637 /
\plot 11.345 22.136  7.216 22.477 /
\plot  6.505 23.332  9.366 23.133 /
\plot 11.360 22.991 11.204 23.004 /
\plot  9.565 19.784  9.565 19.784 /
\plot  9.722 19.757  9.722 19.757 /
\plot  9.851 19.742  9.851 19.742 /
\plot  9.993 19.729  9.993 19.729 /
\plot 10.135 19.715 10.135 19.715 /
\plot 10.279 19.685 10.279 19.685 /
\plot 10.420 19.685 10.420 19.685 /
\plot 10.562 19.657 10.562 19.657 /
\plot 10.706 19.643 10.706 19.643 /
\plot 10.990 19.615 10.990 19.615 /
\plot  7.857 21.609  7.857 21.609 /
\plot  7.999 21.579  7.999 21.579 /
\plot  8.143 21.567  8.143 21.567 /
\plot  8.285 21.567  8.285 21.567 /
\plot  8.426 21.552  8.426 21.552 /
\plot  8.568 21.552  8.568 21.552 /
\plot  8.712 21.537  8.712 21.537 /
\plot  8.854 21.510  8.854 21.510 /
\plot  5.863 22.365  5.863 22.365 /
\plot  6.005 22.392  6.005 22.392 /
\plot  6.149 22.435  6.149 22.435 /
\plot  6.433 22.521  6.433 22.521 /
\plot  6.574 22.521  6.574 22.521 /
\plot  6.718 22.521  6.718 22.521 /
\plot  6.860 22.507  6.860 22.507 /
\plot  7.002 22.492  7.002 22.492 /
\plot  9.565 23.118  9.565 23.118 /
\plot  9.709 23.103  9.709 23.103 /
\plot  9.851 23.091  9.851 23.091 /
\plot  9.993 23.091  9.993 23.091 /
\plot 10.135 23.076 10.135 23.076 /
\plot 10.279 23.061 10.279 23.061 /
\plot 10.420 23.061 10.420 23.061 /
\plot 10.562 23.048 10.562 23.048 /
\plot 10.706 23.034 10.706 23.034 /
\plot 10.848 23.034 10.848 23.034 /
\plot 10.990 23.019 10.990 23.019 /
\plot  6.077 22.777  6.077 22.320 /
\plot  6.291 22.477  6.291 22.477 /
\plot  6.219 22.777  6.219 22.293 /
\plot  6.363 22.777  6.363 22.278 /
\plot  5.080 22.136  5.791 22.335 /
\plot  5.080 19.571  6.505 20.142 /
\plot  7.074 20.968  7.074 20.540 /
\plot  5.791 20.682  7.074 20.540 /
\plot  6.932 20.983  6.932 20.555 /
\plot  6.788 20.995  6.788 20.570 /
\plot  6.646 20.995  6.646 20.582 /
\plot  6.505 21.010  6.505 20.597 /
\plot  6.363 21.025  6.363 20.625 /
\plot  6.219 21.040  6.219 20.640 /
\plot  6.077 21.052  6.077 20.654 /
\plot  5.935 21.067  5.935 20.712 /
\plot  5.791 21.040  5.791 20.769 /
\plot  5.649 20.995  5.649 20.826 /
\plot  5.863 20.712  5.863 20.712 /
\plot  6.005 20.769  6.005 20.769 /
\plot  6.149 20.811  6.149 20.811 /
\plot  6.291 20.868  6.291 20.868 /
\plot  6.433 20.925  6.433 20.925 /
\plot  6.574 20.940  6.574 20.940 /
\plot  6.718 20.925  6.718 20.925 /
\plot  6.860 20.896  6.860 20.896 /
\plot  7.002 20.883  7.002 20.883 /
\plot  5.080 20.426  5.791 20.682 /
\plot  9.259 21.342  9.017 21.241  9.278 21.216 /
\plot  9.017 21.241 12.368 21.749 /
\plot  9.091 21.484  9.832 21.421 /
\plot 10.382 21.368 11.345 21.283 /
\plot 11.184 21.050 10.930 20.987 11.184 20.923 /
\plot 10.930 20.987 12.526 20.987 /
\setdashes < 0.0677cm>
\plot  9.779 20.170  9.779 19.941 /
\plot 10.065 20.127 10.065 19.928 /
\plot 10.336 20.085 10.336 19.971 /
\plot  8.071 22.037  8.071 21.937 /
\plot  8.354 22.022  8.354 21.937 /
\plot  8.640 21.979  8.640 21.865 /
\plot  6.788 21.311  6.788 21.196 /
\plot  6.505 21.338  6.505 21.224 /
\plot  6.219 21.366  6.219 21.253 /
\plot  6.788 23.004  6.788 22.947 /
\plot  6.505 23.034  6.505 22.962 /
\plot  6.219 23.048  6.219 22.976 /
\plot  5.935 23.061  5.935 23.004 /
\plot  9.779 23.660  9.779 23.603 /
\plot 10.065 23.630 10.065 23.575 /
\plot 10.348 23.618 10.348 23.588 /
\plot 10.848 19.628 10.848 19.628 /
\plot  5.935 21.380  5.935 21.296 /
\setsolid
\put{$\biggr\lmoustache$} [lB] at  4.255 20.733
\put{$\biggr\rmoustache$} [lB] at  4.255 19.907
\put{$G_k$} [lB] at  3.524 20.288
\put{$\biggr\rmoustache$} [lB] at  4.255 22.479
\put{$\biggr\lmoustache$} [lB] at  4.255 23.305
\put{$G_q^{-1}$} [lB] at  3.524 22.860
\put{physical sheet} [lB] at 12.764 20.892
\put{cut of $e^{H(h)}$} [lB] at 12.732 21.685
\linethickness=0pt
\putrectangle corners at  3.524 24.155 and 12.764 18.834
\endpicture
\vskip 20pt
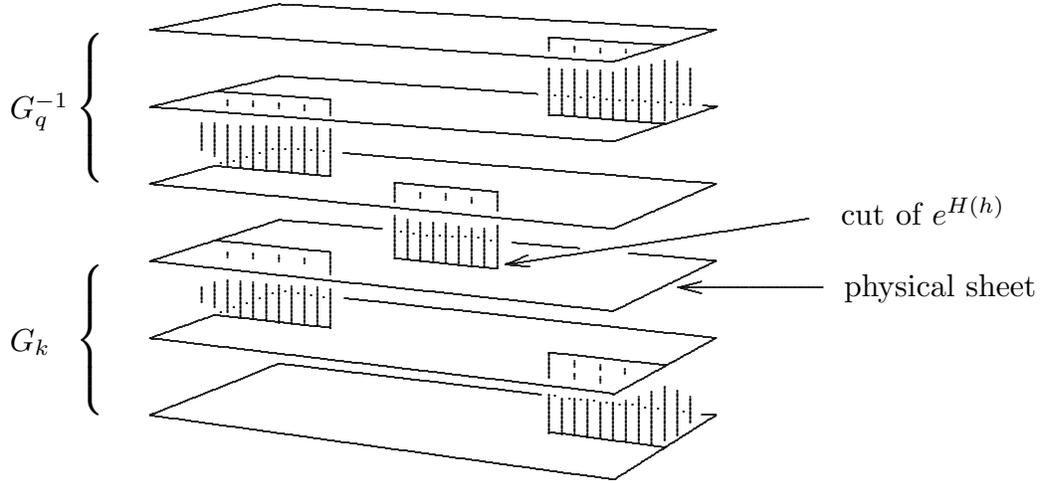
\centerline{{\bf Fig. 10:} Analytic structure of $G(h)$ for
$V_B(MA)=(MA)^2$}
\vskip 20pt
Again, what was a pole at h=1 has opened into a cut connecting
the physical sheet to the mirror image inverses of the bottom sheets.

The above three examples clarify the meaning of
equations \hGexp\ and \HGprod. A simple functional inversion
developed in \KSW\ allows us to relate $H(h)$ to the resolvent,
$\langle \Tr \bigl[{1\over P-M}\bigr]\rangle$, of the matrix model.
To verify the methods of section 3, we have directly
calculated the matrix resolvent of these models using loop equations
and simple diagrammatic arguments.

\listrefs
\end